\PassOptionsToPackage{pdfpagelabels=false}{hyperref} 
\documentclass[draft]{agujournal2019}

\usepackage{url}
\usepackage{lineno}
\usepackage{soul}
\usepackage{siunitx}
\usepackage{amsmath}
\usepackage{bm}
\usepackage{amsfonts}
\usepackage[algoruled,noend]{algorithm2e}
\usepackage{comment}
\numberwithin{equation}{section}

\DeclareMathOperator*{\argmin}{arg\,min}

\hypersetup{final}

\draftfalse
\journalname{JGR: Earth Surface}
\begin{document}

\title{Simulation-Based Inference of Surface Accumulation and Basal Melt Rates of an Antarctic Ice Shelf from Isochronal Layers}

\authors{Guy Moss\thanks{\tt{guy.moss@uni-tuebingen.de}}\affil{1},  Vjeran Višnjević\affil{2}, Olaf Eisen\affil{3,4}, Falk M. Oraschewski\affil{2}, Cornelius Schröder\affil{1}, Jakob H. Macke\thanks{Equal supervision}\affil{1,5}, Reinhard Drews$^\dagger$\affil{2}}

\affiliation{1}{Machine Learning in Science, University of Tübingen and Tübingen AI Center, Germany}
\affiliation{2}{Department of Geosciences, University of Tübingen, Germany}
\affiliation{3}{Alfred-Wegener-Institut, Helmholtz-Zentrum für Polar und Meeresforschung, Bremerhaven, Germany}
\affiliation{4}{Department of Geosciences, Universität Bremen, Germany}
\affiliation{5}{Max Planck Institute for Intelligent Systems, Tübingen, Germany}

\preprint

\begin{abstract}
    The ice shelves buttressing the Antarctic ice sheet determine the rate of ice-discharge into the surrounding oceans. The geometry of ice shelves, and hence their buttressing strength, is determined by ice flow as well as by the local surface accumulation and basal melt rates, governed by atmospheric and oceanic conditions. Contemporary methods resolve one of these rates, but typically not both. Moreover, there is little information of how they changed in time. We present a new method to simultaneously infer the surface accumulation and basal melt rates averaged over decadal and centennial timescales. We infer the spatial dependence of these rates along flow line transects using internal stratigraphy observed by radars, using a kinematic forward model of internal stratigraphy. We solve the inverse problem using simulation-based inference (SBI). SBI performs Bayesian inference by training neural networks on simulations of the forward model to approximate the posterior distribution, allowing us to also quantify uncertainties over the inferred parameters. We demonstrate the validity of our method on a synthetic example, and apply it to Ekström Ice Shelf, Antarctica, for which newly acquired radar measurements are available. We obtain posterior distributions of surface accumulation and basal melt averaging over 42, 84, 146, and 188 years before 2022. Our results suggest stable atmospheric and oceanographic conditions over this period in this catchment of Antarctica. Use of observed internal stratigraphy can separate the effects of surface accumulation and basal melt, allowing them to be interpreted in a historical context of the last centuries and beyond.
\end{abstract}

\section{Introduction} \label{sec: intro}
The majority of the Antarctic Ice Sheet is buttressed by floating ice shelves \cite{Bindschadler2011} which provide large contact areas for ice-ocean interactions. Approximately half of the ice shelves' total mass loss is attributed to ocean-induced melting at the underside of ice shelves \cite{Depoorter2013}, and its spatiotemporal variability imprints ice flow dynamics farther upstream \cite{Reese2017,Gudmundsson2019}. Consequently, ice flow and ocean models need to be coupled for future projections; frameworks \cite{Goldberg2019, Gladstone2021}, parameterizations \cite{Burgard2022, Goldberg2022}, and benchmarks \cite{Asay-Davis2016} for this task have been developed. Similarly, the local snow accumulation is influenced by atmospheric conditions and is crucial in determining ice shelf thickness \cite{Winkelmann2012}. As a result, ice flow models are also coupled to climate models for future projections \cite{Goezler2016, Pattyn2017}.  It is crucial to confront ice flow models with observations to validate them and investigate their ability to explain observed phenomena. Here, we present a new method that infers surface accumulation and basal melt rates (collectively, the mass balance parameters) from the ice shelves’ internal stratigraphy, which can be routinely mapped by radio-echo sounding. 

Previous studies provided much progress in deriving the mass balance parameters from observations. Typically, accumulation is more accessible \cite{Eisen2008Review}; it is measured in-situ using stake farms, and is also derived from multiple firn cores \cite{Lenaerts2019}. Many of these observations validate atmospheric models such as RACMO \cite{VanWessem2018} and MAR \cite{Gallee1994,Agosta2019}, which estimate surface accumulation on 35~\unit{km} grids \cite{Lenaerts2019} (with few locations being estimated at a higher resolution of 5.5~\unit{km}). Estimating the basal melt is more challenging, and is typically dependent on knowledge of surface accumulation. For example, estimates of surface accumulation have been used along with mass conservation arguments to estimate basal melt \cite{Neckel2012, Depoorter2013, Berger2017, Adusumilli2020}. These approaches have provided Antarctic-wide time series of the last few decades of basal melt rates \cite{Adusumilli2020}. The spatial resolution is currently limited to the kilometer scale, which may miss fine grained processes occurring within ice shelf channels \cite{Drews2015channels,Marsh2016} or near basal terraces \cite{Dutrieux2014}. Independent estimates of basal melt are also available, but typically only on short temporal scales; for example, with time-lapse radar measurements of ice thickness change \cite{Zeising2022}. Using phase-coherent data acquisition, these measurements can disentangle the observed thickness change into strain thinning and basal melt \cite{Nicholls2015}. This has provided much insights, e.g., in terms of relevant tidal \cite{Sun2019} and seasonal timescales \cite{Vankova2022}.

\begin{figure}[ht]
    \centering
    \includegraphics[width=\textwidth]{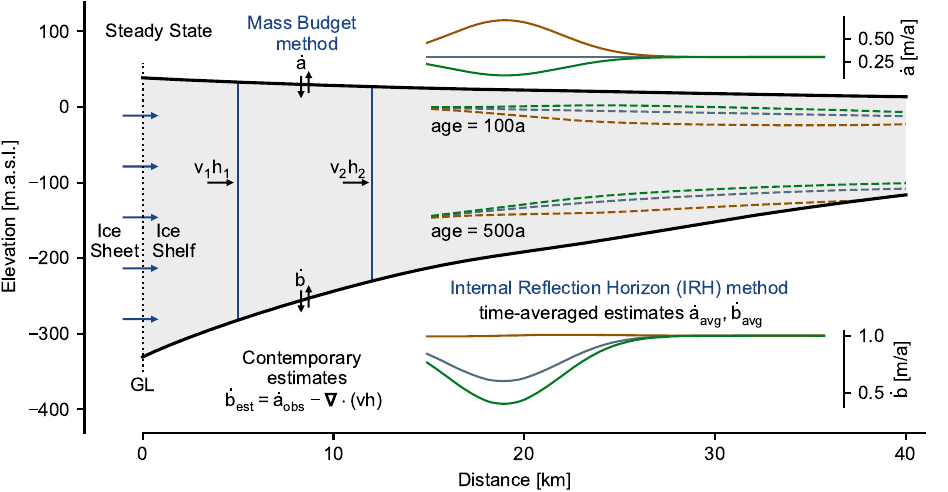}
    \caption{\textbf{Estimation of mass balance parameters from a steady state ice shelf with two methods.} The Eulerian Mass Budget method (left) detects the difference of surface accumulation and basal melt within two flux gates (blue vertical lines) by considering flux divergence $\nabla\cdot(\bf{v}H)$. Often, the basal melt rates $\dot b$ are inferred assuming that the surface accumulation ($\dot{a}_{\text{obs}}$) is known. In the Internal Reflection Horizon (IRH) method we are given information on the internal stratigraphy of the shelf. This information is used to decouple the known total mass balance into individual estimates of surface accumulation and basal melt ($\dot{a}_{\text{avg}},\dot{b}_{\text{avg}}$ respectively). These estimates correspond to the time-averaged value over the age of the IRH. The inset plots show different surface accumulation and basal melt parameterizations which give rise to the same total mass balance and overall shape of the ice shelf, but different internal stratigraphy.}
    \label{fig: motivation}
\end{figure}

Here, we investigate to what extent the radar-imaged isochronal ice stratigraphy \cite{Eisen2004} can provide additional information for inferring mass balance parameters. On grounded ice, radar-imaged Internal Reflection Horizons (IRHs) have been used in multiple ways, for example to infer the surface accumulation history \cite{Waddington2007,Macgregor2009, Catania2010, Steen-Larsen2010,Wolovick2021,Theofilopoulos2023}, velocity patterns of the ice flow \cite{Eisen2008Inference,Holschuh2017}, ice-rise evolution \cite{Drews2015evolution,Henry2023}, or large-scale model calibration \cite{Sutter2021}. On ice shelves, accumulation is also derived from the radar-measured shallow stratigraphy \cite{Pratap2022}, but not from intermediate depths and below where the stratigraphy is also influenced by basal melt and ice flow. The stratigraphy of ice shelves differs for various combinations of surface accumulation and basal melt \cite{Visnjevic2022}. This suggests that given an ice flow model of the internal stratigraphy that accounts for the mass balance parameters, we can use observed IRHs to recover the surface accumulation and basal melt histories (Fig. \ref{fig: motivation}). Thus, our goal is to solve the inverse problem of inferring the surface accumulation and basal melt rates that can explain the observed IRHs under the physical constraints of the ice flow model.

Inverse problems, also known as \textit{inversion}, \textit{data assimilation} or \textit{inference} problems in the literature, denote the task of finding the model parameters that are compatible both with empirical observations and prior knowledge. This problem is widespread in the geosciences, e.g.~in hydrogeology \cite{Linde2015}, seismology \cite{Symes2009}, or in climate science \cite{Tebaldi}. Bayesian inference provides a powerful framework for solving inference problems, but conventional Bayesian approaches are restricted to models for which the so-called "likelihood function" is computationally tractable. However, this is not the case for our model, and many other models in the geosciences. We therefore use simulation-based inference (SBI, \citeA{Papamakarios2016, Lueckmann2017,  Cranmer2020}) to tackle this problem. In SBI, we evaluate the forward model under different values of a model's parameters from a prior distribution. We use the resulting simulated dataset to train a neural network that performs conditional density estimation. For example, in the Neural Posterior Estimation (NPE) variant, the network approximates the Bayesian posterior distribution directly.  A key advantage of NPE is the amortization of simulation cost: once the network is trained, inference for unseen measurements can be performed without performing new simulations. Importantly, SBI does not require the forward model to be differentiable, and can work with "blackbox" models. Therefore, our approach can be extended to a variety of pre-existing forward models. To the authors' knowledge, this work is the first application of SBI in glaciology, but we note that it has already been applied in other geoscientific disciplines such as geothermics \cite{Omagbon2021}, hydrogeology \cite{Allgeier2023}, hydrology \cite{Hull2022}, and molecular ecology \cite{Overcast2021}. 

We limit our study to steady-state ice shelves and to IRHs in the local meteoric ice body of ice shelves \cite{Das2020}. This work is a test case for inferring atmospheric and oceanographic boundary conditions from the ice stratigraphy with a novel inference technique that provides uncertainty estimates. Our approach can be transferred to other ice flow regimes (e.g., flank flow on grounded ice) where similar scientific questions can be explored. Our approach can similarly be adapted to ice shelves exhibiting marine ice formation. Moreover, the isochronal stratigraphy of ice shelves is currently the only archive of surface accumulation and basal melt over the past hundreds of years. Our approach is capable of testing this. Thus, this study provides one link between observational initiatives (such as AntArchitecture, \citeA{AntArchitecture}) for Antarctica-wide internal stratigraphy datasets, and the modeling community. 

The paper is structured as follows: In section \ref{sec:methodology} we describe our forward model of the internal stratigraphy of an ice shelf and introduce our inference approach. In section \ref{sec: synthetic} we detail the synthetic ice shelf construction. We also present the results of inferring the mass balance parameters from this synthetic stratigraphy and compare the posterior distribution to a known ground truth. In section \ref{sec: Ekström} we describe the setting of the Ekström Ice Shelf (EIS) and the dataset of observed IRHs along the central flow line transect. We then provide the results of our inference framework and compare them to independent measurements of surface accumulation uniquely available in this location. In section \ref{sec: discussion} we interpret our results and evaluate our approach. We finally conclude and discuss future perspective in section \ref{sec:conclusion}.

\section{Methodology} \label{sec:methodology}
\subsection{Forward Model} \label{sec: forward model}
We denote spatially varying parameters as functions, e.g. $\dot{a}(x)$ or at times $\dot{a}$ for brevity, while bold-faced characters denote the discretized values of this function on a specified grid, e.g. $\mathbf{\dot{a}} = [\dot{a}(x_{1}),\hdots, \dot{a}(x_{n})]^{\top}$.

\subsubsection{Ice Flow Model} \label{sec: ice flow model}
We model ice shelves using the Shallow Shelf Approximation (SSA) \cite{Morland2007}. Throughout this study, we consider ice shelves in steady state. Consequently, the ice surface $s$, base $f$, thickness $h = s-f$ and velocity $v$ are all fixed throughout our simulations.  We assume plug flow for the ice shelf regime, meaning that the horizontal velocity profile does not change in the vertical direction $z$. These assumptions results in the mass balance condition
\begin{equation*}
   \nabla \cdot (hv) = \dot{m} ,
   \label{eq: mass balance}
\end{equation*}
where $hv$ is the total mass flux, $\nabla \cdot$ is the divergence operator, and $\dot{m} = \dot{a} -\dot{b}$ is the total mass balance rate. Here we use the convention that the surface accumulation rate $\dot{a}$ is positive for mass gain of the ice shelf and the basal melt rate $\dot{b}$ is positive for mass loss. In this exploratory study, we focus on flow lines. We re-parameterize our domain such that $x$ denotes the distance along the flow line, and $v_{x}$ now denotes the velocity parallel to the flow line. In order to account for ice flux into or out of our modelling domain we first calculate the total mass balance $\dot{m}$ on a flow-parallel band defined by ice-flow stream lines (Appendix \ref{sec: advection equation}).

We seek to predict the steady-state internal stratigraphy for a given flow line surface, base, horizontal velocity profile, and surface accumulation and basal melt rates. We define the internal stratigraphy to be a set of isochronal layer elevations $\{e_{1}(x),...e_{L}(x)\}$, with $f(x) \leq e_{1}(x)\leq e_{2}(x)\leq \hdots \leq e_{L}(x) \leq s(x)$. One approach to calculate the internal stratigraphy uses the SSA expression for the vertical component of the velocity \cite{Greve2009} to have a fully specified velocity field. This can then be used to calculate the age field $\mathcal{A}(x,z)$ of the shelf. Contours of constant age (isochrones) then define the internal stratigraphy. However, these methods suffer from numerical diffusion, and can be computationally expensive \cite{Visnjevic2022}.

The computational efficiency of the forward model is crucial for a tractable inference algorithm. As a result, we opt instead to use an implementation of the tracer method \cite{Born2017,Born2021}. The model is seeded with vertical segments each with a thickness profile $\{h_{1}(x),..., h_{L}(x)\}$, such that the sum matches the ice geometry $\sum_{l=1}^{L}h_{l}(x) = h(x)$. The horizontal velocity $v_{x}(x)$ is used to advect mass within segments and to thin or thicken the segments as a function of the prescribed strain rates. The accumulation $\dot{a}(x)$ and melt $\dot{b}(x)$ rates are used to add new segments or take away mass from the two boundary segments at the top and bottom of the shelf respectively. The (isochronal) layer elevations are then the boundaries between our modelled segments. We use the convention that $e_{l}$ corresponds to the top of segment $l$, which can be calculated using the cumulative thicknesses of the segments below,
\begin{equation*}
    e_{l}(x) = f(x) + \sum_{l'=1}^{l} h_{l'}(x).
\end{equation*} 
In our simulations, we used a high temporal resolution of one isochronal layer per year. Despite the high resolution, the layer tracing method allows for determining the internal stratigraphy in a computationally efficient manner. For the domains and timescales considered in our study, the complete forward model can be evaluated on the order of 60 seconds on a single CPU core, enabling the application of simulation-based inference methods (see Appendix \ref{sec: computational cost} for details).

In order to uniquely determine the layer thicknesses in such a scheme, we need to specify the boundary conditions on the layer thicknesses $h_{l}$ at the inflow boundary $x=0$ (here corresponding to the grounding line). The true boundary conditions are typically not known. However, the stratigraphy in a large part of the domain is still independent of the boundary conditions. This zone corresponds to the Local Meteoric Ice (LMI) body of ice shelves \cite{Das2020}. When inferring from observed stratigraphy data, we use only data within the LMI body. We detail our model of the LMI body in Appendix \ref{sec: LMI Body}.

\subsubsection{Noise Model}
The ice flow model predicts isochronal layers with varying depth over spatial scales of kilometers. Observed IRHs, however, also show  variability on sub-kilometer scales. This systematic model-data misfit is the total cause of measurement errors in all input datasets, discretization errors of the forward model, and omission of higher order processes that are not included in the shallow shelf approximation. For inference, it is important that the predicted isochrones have consistent statistical properties with the observed IRHs. This is done through the definition of a noise model.

The ice flow model predicts isochronal layer elevations $\{\mathbf{e}_{1}(\mathbf{x}),\hdots,\mathbf{e}_{L}(\mathbf{x})\}$ on a fixed grid $\mathbf{x}\in \mathbb{R}^{N}$ where $N$ is the number of grid points. The noise model should have the property that the errors of different layers $l$ are spatially correlated, and amplified for deeper layers. Guided by these physical constraints, we define a layer-wise noise model as the product of an $x$-dependent baseline noise function and a $z$-dependent vertical amplification factor. More precisely, the additive noise $\bm{\delta}_{l}\in\mathbb{R}^{N}$ of layer $l$ is defined as
\begin{equation*}
    \bm{\delta}_{l} = \bm{\epsilon} \odot \mathbf{T}(\mathbf{e}_{l}),
    \label{eq: noise model}
\end{equation*}
where $\bm{\epsilon} = [\epsilon_{1},...,\epsilon_{N}]^{\top}$ is a  $\mathbf{x}$-dependent noise profile, which is shared for all layers, $\mathbf{T}(\mathbf{e}_{l}) = [T(e_{l,1}),\hdots,T(e_{l,N})]^{\top}$ is a deterministic function of elevation (increasing with depth), and $\odot$ denotes an element-wise product. The vertical scaling $\mathbf{T}(\cdot)$ mimics uncertainties in the traveltime-to-depth conversion which depend on the density $\rho(z)$. Here, this is done using $\rho(z)$ as in \citeA{Drews2016} and an empirical density-permittivity relation \cite{Looyenga1965} to calculate the radio-wave speed $c(z)$. This results in the factor
\begin{equation*}
    T(z) = \int_{z}^{s}\frac{dz'}{c(z')},
    \label{eq: travel time}
\end{equation*}
which we then discretize on the set of layer elevations.

The sub-kilometer variability of the observed IRHs are modelled with power spectral densities $\bm{\epsilon}$:
\begin{equation*}
    \bm{\epsilon} = A_{\bm{\epsilon}}\sum_{n=1}^{N} \sqrt{\exp^{\beta_{n}}}\cos(2\pi\omega_{n} + \chi_{n}),
    \label{eq: PSD noise}
\end{equation*}
where the log power spectral densities $\beta_n$ and offsets $\chi_n$ are randomly sampled from normal and uniform distributions respectively: $\beta_{n}\sim\mathcal{N}(\mu_{\beta_{n}},\sigma_{\beta_{n}}^{2})$ and $\chi_{n}\sim U([-\pi,\pi))$. The frequencies $\omega_{n}$ are the corresponding Fourier frequencies of the simulation grid $\mathbf{x}$ and $A_{\bm{\epsilon}}$ is a global scale factor (set to $4\cdot10^{-10}$). In the synthetic ice shelf (Sec. \ref{sec: synthetic}), we define the distribution of the log power densities using $\sigma_{\beta_{n}}^{2} = 0.5$ and
\begin{equation*}
    \mu_{\beta_{n}} = -8\left( 1-\exp^{-200\omega_{n}}\right).
\end{equation*}

For Ekström Ice Shelf, the distribution means $\mu_{\beta_{n}}$ and variances $\sigma_{\beta_{n}}^{2}$ were calibrated given the observed IRHs on a separate set of calibration simulations (full details in Appendix \ref{sec: sim calibration}).

By combining the ice flow model with the robust noise model, we have arrived at a physically-motivated forward model to sample a plausible observed internal stratigraphy of an ice shelf from the mass balance rate parameters $\dot{a}$ and $\dot{b}$.

\subsection{Inference} \label{sec: inference}
\begin{figure*}[ht]
    \centering
    \includegraphics[width=\textwidth]{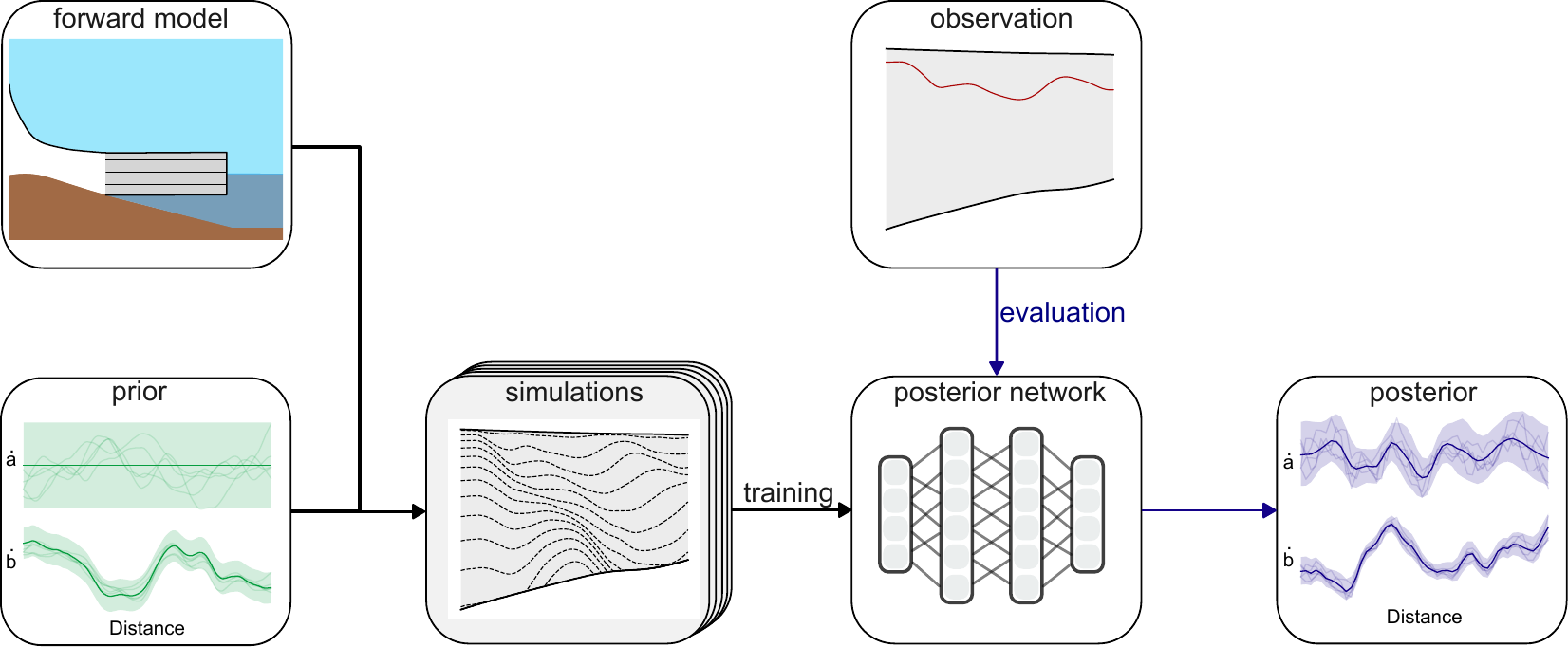}
    \caption{\textbf{Simulation-based inference workflow.}
    In the training phase, accumulation rates are randomly sampled from a prior distribution, the corresponding basal melt rates are obtained using total mass balance, and the resulting internal stratigraphy is calculated using the forward model. These simulations from the prior are used to train a neural network which parameterizes conditional distributions. In the evaluation phase, the trained network is conditioned on the observed IRH and outputs the Bayesian posterior distribution over the parameters (without any additional calls to the forward model). 
    }
    \label{fig:sketch SBI workflow}
\end{figure*}

Having established the forward model, we arrive at the \textit{inverse problem} of finding the surface accumulation and basal melt rates that best explain the observed internal stratigraphy. We use Bayes theorem with model parameters $\theta$ and observations $X$:
\begin{equation}
    p(\theta|X) = \frac{p(X|\theta)p(\theta)}{p(X)} .
    \label{eq: Bayes Theorem}
\end{equation}
Here, $p(\theta|X)$ is the posterior distribution of the parameters given an observation, $p(X|\theta)$ is the likelihood function of the model, $p(\theta)$ is the prior distribution encoding our existing knowledge on the plausible values of $\theta$, and $p(X)$ is the model evidence. The goal of Bayesian inference is to find the posterior $p(\theta|X_{o})$, where $X_{o}$ is \textit{observed data} obtained experimentally or otherwise.

\subsubsection{Simulation-Based Inference} \label{sec: sbi}
It is generally not possible to analytically solve for the Bayesian posterior distribution (Eq. \ref{eq: Bayes Theorem}), as the evidence term $p(X)$ involves the calculation of an intractable integral. Approximate methods exist to solve Eq. \ref{eq: Bayes Theorem} using knowledge of only the likelihood function and prior distribution. However, we opt to approximate Bayesian inference using only samples from our forward model (a \textit{likelihood-free} approach) using simulation-based inference (SBI). In SBI, we use artificial neural networks (ANNs) to approximate conditional probability distributions. While there exist different variants of SBI which target either the likelihood $p(X|\theta)$ or the likelihood ratio (see \citeA{Cranmer2020} for an overview), we focus on neural posterior estimation (NPE), which approximates the posterior distribution directly \cite{Papamakarios2016,Greenberg2019}.

In NPE, we generate a training dataset $\{(\theta_{k},X_{k})\}_{k=1}^{K}$ (Fig. \ref{fig:sketch SBI workflow}) by sampling parameters from the prior $\theta_{k}\sim p(\theta)$ and evaluating the forward model $X_{k} \sim p(X|\theta_{k})$. A family of distributions $q_{\phi}(\theta|X)$ is typically defined in terms of a neural network with learnable weights $\phi$. We represent $q_{\phi}$ as a normalizing flow \cite{Kobyzev2019,Papamakarios2019,Durkan2019}.
The neural network is trained by minimizing the expected negative log-probability
\begin{equation*}
    \mathcal{L}(\phi) = \mathbb{E}_{\theta\sim p(\theta),X\sim p(X|\theta)}[ -\log q_{\phi}(\theta|X)]
    \label{eq: NPE Loss}
\end{equation*}
on the training dataset. It has been shown that, under some assumptions, the minimum of this loss is reached when $q_{\phi}(\theta|X) = p(\theta|X)$--i.e. when our estimated distribution matches the true posterior \cite{Papamakarios2016}. 

We also make use of an \textit{embedding network}, which are commonly used in SBI workflows in order to improve performance. Embedding networks learns \textit{summary statistics} $Y(X)$, which are lower-dimensional representations of the data $X$. The embedding network is trained jointly with the normalizing flow. In our setting, $X_{k}$ are spatially-varying IRH elevations, and so we choose a 1-dimensional convolutional neural network (CNN) as our embedding net, resulting in 50-dimensional embeddings on which the posterior network is conditioned (full details in Appendix \ref{sec: architecture}).

\subsubsection{Details on Model Parameters and Observations}
\label{sec: inference definitions}
We define $\theta = \mathbf{\dot{a}}_{\text{inf}}= [\dot{a}_{i_{1}},\hdots,\dot{a}_{i_{J}}]^{\top}$, the values of the accumulation rate on a discretized grid $\mathbf{\tilde{x}}$. In our experiment, we choose the number of inference grid points $J=50$ as a compromise between computational complexity of the inference problem while still inferring accumulation rate at a high resolution of approximately $2.5$~km. This is smaller than the discretized grid $\mathbf{x}$ we use for our simulations, which has 500 gridpoints in our experiments. In practice, we take $\mathbf{\tilde{x}}$ to be a regularly spaced subset of $\mathbf{x}$, so that $\mathbf{\dot{a}}_{\text{inf}}$ can also be taken as a subset of $\mathbf{a}$. However, $\mathbf{\tilde{x}}$ can be any discretization of the flow line, and need not be a subset of $\mathbf{x}$. Furthermore, despite defining $\theta$ to only represent the surface accumulation, any inference of the surface accumulation automatically extends to inference of the basal melt rates. This is because for any probability distribution $q(\mathbf{\dot{a}}_{\text{inf}})$, the total mass balance relationship implies that $\mathbf{\dot{b}}_{\text{inf}} \sim q(\mathbf{\dot{a}_{\text{inf}} - \dot{m}}_{\text{inf}})$, where $\mathbf{\dot{b}}_{\text{inf}},\mathbf{\dot{m}}_{\text{inf}}$ are the respective discretizations of $\dot{b},\dot{m}$ onto $\mathbf{\tilde{x}}$.

We now turn to describing the observation, $X$. The observed data is a set of different IRHs, $\{\mathbf{e}_{m}(\mathbf{x})\}_{m=1}^{M}$, where $e_{m}(x_{i})$ is the elevation of the $m^{\text{th}}$ IRH in our dataset at grid position $i$. The IRH elevations need not and typically are not observed at the same locations as the simulation gridpoints, and so we first interpolate the IRH elevations onto the simulation grid $\mathbf{x}$ using linear spline interpolation (as implemented in \texttt{Scipy} \cite{scipy}). Therefore, w.l.o.g.~we assume $\{\mathbf{e}_{m}(\mathbf{x})\}_{m=1}^{M}$ is already defined on $\mathbf{x}$. In our work, we choose to separately infer the mass balance from each IRH in our observed dataset. This has two advantages:
first, the number and average depths of picked IRHs in datasets varies between measurements, and cannot be predicted a-priori. Thus, learning some \textit{global} property about the ice shelf from a set of IRHs is highly challenging. Second, ordering IRHs by depth also corresponds to their reverse age order, with the oldest IRHs being the deepest. Thus, inferring the surface accumulation and basal melt rates for deeper IRHs corresponds to inferring the average rates over longer periods of time. By comparing the inferred mass balance parameters obtained with different IRHs, we can reason about how they changed over time.

Thus, given a dataset of $M$ observed IRHs, we have $M$ inference problems to solve, where each observation corresponds to one IRH. It is therefore reasonable to take one isochronal layer of the simulated stratigraphy as the output of the forward model. For the $m^{\text{th}}$ inference problem, we define the outcome of forward model as the isochronal layer $\mathbf{e}_{l}$ that is closest to IRH $m$ (in the mean square sense). More precisely, for inference problem $m$ and simulation $k$, we define the observation of the forward model to be $X_{k}^{m} = \mathbf{e}_{l^{*}}(\mathbf{x}_{i\geq i(m)})$, where
\begin{equation*}
    l^{*} = \argmin\limits_{l} ||\mathbf{e}_{l}(\mathbf{x}_{i\geq i(m)})-\mathbf{e}_{m}(\mathbf{x}_{i\geq i(m)})||_{2}^{2}.
    \label{eq: Selecting Best Layer}
\end{equation*}
Here, $i(m)$ is the index of the boundary of the LMI body for IRH $\mathbf{e}_{m}(\mathbf{x})$. For $i<i(m)$ the IRH $e_m(x_i)$ is outside the LMI body and for $i\geq i(m)$ within the LMI body (see Appendix \ref{sec: practical LMI} for details). We also define $\mathbf{x}_{i\geq i(m)} = [x_{i(m)}, x_{i(m)+1}, ..., x_{N}]^{\top}$ as the restriction of the gridpoints $\mathbf{x}$ to within the LMI body of $\mathbf{e}_{m}(\mathbf{x})$. We correspondingly set the observation for IRH $m$ to $X_{o}^{m} = \mathbf{e}_{m}(\mathbf{x}_{i\geq i(m)})$.

\subsubsection{Choice of Prior Distribution} \label{sec: prior choice}
We aim to approximate the posterior distribution $p(\mathbf{\dot{a}}| X_{o}^{m}) \propto p(X_{o}^{m}|\mathbf{\dot{a}})p(\mathbf{\dot{a}})$. The likelihood $p(X_{o}^{m}|\mathbf{\dot{a}})$ is not analytically tractable, but can be sampled from using the forward model. For Ekström Ice Shelf we use the long-term snow accumulation observations of the Neumayer stations \cite{Neumayer,Wesche2022} over more than 30 years to define an empirically-motivated prior. We first assume that localized surface melt ($\dot{a}<0$), is possible, but rare. We also observe that average rate of accumulation is approximately 0.5~\unit{m}\unit{a^{-1}}, and that the accumulation rate is almost everywhere under 2~\unit{m}\unit{a^{-1}}. Finally, we take the accumulation rate to vary smoothly in space. As a result of these empirical observations, we define the following generative process as a prior over $\mathbf{\dot{a}}$: first, we draw a sample $\mathbf{\alpha} = [\alpha_{1},...,\alpha_{N}]^{\top}$ from a Gaussian process with mean function $\mu = 0$ and a Matérn kernel with a Matérn-$\nu$ of 2.5 and a lengthscale of $2500$~m \cite{RasmussenGPs}. We then independently sample an offset $\mu_{\text{off}}\sim \mathcal{N}(0.5,0.25^{2})$ and scale $\sigma_{\text{sc}}\sim U(0.1,0.3)$ parameter. Finally, we set $\mathbf{\dot{a}} = \sigma_{\text{sc}}\mathbf{\dot{\alpha}} + \mu_{\text{off}}\mathbf{1}$. We use this prior distribution for both synthetic and Ekström ice shelves.

Defining the prior in this way is sufficiently expressive to capture numerous accumulation rate profiles, while also restricting the samples to conform to empirical knowledge. Additionally, the prior is shared for all $M$ inference problems we have defined, and one evaluation of the forward model provides an observation $X_{k}^{m}$ for each of the inference problems. Thus, the same training dataset can be used for all posterior networks in our SBI approach, significantly reducing the computational costs.

\subsection{Implementation Details} \label{sec: implementation details}
First, we used \texttt{Antarctic Mapping Tools} \cite{AntarcticMappingTools}, BedMachine Antarctica \cite{Morlighem2017}, and \texttt{ITS\textunderscore LIVE} \cite{ITS_LIVE} to  obtain the surface elevation $s$, thickness $h$ and velocity $\mathbf{v}$ for Ekström Ice Shelf. In order to define the flow tube domain for Ekström Ice Shelf, we also used the \texttt{itslive\textunderscore flowline} tool to find two flow lines which formed the side-boundaries of the domain. The other two boundaries of the domain were the grounding line, and a straight line connecting the two flow lines. The straight line was chosen to ensure that the radar transect where data was measured is wholly contained within the flow tube domain.

For Ekström ice shelf (Sec. \ref{sec: Ekström}) we preprocessed the raw ice shelf geometry and velocity data prior to evaluating the model. This ensured numerical stability of the forward model. Using the \texttt{icepack} package for Python \cite{icepack}, we first smoothed the raw thickness data by solving a regularized minimization problem. We then solved for the best-fitting velocity by fitting a fluidity parameter in an SSA model to the observed velocity and smoothed thickness. In the synthetic example (Section \ref{sec: synthetic}), \texttt{icepack} was also used in order to create a steady-state ice shelf. The hyperparameters used for preprocessing are given in Appendix \ref{sec: preprocessing appendix}. Finally, once the training dataset was created, the \texttt{sbi} package for Python \cite{Tejero-Cantero2020} was used for the inference procedure described in Section \ref{sec: sbi}.

\subsection{Radar Measurements of Internal Stratigraphy} \label{sec: Ekström Measurements}
\begin{figure*}[ht]
    \centering
    \includegraphics[width=\textwidth]{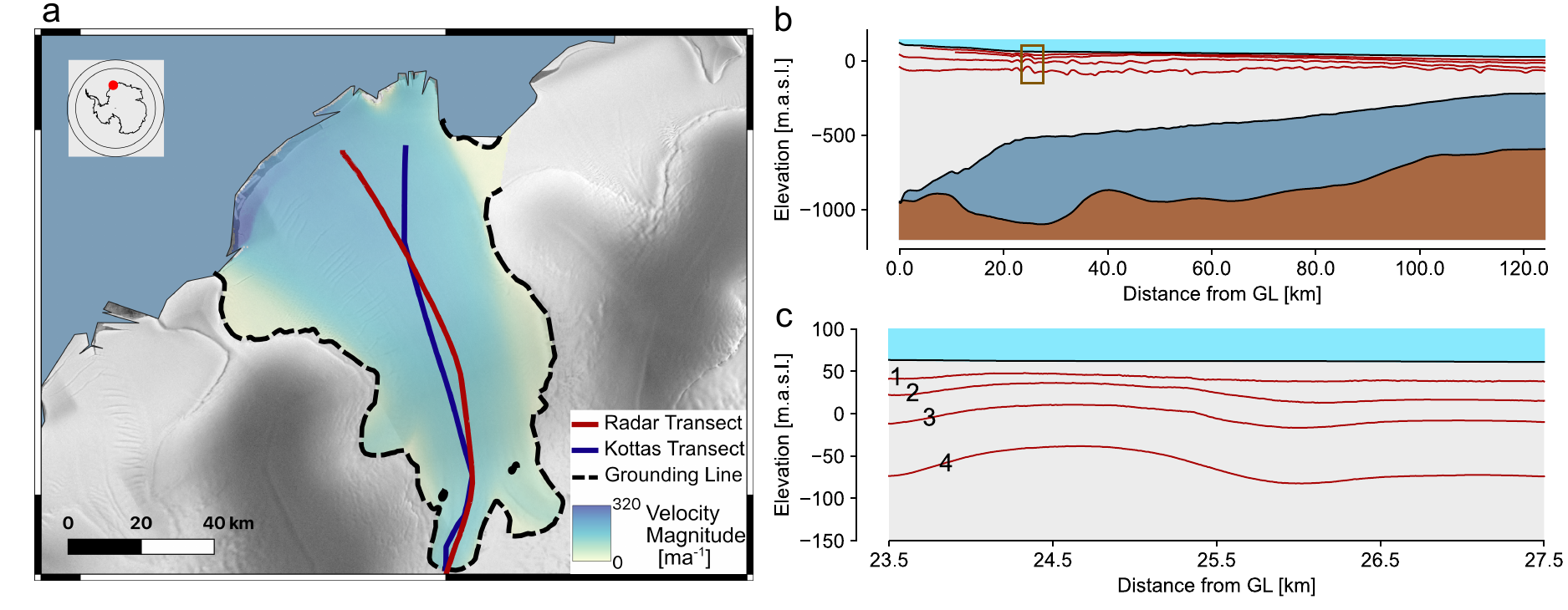}
    \caption{\textbf{Overview of the Ekström Ice Shelf.} \textbf{a:} Satellite view of Ekström Ice Shelf along with location of the radar transect along the central flow line (red line) and the Kottas traverse (blue line). An independent estimate of surface accumulation via stake arrays is available on Kottas traverse, which we use to validate our results. \textbf{b:} Vertical cross-section view of the radar transect, along with ice surface and base take from BedMachine Antarctica \cite{Morlighem2017}, starting at the grounding line (GL). Red lines indicate four picked Internal Reflection Horizons (IRHs). \textbf{c:} Zoom in on box in B. The IRHs are numbered 1--4 in order of increasing depth.}
    \label{fig: Ekstrom_data}
\end{figure*}
Internal stratigraphy data along the central flow line of Ekstr\"om Ice Shelf (Fig. \ref{fig: Ekstrom_data}a) were acquired using a ground-based ground-penetrating radar with a center frequency of 50~\unit{MHz}  (pulseEKKO\textsuperscript{TM} from Sensors \& Software) in two consecutive field seasons (2021/22 and 2022/23) with logistic support from the Neumayer III station \cite{Neumayer,Wesche2022}. Radar processing was done with ImpDAR \cite{Lilien2020} and included trace averaging to equidistant spacing (10~\unit{m}), bandpass filtering (with cut-off frequencies of 20 and 75~\unit{MHz}), and a topographic correction using the REMA surface elevation \cite{Howat2019}. The latter provides observations consistent with the modeling setup. The radar detects the ice-ocean interface and continuous internal reflection horizons (IRHs) down to approximately 200 m depth (Fig. \ref{fig: Ekstrom_data}b and c). Four IRHs were digitized along the entire 130 km long profile using a semi-automatic maximum tracking scheme. The vertical offset off IRHs at the profile junction in the mid-shelf region between both years is much smaller than the radar system's wavelength in ice ($\sim$ 3.4~\unit{m}). Consequently IRHs were connected without adjustments. For the travel-time to depth conversion we used a depth-density profile representative for ice shelves of the Dronning Maud Land Coast (\citeA{Hubbard2013}, Eq. 1). 

\section{Synthetic Test Case} \label{sec: synthetic}
Before we apply the presented workflow to a real example in Ekström Ice Shelf, we showcase its applicability in a synthetic test case in which the ground truth parameters are known. 
\subsection{Configuration of Shelf and flow line} \label{sec: synthetic domain}
\begin{figure}[ht]
    \centering
    \includegraphics[width=\textwidth]{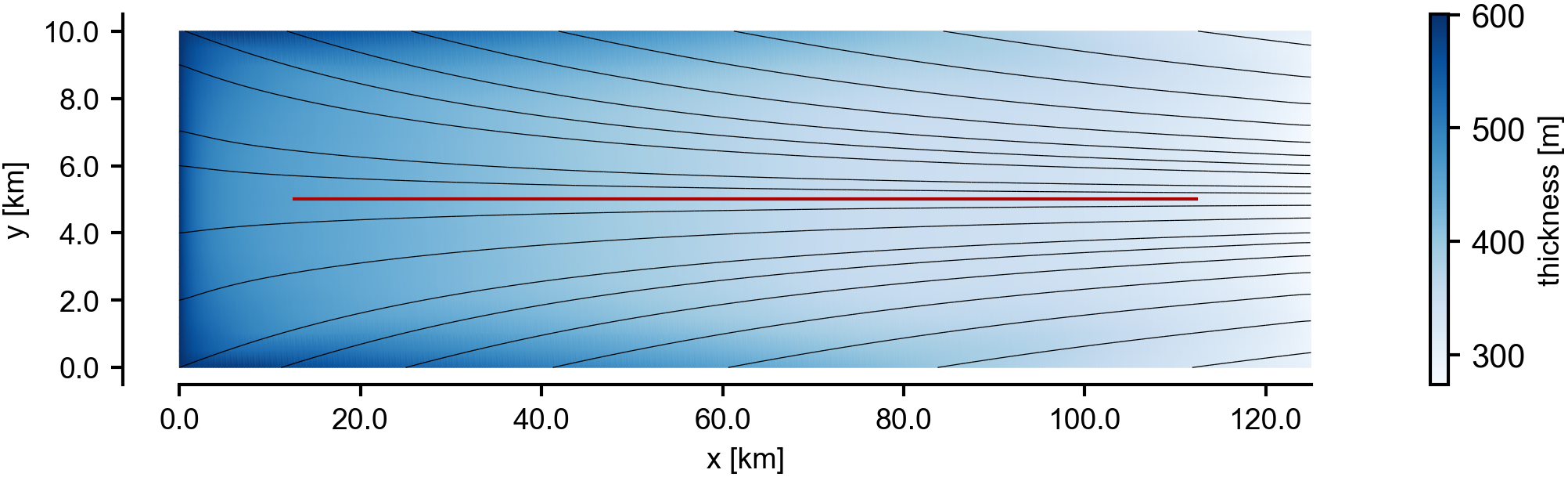}
    \caption{\textbf{Two dimensional flow tube domain setup for the synthetic example.} Map view of the simulated ice shelf's surface. Flow lines (grey lines) converge to the central flow line (red). Colour indicates ice thickness. The input variables for the internal stratigraphy model are evaluated on the central flow line.}
    \label{fig: synthetic setup}
\end{figure}

We present a case study on a synthetically-generated flow line. We begin by creating a two dimensional flow tube on a grid $L_{x}$~\unit{km}$\times L_{y}$~\unit{km}, with the along-flow direction $x$ and across-flow direction $y$. We instantiate the flow tube with a Dirichlet boundary condition at the inflow and lateral boundaries, with a constant thickness of $h_{0}$, and a constant along-flow velocity of $v_{0x}$. We mimic lateral boundary conditions with some friction by initializing a zero centered, longitudinally symmetric across-flow velocity $v_{0y}$ on the lateral boundaries, resulting in a flow field that has convergence (i.e. mass input) on the center flow line.

We fix a total mass balance $\dot{m}$ of the shelf, and perform a transient simulation of the resulting flow tube using the SSA approximation, until steady state is reached. From the steady state ice shelf, we choose a discretization of the central flow line, $\mathbf{x}$, and extract the relevant variables along this flow line to define the internal stratigraphy model (Fig. \ref{fig: synthetic setup}). The variables we need are the surface $\mathbf{s}$ and base $\mathbf{f}$ elevations, the along-flow velocities $\mathbf{v}_{x}$, and the along- and across-flow flux divergences $d(\mathbf{v}_{x}\mathbf{h})/dx$, $d(\mathbf{v}_{y}\mathbf{h})/dy$. These define the total mass balance, since:
\begin{equation*}
    \mathbf{\dot{a}}-\mathbf{\dot{b}} = \frac{d(\mathbf{v}_{x}\mathbf{h})}{dx} + \frac{d(\mathbf{v}_{y}\mathbf{h})}{dy}.
\end{equation*}
We also use the flux divergences to account for the flow tube correction (Appendix \ref{sec: advection equation}).

 We choose an arbitrary sample from the prior distribution as the ground truth, $\mathbf{\dot{a}}_{\text{GT}}$. We calculate the corresponding ground truth basal melt rate profile $\mathbf{\dot{b}}_{\text{GT}}$. The forward model is then sampled to obtain a set of ground truth layer elevations, $\mathbf{e}_{o}(\mathbf{x})$. From these layer elevations, we choose to perform inference for four layers of ages 50, 100 and 150, and 300 years (labelled 1 to 4 in ascending order of age). These ages roughly correspond to the range of ages of the IRHs that we expect to observe on ice shelves.

 \subsection{Inference Results} \label{sec: Synthetic results}
\begin{figure*}[ht]
    \centering
    \includegraphics[width=\textwidth]{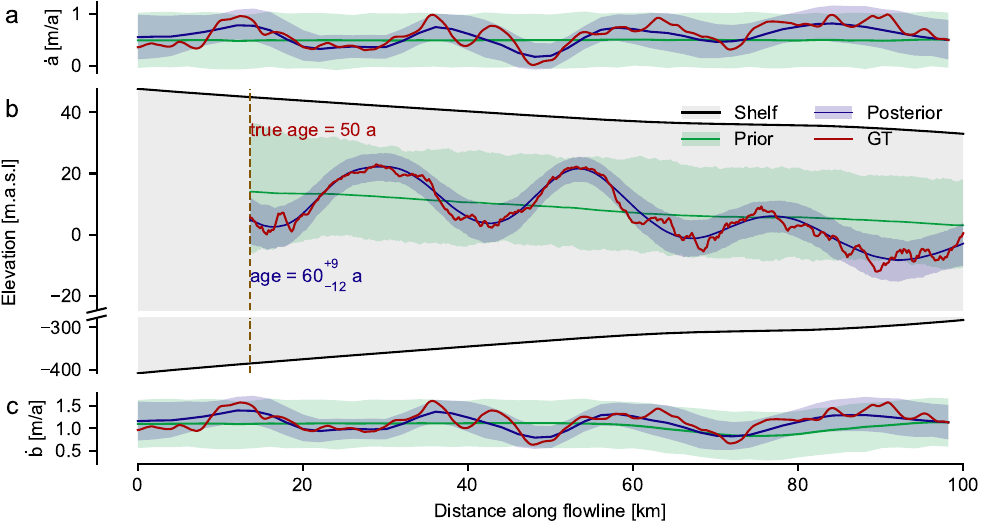}
    \caption{\textbf{Prior and posterior (predictive) for the synthetic dataset. a and c:} Prior and posterior over surface accumulation and basal melt rates respectively for layer 1 of the synthetic ice shelf, of age 50 years. Solid line is the distribution mean, the shaded region represents the 5th and 95th percentiles. The ground truth (GT) parameters used to generate the reference isochronoal layer are also shown. \textbf{b:} Cross section of the ice shelf. Prior and posterior predictive distributions for the layer closest matching the ground truth isochronal layer. The vertical dashed line represents the LMI boundary for this isochronal layer. The posterior predictive reconstructs the observed layer with higher accuracy and lower uncertainty. The posterior predictive distribution of the age of the isochronal layer is $60^{+9}_{-12}$ years (meaning a median of 60 years, and 16th and 84th percentiles of 48 and 69 years respectively). The average root mean square error relative to the GT isochronal layer is 3.9~\unit{m} for the posterior and 11.5~\unit{m} for the prior.}
    \label{fig:posterior Synthetic}
\end{figure*}
We evaluate the trained neural posterior network on the ground truth isochronal layer of age 50 years. The inferred posterior mean surface accumulation rate is close to the ground truth accumulation rate (Fig. \ref{fig:posterior Synthetic}a,c) and the ground truth lies within the 95\% confidence intervals of the posterior distribution. 

Next, we evaluate the forward model on samples from the posterior (and prior) distribution to get the respective \textit{predictive distributions}. This shows the prior and posterior predictions of the isochronal layer (Fig. \ref{fig:posterior Synthetic}b). The posterior predictive matches the ground truth isochronal layer with high fidelity. We calculate the root mean square error (RMSE) of the predictive simulations relative to the ground truth layer elevations for 1000 simulations using prior and posterior samples. The average RMSE for the posterior is 3.9~\unit{m}, compared to 11.5~\unit{m} for the prior. Uncertainties in the layer elevations are much smaller than those of the prior predictive distribution. This is in contrast to the posterior uncertainty over the mass balance rates, which is still considerable. This showcases the importance of our uncertainty-aware approach: there is more than one parameterization of accumulation and basal melt rates that can lead to similar isochronal layers. 

The posterior uncertainty is also reflected in the inferred age of the isochronal layer. We infer an age of $60^{+9}_{-12}$ years for this layer (meaning a median of 60 years, and 16th and 84th percentiles of 48 and 69 years respectively). This value matches the age of the ground truth isochronal layer, which was not used during inference. Thus we have produced an estimate of the age of the layer without requiring invasive measurements such as ice cores. We report the posterior distributions for deeper synthetic layers in Appendix \ref{sec: additional results}.

\section{Ekström Ice Shelf} \label{sec: Ekström}
Ekström Ice Shelf is a medium-sized ice shelf located between the Sörasen and Halvfarryggen Ice Rises in Dronning
Maud Land, East Antarctica (Fig. \ref{fig: Ekstrom_data} C). Ekström Ice Shelf makes for an appropriate study site because the steady-state assumption likely holds \cite{Drews2013}, and because it is well modelled by the shallow shelf approximation \cite{Schanwell2019}. Moreover, because of the proximity of the Neumayer station III numerous observations are available, e.g.~ice thickness, surface velocities and most importantly surface accumulation rates, which we will use later for validation. 

\subsection{Inference Results} \label{sec: Ekström results}
\begin{figure*}[ht]
    \centering
    \includegraphics[width=\textwidth]{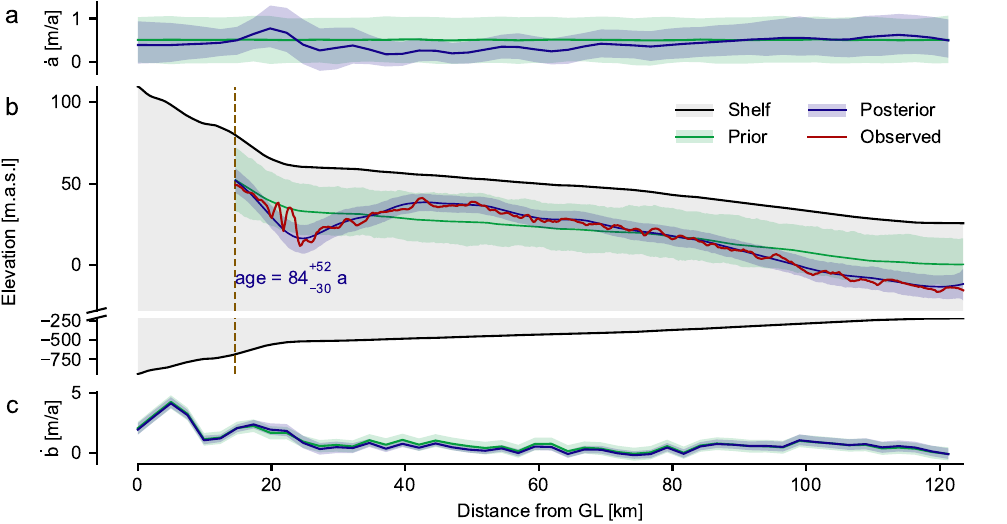}
    \caption{\textbf{Prior and posterior (predictive) for the Ekström dataset, IRH 2, of average depth 30 m. a and c:} Prior and posterior over surface accumulation and basal melt rates respectively, starting at the grounding line (GL). Solid line is the distribution mean, the shaded region represents the 5th and 95th percentiles. \textbf{b:} Cross section of the ice shelf. Prior and posterior predictive distributions for the layer closest matching the observed IRH.  The vertical dashed line represents the LMI boundary for this IRH. The posterior predictive reconstructs the observed IRH with higher accuracy and lower uncertainty. The posterior predictive distribution of the age of the IRH is $84^{+52}_{-30}$ years. The average root mean square error relative to the observed IRH is 4.6~\unit{m} for the posterior and 11.8~\unit{m} for the prior.}
    \label{fig:posterior Ekstrom}
\end{figure*}

We inspect the prior over the basal melt rates as a validation of our modeling choices. The implicit prior is the same as the prior defined for the surface accumulation, with the mean shifted by the total mass balance on the flow line, $\mathbf{\dot{m}}$. The basal melt rate is larger (up to 4~\unit{m}\unit{a^{-1}}) near the grounding line, and gradually stabilises in the along-flow direction to values between 0 and 1~\unit{m}\unit{a^{-1}} downstream. This is in agreement with previous estimates for basal melt profiles on this particular ice shelf \cite{Neckel2012}.

We infer the surface accumulation and basal melt rates from IRH 2 in our dataset, which has an average (ice equivalent) depth of 30~\unit{m} (Fig. \ref{fig:posterior Ekstrom}). The posterior over the surface accumulation rate has uncertainty comparable to that of the prior. However, there is a shift in the overall spatial trend of the accumulation rate; particularly, there is higher surface accumulation rate at approximately 20~\unit{km} from the grounding line. Accumulation rate also increases steadily downstream the flow line. As in the synthetic case, the posterior predictive distribution reproduces the observed IRH with much higher fidelity and confidence than the prior predictive distribution. The average RMSE relative to the observed IRH is 4.6~\unit{m} for 1000 posterior predictive simulations, compared to 11.8~\unit{m} for 1000 prior predictive simulations. The posterior predictive produces an independent estimate of the unknown age of the IRH of $84^{+52}_{-30}$ years.

\begin{figure*}[ht]
    \centering
    \includegraphics[width=\textwidth]{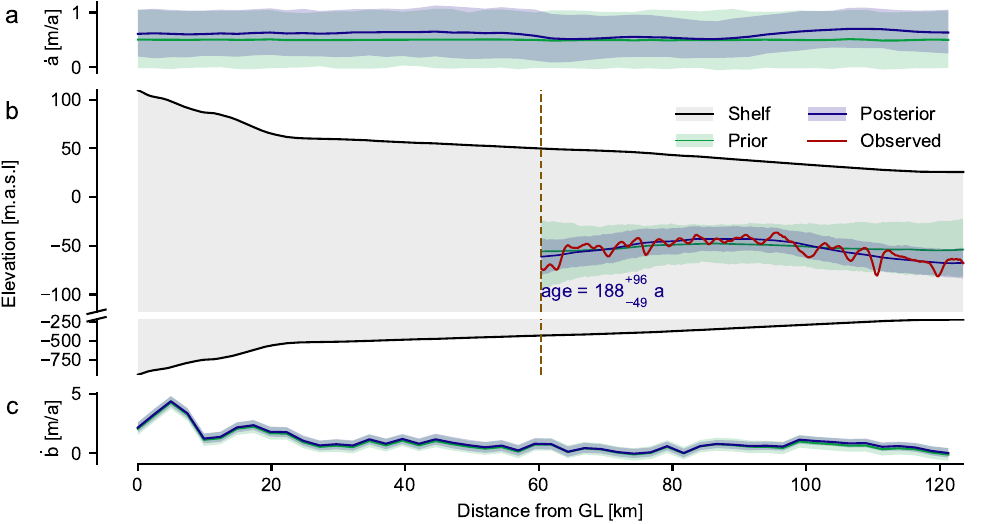}
    \caption{\textbf{Prior and posterior (predictive) for the Ekström dataset, IRH 4, of average depth 113 m.} Same as Fig. \ref{fig:posterior Ekstrom} for the deeper IRH. The posterior predictive distribution of the age of the IRH is $188^{+96}_{-49}$ years. The average root mean square error relative to the observed IRH is 10.0~\unit{m} for the posterior and 16.4~\unit{m} for the prior.}
    \label{fig:Ekstrom deep}
\end{figure*}
Our method can use much deeper IRHs for the inference of accumulation and basal melt rates. For IRH 4 of the observed dataset (of average depth 131 m), the proportion of the IRH that is within the LMI body is smaller. This is due to the unknown boundary condition influencing the IRH elevation at much further points along the flow line. This discarding of data has visible effects on our parameter posterior (Fig. \ref{fig:Ekstrom deep}), which now is more similar to the prior near the grounding line, and only diverges at points further down the ice shelf, where the values of accumulation and basal melt rates affect the dynamics of the IRH. Regardless, the posterior predictive reconstructs the observed IRH at higher fidelity and precision than the prior predictive. The average RMSE relative to the observed IRH is 10.0~\unit{m} for the posterior and 16.4~\unit{m} for the prior. The estimated age of this IRH by our method is $188^{+96}_{-49}$ years. The uncertainty of the age estimates reasonably increases for deeper IRHs.

\section{Discussion} \label{sec: discussion}

\subsection{Ekström Ice Shelf is in Steady State}
\label{sec: Ekstrom steady state}

We compare the four posteriors over the surface accumulation obtained from the Ekström IRH dataset (Fig.~\ref{fig: Ekstrom smb comparison}). The posteriors for the shallower IRHs 1--3 all show a similar qualitative relationship: a local maximum of the accumulation at a distance of approximately 20~\unit{km} from the grounding line, followed by a steady increase in the accumulation downstream. This supports our initial assumption that Ekström Ice Shelf is in steady state. Additionally, the increase in accumulation at $\sim$20~\unit{km} is even identified in the posterior for IRH 3, despite the LMI boundary being downstream of it, at approximately 30~\unit{km} from the grounding line. This is reasonable, as the mass balance parameters at a given location affect the flow field downstream of this location, and consequently, the formation of isochronal layers. For IRH 4, the LMI boundary is much further downstream at $\sim$60 \unit{km}. Thus, the local surface accumulation maximum at $\sim$20~\unit{km} is not found, however the overall trend of increasing surface accumulation downstream is still identified. The corresponding trend in the basal melt rate is dominated by the total mass balance, $\dot{m}$. However, the basal melt rate still exhibits a local maximum at $\sim$20 \unit{km}. This could be related to increased tidal forces in the flexure zone of the ice shelf.

The inferred posteriors also allow us to estimate the age of the IRHs. By sampling from the posterior distribution, and evaluating the forward model with the resulting mass balance parameter samples, we obtain a distribution of isochronal layers similar to the observed IRH, with known ages. Thus, we estimate the ages of the four IRHs as 42, 84, 146 and 188 years. Our results therefore support the assumption that Ekström Ice Shelf has been in steady state for the past 188 years. Crucially, we are able to estimate the age of the IRHs without invasive measurements such as ice cores. This estimate depends on a realistic prior for the surface and basal mass balance, as defined in Sec.~\ref{sec: prior choice}. Given a miscalibrated prior, the estimated ages would not be reliable (see Appendix \ref{sec: miscalibrated prior} for an example). We hypothesise that given an independent measurement of the age of the IRH, our approach could constrain the posterior distributions over the mass balance parameters further.

\subsection{Comparison to Shallow and Local Layer Approximations, and Kottas Traverse Accumulation Measurements}
\label{sec: result comparison}

\begin{figure*}[ht]
    \centering
    \includegraphics[width=\textwidth]{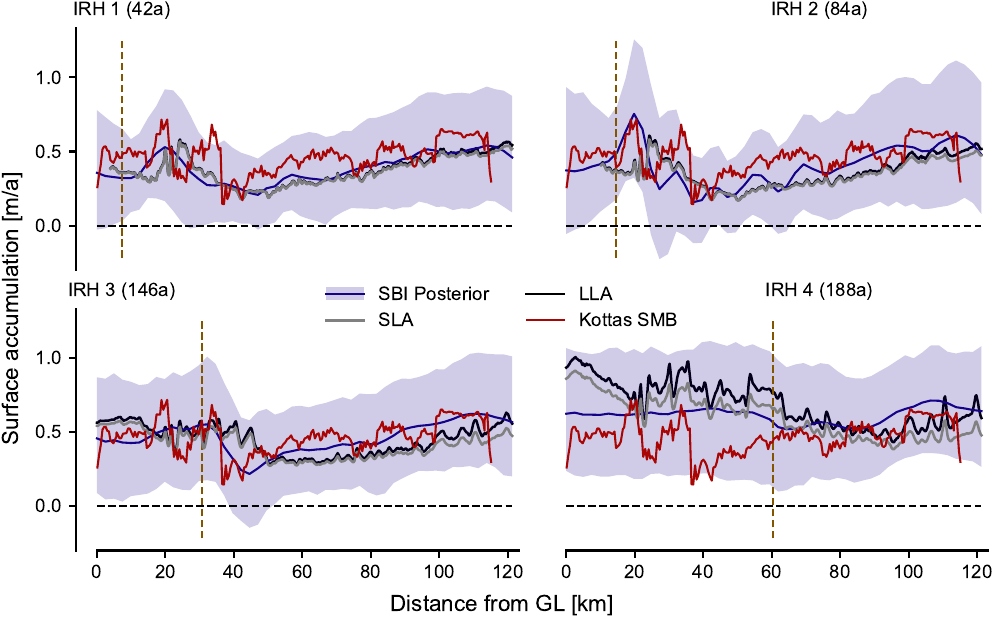}
    \caption{\textbf{Ekström Ice Shelf - Dependence of posterior surface accumulation rate on depth of IRH used for inference}. The posteriors are compared to the Shallow Layer Approximation (SLA) and Local Layer Approximation (LLA) \cite{Waddington2007}, and an estimate of the distribution of the accumulation rate based on measurements along the Kottas traverse. See Fig. \ref{fig:Kottas Uncertainty} for yearly Kottas measurements. As the real age of the IRHs is not known, the SBI-derived median age is used for the SLA and LLA approximations. Median ages for IRH~1--4 are 42, 84, 146, 188 years. The LMI boundary, representing where the IRH data was masked, is shown with the brown dashed lines.}
    \label{fig: Ekstrom smb comparison}
\end{figure*}

To validate our approach, we compare the inferred surface accumulation rate of our experiments with estimates from other methods. First, we computed the Shallow Layer Approximation (SLA) and Local Layer Approximation (LLA) as described in \citeA{Waddington2007}. Given the depth and age of IRH $m$, the SLA and LLA approximations for the accumulation rate $\mathbf{\dot{a}}$ are defined as
\begin{equation*}
    \mathbf{\dot{a}}_{\text{SLA}}^{m} = \frac{1}{\mathcal{A}_{m}}(\mathbf{s} - \mathbf{e}_{m}(\mathbf{x})), \quad \mathbf{\dot{a}}_{\text{LLA}}^{m} = -\ln \left( 1- \frac{\mathbf{
    s}-\mathbf{e}_{m}(\mathbf{x})}{\mathbf{h}}\right)\frac{\mathbf{h}}{\mathcal{A}_{m}}~,
    \label{eq: Shallow Layer Approximation}
\end{equation*}
where $\mathcal{A}_{m}$ is the age of IRH $m$. Intuitively, the SLA takes the ice thickness above layer $m$ and divides it by the layer age, whereas LLA accounts for strain thinning assuming a linear vertical velocity profile (which is often the case for ice-shelf flow). Since the age of the observed IRHs is not known, we use the median age of the posterior predictive distribution results. As expected, we observe that both SLA and LLA closely match the SBI posterior mean accumulation rate for the shallow IRHs of estimated ages 42, 84 years (Fig. \ref{fig: Ekstrom smb comparison}). As the strain rates of the flow are small, the relatively shallow IRHs (mean ice equivalent depth of 30~\unit{m}) have not notably deformed, and hence the assumptions of SLA and LLA are appropriate. However, for the deeper IRHs 3 and 4 of estimated ages 146 and 188 years, we see that both SLA and LLA estimates diverge from our posterior mean accumulation rate. This shows that more involved approaches are required when using deeper IRHs for inference. For deeper IRHs where the SLA and LLA no longer applied, \citeA{Steen-Larsen2010} inferred the surface accumulation rates on grounded ice using a Monte Carlo approach. By treating the age of the IRH as an additional parameter to infer, they were able to identify the age of the IRH with high confidence. Extensions of our approach could incorporate this parameterization to reduce the uncertainty of the inferred IRH.

For the Ekström transect comparison data is provided by repeat readings of accumulation stakes in 500~\unit{m} spacing along the nearby Kottas traverse (Fig.~\ref{fig: Ekstrom smb comparison}). Yearly readings are available in the period 1996--2005 and on a yearly to three-yearly interval between 2014--2023 \cite{Mengert2018}. We use this dataset to construct a direct estimate of time-averaged surface accumulation rate along the central flow line transect. For this, we project the measurements from the Kottas traverse to the flow line transect, taking into account an increased uncertainty for increasing projection distance (see Appendix \ref{sec: Kottas data} for details). 

The Kottas traverse accumulation measurements closely match the posterior means of our approach (Fig.~\ref{fig: Ekstrom smb comparison}) for IRHs 1 and 2. As the accumulation rate measurements on the Kottas traverse span the past 26 years it may not be a good validation for the deeper IRHs. Regardless, the Kottas accumulation rate measurements lie within the posterior uncertainty for IRHs 3 and 4.  These comparisons further corroborate our approach and highlight the advantages of uncertainty-aware methods, especially as the measured accumulation rates also varied considerably year-to-year (Appendix \ref{sec: Kottas data}).

\subsection{Future Directions for Forward Modeling Approach} \label{sec: future modeling directions}

\begin{figure*}[t]
    \centering
    \includegraphics[width=\textwidth]{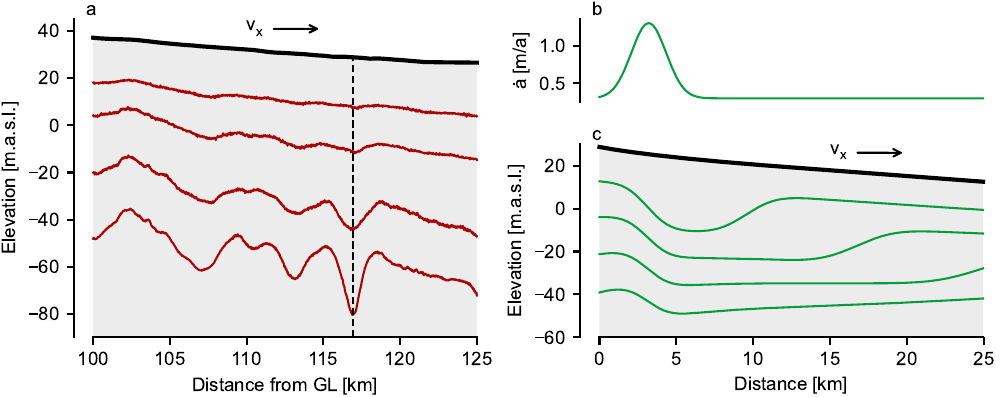}
    \caption{\textbf{IRHs from Ekström Ice Shelf motivate use of Lagrangian parameterization of mass balance parameters}.  \textbf{a:} Zoom in on observed IRHs from Ekström Ice Shelf dataset. Deeper IRHs show perturbations at the same location (dashed black line), but the magnitude of the perturbation increases with depth. This suggests the local perturbation gets reinforced with time, not only transported by ice flow. \textbf{b:} Surface accumulation rate parameterization for an idealized ice shelf example (basal melt rate is constant across domain). \textbf{c}: Selected isochronal layers from noiseless internal stratigraphy model. The local perturbation in the layer elevation is transported by ice flow. Arrows mark flow direction.}
    \label{fig: Lagrangian MB}
\end{figure*}

The fidelity at which the posterior predictive distributions reproduce the observed IRHs of Ekström Ice Shelf (Figs. \ref{fig:posterior Ekstrom}, \ref{fig:Ekstrom deep}) justify our modeling choices for this ice shelf, as the combination of the forward model and accumulation rate prior distribution are sufficiently expressive to reproduce the IRHs. However, we see that local perturbations in the IRH elevations are amplified for deeper IRHs in the observed dataset (Fig.~\ref{fig: Lagrangian MB}c). This is in contrast to the layer tracing model (Fig.~\ref{fig: Lagrangian MB}b), where the layer elevation perturbations are transported by the flow without amplification. Thus, a Lagrangian scheme, which takes the flow velocity into account, could be beneficial to parameterize accumulation rates and capture this behaviour in future work. In such a scheme, the surface accumulation profile would be coupled to the ice shelf surface topography. This approach is supported by studies of the atmospheric forcing that determines surface accumulation \cite{Gow1965,Dattler2019}, which suggest that local features in the surface topography can cause local surface accumulation perturbations.  In the plug flow regime of the shelf, the isochronal layers are transported at the same speed as the surface. Hence a local perturbation in the surface accumulation would continue to amplify local perturbations in the isochronal layer elevations. This would be captured by a Lagrangian parameterization scheme.

Although the steady state assumption was appropriate for Ekström Ice Shelf, it will need to be relaxed in order to apply our methodology to non-steady state flow regimes. While it is possible to couple the layer tracing scheme to an external solver of ice dynamics (e.g., \citeA{Born2021}), the main challenge will be the added computational cost of modeling ice flow as opposed to only the steady state internal stratigraphy. Here, a new family of fast ice flow solvers could be of great use \cite{Sandip2023}. An alternative approach is \textit{probabilistic numerics}, in which the forward and inverse problems are solved simultaneously \cite{Kramer2022,HennigPN}. Finally, the forward model can be extended to remove the plug flow assumption, to allow for depth-dependent velocities. This would allow our approach to be applied to grounded ice.

\subsection{Simulation Based Inference as a Tool for Geoscientific Inversion Problems}
 The inverse problem tackled in this work typifies geoscientific inverse problems, as the forward model is defined in terms of a partial differential equation (PDE), and the parameters are high dimensional and vary in space. Hence, it is valuable to compare the SBI approach in this case to the wide variety of methods and algorithms that have been developed to solve geoscientific inverse problems. 

The SBI approach as presented here has two key features. First, we estimate the Bayesian posterior distribution, providing quantitative uncertainty estimates. Modeling uncertainty is important as it can highly influence and propagate to future modeling predictions. Additionally, locations of high uncertainty show areas requiring further study, helping to guide future work. In contrast, deterministic inference methods typically only attempt to infer a single estimate of best-fitting parameters, e.g., the maximum likelihood estimate (MLE), and provide no measure of uncertainty. Estimates of uncertainty are also present in other inference approaches, for example in Markov Chain Monte Carlo (MCMC, \citeA{Gallagher2009}), Bayesian filtering approaches \cite{Stordal2011, VanLeeuwen2019}, and variational inference \cite{Zhang2021}. However, these methods require knowledge of the likelihood function of the forward model. For computationally complex forward models, such as PDE solvers, the likelihood is often computationally intractable. The SBI approach performs approximate Bayesian computation (ABC, \citeA{Rubin1984,Sisson2018}), and only requires us to be able to evaluate the forward model. Thus, ABC methods can be applied to a larger class of inference problems. Second, a unique advantage of \textit{single round} SBI methods \cite{Cranmer2020} such as NPE (as used in our study) is \textit{amortization}. An amortized inference framework is one that, once trained, can be applied to find the posterior distribution $p(\theta|X)$ for any measurement $X$ without any additional simulation or training costs. Our method as presented here is not yet fully amortized as preprocessing relies on the observed value of $X$: in order to train the density estimator $q_{\phi}(\mathbf{\dot{a}}|X_{k}^{m})$, we first calculate $X_{k}^{m}$ for each simulation dependent on the value of $X_{o}^{m}$. Regardless, our method still amortizes the cost of simulating the forward model many times, which is by far the largest computational cost in the approach. In particular, we solved $M$ parallel inference problems that shared the same prior distribution, and where the forward model could output the $M$ observations in one evaluation. Therefore, the same set of simulations was shared for all inference problems. In the Ekström example, we have evaluated the forward model a total of 190,000 times, accounting for approximately 99\% of the total computation cost (\ref{sec: computational cost}). Thus, we have amortized the majority of the computational cost of inference.

On the other hand, SBI faces some limitations as an inference tool. Primarily, SBI methods are known to require a large number of simulations to be trained, \cite{Lueckmann2021}. Performing on the order of $10^{5}$ simulations would not have been feasible for forward models of higher computational complexity. Moreover, SBI methods scale poorly in number of simulations required as the size of the parameter vector increases. The SBI approach needs to be adapted to more efficiently represent high-dimensional, spatially varying parameters $\theta$ at high resolutions. Some potential approaches are polynomial or spectral representations. Future work should also explore variants of SBI that are better suited to high-dimensional or even continuous  parameters \cite{Ramesh2022, Geffner2022}. Finally, SBI works under the assumption that the forward model is well-specified, meaning that given samples from the prior, it can generate simulations closely resembling the observation. The posteriors obtained by SBI can be strongly biased when this is not the case \cite{Cannon2022}. Work to address this concern has been done, e.g.~by incorporating the model mismatch into the forward model \cite{Ward2022}, as done in our work using the calibrated noise model.

\section{Conclusions} \label{sec:conclusion}
We presented a novel approach for inferring the spatially-varying surface accumulation and basal melt rates along ice shelf flow lines from radar measurements of their internal stratigraphy. We validated the method on a synthetic ice shelf example, and inferred the surface accumulation and basal melt rates along a flow line in Ekström Ice Shelf, Antarctica. We separately inferred the mass balance parameters from four different internal reflection horizons, obtaining posterior distributions consistent with a steady state ice shelf. The inferred distributions were further validated by independent stake array measurements of surface accumulation rates uniquely available in Ekström Ice Shelf. Using our approach, we were able to estimate the otherwise unknown age of the internal reflection horizons as 42, 84, 146 and 188 years. The presented approach can be transferred to other Antarctic ice shelves, and also to other flow regimes such as grounded ice. A strength of our approach is the principled uncertainty estimates in the inferred surface accumulation and basal melt rates. Uncertainty estimates can be integrated in future projections of the Antarctic Ice Sheet \cite{Verjans2022,Ultee2023}. We identified avenues for future work as more can be learned by relaxing the steady state assumption on the ice shelf. The forward model and inference framework should be adapted to account for potential transient signals in the mass balance parameters.

This work was an example use case of SBI for a geoscientific inverse problem. We showcased the strengths of SBI as a likelihood-free approach to approximate the Bayesian posterior, amortizing the cost of simulating the forward model many times.  SBI can become more applicable to such inverse problems involving spatially- (and temporally-) varying parameters if it can be extended to deal with the challenge of high-dimensional parameter inference.

Finally, our approach highlights the value of internal stratigraphy measurements. Initiatives to map the Antarctic-wide internal stratigraphy (e.g. \citeA{AntArchitecture}) can provide invaluable data towards uncovering the history of the Antarctic ice sheet. Sophisticated inference methods could be combined with such a dataset to provide a new, independent, Antarctica-wide parameterization of accumulation and basal melt rate histories.

\section*{Open Research}
\subsection*{Data Availability Statement}
Simulation data available at \url{https://doi.org/doi:10.5281/zenodo.10245153}.
\subsection*{Software Availability}
Code for preprocessing Ekström ice shelf data and generating synthetic ice shelf data available at \url{https://github.com/mackelab/preprocessing-ice-data}.

Code for layer tracing forward model and simulation-based inference workflow available at \url{https://github.com/mackelab/sbi-ice}.

\acknowledgments
The authors would like to thank Daniel Shapero for his inputs on use of icepack. The authors would also like to thank Andreas Born and Therese Riekch for insightful discussions on the implementation of the layer tracing solver for calculating internal stratigraphy. We acknowledge excellent logistic support from staff at Neumayer III station and the GrouZe team on-site.

This work was funded by the German Research Foundation (DFG) under Germany’s Excellence Strategy – EXC number 2064/1 – 390727645 and SFB 1233 'Robust Vision' (276693517) and the German Federal Ministry of Education and Research (BMBF): Tübingen AI Center, FKZ: 01IS18039A. Reinhard Drews and Vjeran Višnjević were supported by an Emmy Noether grant of the Deutsche Forschungsgemeinschaft (DR 822/3-1). We acknowledge the support by the German Academic Scholarship Foundation to Falk M. Oraschewski. Guy Moss is a member of the International Max Planck Research School for Intelligent Systems (IMPRS-IS).

\bibliography{bibliography}

\begin{thebibliography}{}

\bibitem [\protect \citeauthoryear {%
Adusumilli%
, Fricker%
, Medley%
, Padman%
\BCBL {}\ \BBA {} Siegfried%
}{%
Adusumilli%
\ \protect \BOthers {.}}{%
{\protect \APACyear {2020}}%
}]{%
Adusumilli2020}
\APACinsertmetastar {%
Adusumilli2020}%
\begin{APACrefauthors}%
Adusumilli, S.%
, Fricker, H\BPBI A.%
, Medley, B.%
, Padman, L.%
\BCBL {}\ \BBA {} Siegfried, M\BPBI R.%
\end{APACrefauthors}%
\unskip\
\newblock
\APACrefYearMonthDay{2020}{8}{}.
\newblock
{\BBOQ}\APACrefatitle {Interannual variations in meltwater input to the
  Southern Ocean from Antarctic ice shelves} {Interannual variations in
  meltwater input to the southern ocean from antarctic ice shelves}.{\BBCQ}
\newblock
\APACjournalVolNumPages{Nature Geoscience 2020 13:9}{13}{}{616-620}.
\newblock
\begin{APACrefDOI} \doi{10.1038/s41561-020-0616-z} \end{APACrefDOI}
\PrintBackRefs{\CurrentBib}

\bibitem [\protect \citeauthoryear {%
Agosta%
\ \protect \BOthers {.}}{%
Agosta%
\ \protect \BOthers {.}}{%
{\protect \APACyear {2019}}%
}]{%
Agosta2019}
\APACinsertmetastar {%
Agosta2019}%
\begin{APACrefauthors}%
Agosta, C.%
, Amory, C.%
, Kittel, C.%
, Orsi, A.%
, Favier, V.%
, Gall\'ee, H.%
\BDBL {}Fettweis, X.%
\end{APACrefauthors}%
\unskip\
\newblock
\APACrefYearMonthDay{2019}{}{}.
\newblock
{\BBOQ}\APACrefatitle {Estimation of the Antarctic surface mass balance using
  the regional climate model MAR (1979--2015) and identification of dominant
  processes} {Estimation of the antarctic surface mass balance using the
  regional climate model mar (1979--2015) and identification of dominant
  processes}.{\BBCQ}
\newblock
\APACjournalVolNumPages{The Cryosphere}{13}{1}{281--296}.
\newblock
\begin{APACrefDOI} \doi{10.5194/tc-13-281-2019} \end{APACrefDOI}
\PrintBackRefs{\CurrentBib}

\bibitem [\protect \citeauthoryear {%
Allgeier%
\ \BBA {} Cirpka%
}{%
Allgeier%
\ \BBA {} Cirpka%
}{%
{\protect \APACyear {2023}}%
}]{%
Allgeier2023}
\APACinsertmetastar {%
Allgeier2023}%
\begin{APACrefauthors}%
Allgeier, J.%
\BCBT {}\ \BBA {} Cirpka, O\BPBI A.%
\end{APACrefauthors}%
\unskip\
\newblock
\APACrefYearMonthDay{2023}{}{}.
\newblock
{\BBOQ}\APACrefatitle {Surrogate-Model Assisted Plausibility-Check,
  Calibration, and Posterior-Distribution Evaluation of Subsurface-Flow Models}
  {Surrogate-model assisted plausibility-check, calibration, and
  posterior-distribution evaluation of subsurface-flow models}.{\BBCQ}
\newblock
\APACjournalVolNumPages{Water Resources Research}{59}{7}{}.
\newblock
\begin{APACrefDOI} \doi{https://doi.org/10.1029/2023WR034453} \end{APACrefDOI}
\PrintBackRefs{\CurrentBib}

\bibitem [\protect \citeauthoryear {%
Asay-Davis%
\ \protect \BOthers {.}}{%
Asay-Davis%
\ \protect \BOthers {.}}{%
{\protect \APACyear {2016}}%
}]{%
Asay-Davis2016}
\APACinsertmetastar {%
Asay-Davis2016}%
\begin{APACrefauthors}%
Asay-Davis, X\BPBI S.%
, Cornford, S\BPBI L.%
, Durand, G.%
, Galton-Fenzi, B\BPBI K.%
, Gladstone, R\BPBI M.%
, Gudmundsson, G\BPBI H.%
\BDBL {}Seroussi, H.%
\end{APACrefauthors}%
\unskip\
\newblock
\APACrefYearMonthDay{2016}{7}{}.
\newblock
{\BBOQ}\APACrefatitle {Experimental design for three interrelated marine ice
  sheet and ocean model intercomparison projects: MISMIP v. 3 (MISMIP +),
  ISOMIP v. 2 (ISOMIP +) and MISOMIP v. 1 (MISOMIP1)} {Experimental design for
  three interrelated marine ice sheet and ocean model intercomparison projects:
  Mismip v. 3 (mismip +), isomip v. 2 (isomip +) and misomip v. 1
  (misomip1)}.{\BBCQ}
\newblock
\APACjournalVolNumPages{Geoscientific Model Development}{9}{}{2471-2497}.
\newblock
\begin{APACrefDOI} \doi{10.5194/GMD-9-2471-2016} \end{APACrefDOI}
\PrintBackRefs{\CurrentBib}

\bibitem [\protect \citeauthoryear {%
Berger%
, Drews%
, Helm%
, Sun%
\BCBL {}\ \BBA {} Pattyn%
}{%
Berger%
\ \protect \BOthers {.}}{%
{\protect \APACyear {2017}}%
}]{%
Berger2017}
\APACinsertmetastar {%
Berger2017}%
\begin{APACrefauthors}%
Berger, S.%
, Drews, R.%
, Helm, V.%
, Sun, S.%
\BCBL {}\ \BBA {} Pattyn, F.%
\end{APACrefauthors}%
\unskip\
\newblock
\APACrefYearMonthDay{2017}{11}{}.
\newblock
{\BBOQ}\APACrefatitle {Detecting high spatial variability of ice shelf basal
  mass balance, Roi Baudouin Ice Shelf, Antarctica} {Detecting high spatial
  variability of ice shelf basal mass balance, roi baudouin ice shelf,
  antarctica}.{\BBCQ}
\newblock
\APACjournalVolNumPages{Cryosphere}{11}{}{2675-2690}.
\newblock
\begin{APACrefDOI} \doi{10.5194/tc-11-2675-2017} \end{APACrefDOI}
\PrintBackRefs{\CurrentBib}

\bibitem [\protect \citeauthoryear {%
Bindschadler%
\ \protect \BOthers {.}}{%
Bindschadler%
\ \protect \BOthers {.}}{%
{\protect \APACyear {2011}}%
}]{%
Bindschadler2011}
\APACinsertmetastar {%
Bindschadler2011}%
\begin{APACrefauthors}%
Bindschadler, R.%
, Choi, H.%
, Wichlacz, A.%
, Bingham, R.%
, Bohlander, J.%
, Brunt, K.%
\BDBL {}Young, N.%
\end{APACrefauthors}%
\unskip\
\newblock
\APACrefYearMonthDay{2011}{}{}.
\newblock
{\BBOQ}\APACrefatitle {Getting around Antarctica: New high-resolution mappings
  of the grounded and freely-floating boundaries of the Antarctic ice sheet
  created for the International Polar Year} {Getting around antarctica: New
  high-resolution mappings of the grounded and freely-floating boundaries of
  the antarctic ice sheet created for the international polar year}.{\BBCQ}
\newblock
\APACjournalVolNumPages{Cryosphere}{5}{}{569-588}.
\newblock
\begin{APACrefDOI} \doi{10.5194/TC-5-569-2011} \end{APACrefDOI}
\PrintBackRefs{\CurrentBib}

\bibitem [\protect \citeauthoryear {%
Bingham%
\ \protect \BOthers {.}}{%
Bingham%
\ \protect \BOthers {.}}{%
{\protect \APACyear {2019}}%
}]{%
AntArchitecture}
\APACinsertmetastar {%
AntArchitecture}%
\begin{APACrefauthors}%
Bingham, R\BPBI G.%
, Eisen, O.%
, Karlsson, N\BPBI B.%
, MacGregor, J\BPBI A.%
, Ross, N.%
\BCBL {}\ \BBA {} Young, D\BPBI A.%
\end{APACrefauthors}%
\unskip\
\newblock
\APACrefYearMonthDay{2019}{July}{}.
\newblock
{\BBOQ}\APACrefatitle {AntArchitecture: an international project to use
  Antarctic englacial layering to interrogate stability of the Antarctic Ice
  Sheets} {Antarchitecture: an international project to use antarctic englacial
  layering to interrogate stability of the antarctic ice sheets}.{\BBCQ}
\newblock
\BIn{} \APACrefbtitle {IGS Symposium Five Decades of Radioglaciology.} {Igs
  symposium five decades of radioglaciology.}
\PrintBackRefs{\CurrentBib}

\bibitem [\protect \citeauthoryear {%
Born%
}{%
Born%
}{%
{\protect \APACyear {2017}}%
}]{%
Born2017}
\APACinsertmetastar {%
Born2017}%
\begin{APACrefauthors}%
Born, A.%
\end{APACrefauthors}%
\unskip\
\newblock
\APACrefYearMonthDay{2017}{2}{}.
\newblock
{\BBOQ}\APACrefatitle {Tracer transport in an isochronal ice-sheet model}
  {Tracer transport in an isochronal ice-sheet model}.{\BBCQ}
\newblock
\APACjournalVolNumPages{Journal of Glaciology}{63}{}{22-38}.
\newblock
\begin{APACrefDOI} \doi{10.1017/JOG.2016.111} \end{APACrefDOI}
\PrintBackRefs{\CurrentBib}

\bibitem [\protect \citeauthoryear {%
Born%
\ \BBA {} Robinson%
}{%
Born%
\ \BBA {} Robinson%
}{%
{\protect \APACyear {2021}}%
}]{%
Born2021}
\APACinsertmetastar {%
Born2021}%
\begin{APACrefauthors}%
Born, A.%
\BCBT {}\ \BBA {} Robinson, A.%
\end{APACrefauthors}%
\unskip\
\newblock
\APACrefYearMonthDay{2021}{9}{}.
\newblock
{\BBOQ}\APACrefatitle {Modeling the Greenland englacial stratigraphy} {Modeling
  the greenland englacial stratigraphy}.{\BBCQ}
\newblock
\APACjournalVolNumPages{Cryosphere}{15}{}{4539-4556}.
\newblock
\begin{APACrefDOI} \doi{10.5194/TC-15-4539-2021} \end{APACrefDOI}
\PrintBackRefs{\CurrentBib}

\bibitem [\protect \citeauthoryear {%
Burgard%
, Jourdain%
, Reese%
, Jenkins%
\BCBL {}\ \BBA {} Mathiot%
}{%
Burgard%
\ \protect \BOthers {.}}{%
{\protect \APACyear {2022}}%
}]{%
Burgard2022}
\APACinsertmetastar {%
Burgard2022}%
\begin{APACrefauthors}%
Burgard, C.%
, Jourdain, N\BPBI C.%
, Reese, R.%
, Jenkins, A.%
\BCBL {}\ \BBA {} Mathiot, P.%
\end{APACrefauthors}%
\unskip\
\newblock
\APACrefYearMonthDay{2022}{12}{}.
\newblock
{\BBOQ}\APACrefatitle {An assessment of basal melt parameterisations for
  Antarctic ice shelves} {An assessment of basal melt parameterisations for
  antarctic ice shelves}.{\BBCQ}
\newblock
\APACjournalVolNumPages{Cryosphere}{16}{}{4931-4975}.
\newblock
\begin{APACrefDOI} \doi{10.5194/TC-16-4931-2022} \end{APACrefDOI}
\PrintBackRefs{\CurrentBib}

\bibitem [\protect \citeauthoryear {%
{Cannon}%
, {Ward}%
\BCBL {}\ \BBA {} {Schmon}%
}{%
{Cannon}%
\ \protect \BOthers {.}}{%
{\protect \APACyear {2022}}%
}]{%
Cannon2022}
\APACinsertmetastar {%
Cannon2022}%
\begin{APACrefauthors}%
{Cannon}, P.%
, {Ward}, D.%
\BCBL {}\ \BBA {} {Schmon}, S\BPBI M.%
\end{APACrefauthors}%
\unskip\
\newblock
\APACrefYearMonthDay{2022}{{\APACmonth{09}}}{}.
\newblock
{\BBOQ}\APACrefatitle {{Investigating the Impact of Model Misspecification in
  Neural Simulation-based Inference}} {{Investigating the Impact of Model
  Misspecification in Neural Simulation-based Inference}}.{\BBCQ}
\newblock
\APACjournalVolNumPages{arXiv e-prints}{}{}{arXiv:2209.01845}.
\newblock
\begin{APACrefDOI} \doi{10.48550/arXiv.2209.01845} \end{APACrefDOI}
\PrintBackRefs{\CurrentBib}

\bibitem [\protect \citeauthoryear {%
Catania%
, Hulbe%
\BCBL {}\ \BBA {} Conway%
}{%
Catania%
\ \protect \BOthers {.}}{%
{\protect \APACyear {2010}}%
}]{%
Catania2010}
\APACinsertmetastar {%
Catania2010}%
\begin{APACrefauthors}%
Catania, G.%
, Hulbe, C.%
\BCBL {}\ \BBA {} Conway, H.%
\end{APACrefauthors}%
\unskip\
\newblock
\APACrefYearMonthDay{2010}{}{}.
\newblock
{\BBOQ}\APACrefatitle {Grounding-line basal melt rates determined using
  radar-derived internal stratigraphy} {Grounding-line basal melt rates
  determined using radar-derived internal stratigraphy}.{\BBCQ}
\newblock
\APACjournalVolNumPages{Journal of Glaciology}{56}{197}{545–554}.
\newblock
\begin{APACrefDOI} \doi{10.3189/002214310792447842} \end{APACrefDOI}
\PrintBackRefs{\CurrentBib}

\bibitem [\protect \citeauthoryear {%
Cranmer%
, Brehmer%
\BCBL {}\ \BBA {} Louppe%
}{%
Cranmer%
\ \protect \BOthers {.}}{%
{\protect \APACyear {2020}}%
}]{%
Cranmer2020}
\APACinsertmetastar {%
Cranmer2020}%
\begin{APACrefauthors}%
Cranmer, K.%
, Brehmer, J.%
\BCBL {}\ \BBA {} Louppe, G.%
\end{APACrefauthors}%
\unskip\
\newblock
\APACrefYearMonthDay{2020}{}{}.
\newblock
{\BBOQ}\APACrefatitle {The frontier of simulation-based inference} {The
  frontier of simulation-based inference}.{\BBCQ}
\newblock
\APACjournalVolNumPages{Proceedings of the National Academy of
  Sciences}{117}{48}{30055-30062}.
\newblock
\begin{APACrefDOI} \doi{10.1073/pnas.1912789117} \end{APACrefDOI}
\PrintBackRefs{\CurrentBib}

\bibitem [\protect \citeauthoryear {%
Das%
\ \protect \BOthers {.}}{%
Das%
\ \protect \BOthers {.}}{%
{\protect \APACyear {2020}}%
}]{%
Das2020}
\APACinsertmetastar {%
Das2020}%
\begin{APACrefauthors}%
Das, I.%
, Padman, L.%
, Bell, R\BPBI E.%
, Fricker, H\BPBI A.%
, Tinto, K\BPBI J.%
, Hulbe, C\BPBI L.%
\BDBL {}Siegfried, M\BPBI R.%
\end{APACrefauthors}%
\unskip\
\newblock
\APACrefYearMonthDay{2020}{}{}.
\newblock
{\BBOQ}\APACrefatitle {Multidecadal Basal Melt Rates and Structure of the Ross
  Ice Shelf, Antarctica, Using Airborne Ice Penetrating Radar} {Multidecadal
  basal melt rates and structure of the ross ice shelf, antarctica, using
  airborne ice penetrating radar}.{\BBCQ}
\newblock
\APACjournalVolNumPages{Journal of Geophysical Research: Earth
  Surface}{125}{3}{e2019JF005241}.
\newblock
\begin{APACrefDOI} \doi{https://doi.org/10.1029/2019JF005241} \end{APACrefDOI}
\PrintBackRefs{\CurrentBib}

\bibitem [\protect \citeauthoryear {%
Dattler%
, Lenaerts%
\BCBL {}\ \BBA {} Medley%
}{%
Dattler%
\ \protect \BOthers {.}}{%
{\protect \APACyear {2019}}%
}]{%
Dattler2019}
\APACinsertmetastar {%
Dattler2019}%
\begin{APACrefauthors}%
Dattler, M\BPBI E.%
, Lenaerts, J\BPBI T\BPBI M.%
\BCBL {}\ \BBA {} Medley, B.%
\end{APACrefauthors}%
\unskip\
\newblock
\APACrefYearMonthDay{2019}{}{}.
\newblock
{\BBOQ}\APACrefatitle {Significant Spatial Variability in Radar-Derived West
  Antarctic Accumulation Linked to Surface Winds and Topography} {Significant
  spatial variability in radar-derived west antarctic accumulation linked to
  surface winds and topography}.{\BBCQ}
\newblock
\APACjournalVolNumPages{Geophysical Research Letters}{46}{22}{13126-13134}.
\newblock
\begin{APACrefDOI} \doi{https://doi.org/10.1029/2019GL085363} \end{APACrefDOI}
\PrintBackRefs{\CurrentBib}

\bibitem [\protect \citeauthoryear {%
Depoorter%
\ \protect \BOthers {.}}{%
Depoorter%
\ \protect \BOthers {.}}{%
{\protect \APACyear {2013}}%
}]{%
Depoorter2013}
\APACinsertmetastar {%
Depoorter2013}%
\begin{APACrefauthors}%
Depoorter, M\BPBI A.%
, Bamber, J\BPBI L.%
, Griggs, J\BPBI A.%
, Lenaerts, J\BPBI T\BPBI M.%
, Ligtenberg, S\BPBI R.%
, Broeke, M\BPBI R\BPBI V\BPBI D.%
\BCBL {}\ \BBA {} Moholdt, G.%
\end{APACrefauthors}%
\unskip\
\newblock
\APACrefYearMonthDay{2013}{9}{}.
\newblock
{\BBOQ}\APACrefatitle {Calving fluxes and basal melt rates of Antarctic ice
  shelves} {Calving fluxes and basal melt rates of antarctic ice
  shelves}.{\BBCQ}
\newblock
\APACjournalVolNumPages{Nature 2013 502:7469}{502}{}{89-92}.
\newblock
\begin{APACrefDOI} \doi{10.1038/nature12567} \end{APACrefDOI}
\PrintBackRefs{\CurrentBib}

\bibitem [\protect \citeauthoryear {%
Drews%
}{%
Drews%
}{%
{\protect \APACyear {2015}}%
}]{%
Drews2015channels}
\APACinsertmetastar {%
Drews2015channels}%
\begin{APACrefauthors}%
Drews, R.%
\end{APACrefauthors}%
\unskip\
\newblock
\APACrefYearMonthDay{2015}{6}{}.
\newblock
{\BBOQ}\APACrefatitle {Evolution of ice-shelf channels in Antarctic ice
  shelves} {Evolution of ice-shelf channels in antarctic ice shelves}.{\BBCQ}
\newblock
\APACjournalVolNumPages{Cryosphere}{9}{}{1169-1181}.
\newblock
\begin{APACrefDOI} \doi{10.5194/TC-9-1169-2015} \end{APACrefDOI}
\PrintBackRefs{\CurrentBib}

\bibitem [\protect \citeauthoryear {%
Drews%
\ \protect \BOthers {.}}{%
Drews%
\ \protect \BOthers {.}}{%
{\protect \APACyear {2016}}%
}]{%
Drews2016}
\APACinsertmetastar {%
Drews2016}%
\begin{APACrefauthors}%
Drews, R.%
, Brown, J.%
, Matsuoka, K.%
, Witrant, E.%
, Philippe, M.%
, Hubbard, B.%
\BCBL {}\ \BBA {} Pattyn, F.%
\end{APACrefauthors}%
\unskip\
\newblock
\APACrefYearMonthDay{2016}{}{}.
\newblock
{\BBOQ}\APACrefatitle {Constraining variable density of ice shelves using
  wide-angle radar~measurements} {Constraining variable density of ice shelves
  using wide-angle radar~measurements}.{\BBCQ}
\newblock
\APACjournalVolNumPages{The Cryosphere}{10}{2}{811--823}.
\newblock
\begin{APACrefDOI} \doi{10.5194/tc-10-811-2016} \end{APACrefDOI}
\PrintBackRefs{\CurrentBib}

\bibitem [\protect \citeauthoryear {%
Drews%
, Martín%
, Steinhage%
\BCBL {}\ \BBA {} Eisen%
}{%
Drews%
\ \protect \BOthers {.}}{%
{\protect \APACyear {2013}}%
}]{%
Drews2013}
\APACinsertmetastar {%
Drews2013}%
\begin{APACrefauthors}%
Drews, R.%
, Martín, C.%
, Steinhage, D.%
\BCBL {}\ \BBA {} Eisen, O.%
\end{APACrefauthors}%
\unskip\
\newblock
\APACrefYearMonthDay{2013}{}{}.
\newblock
{\BBOQ}\APACrefatitle {Characterizing the glaciological conditions at
  Halvfarryggen ice dome, Dronning Maud Land, Antarctica} {Characterizing the
  glaciological conditions at halvfarryggen ice dome, dronning maud land,
  antarctica}.{\BBCQ}
\newblock
\APACjournalVolNumPages{Journal of Glaciology}{59}{213}{9–20}.
\newblock
\begin{APACrefDOI} \doi{10.3189/2013JoG12J134} \end{APACrefDOI}
\PrintBackRefs{\CurrentBib}

\bibitem [\protect \citeauthoryear {%
Drews%
\ \protect \BOthers {.}}{%
Drews%
\ \protect \BOthers {.}}{%
{\protect \APACyear {2015}}%
}]{%
Drews2015evolution}
\APACinsertmetastar {%
Drews2015evolution}%
\begin{APACrefauthors}%
Drews, R.%
, Matsuoka, K.%
, Martín, C.%
, Callens, D.%
, Bergeot, N.%
\BCBL {}\ \BBA {} Pattyn, F.%
\end{APACrefauthors}%
\unskip\
\newblock
\APACrefYearMonthDay{2015}{3}{}.
\newblock
{\BBOQ}\APACrefatitle {Evolution of Derwael Ice Rise in Dronning Maud Land,
  Antarctica, over the last millennia} {Evolution of derwael ice rise in
  dronning maud land, antarctica, over the last millennia}.{\BBCQ}
\newblock
\APACjournalVolNumPages{Journal of Geophysical Research: Earth
  Surface}{120}{}{564-579}.
\newblock
\begin{APACrefDOI} \doi{10.1002/2014JF003246} \end{APACrefDOI}
\PrintBackRefs{\CurrentBib}

\bibitem [\protect \citeauthoryear {%
{Durkan}%
, {Bekasov}%
, {Murray}%
\BCBL {}\ \BBA {} {Papamakarios}%
}{%
{Durkan}%
\ \protect \BOthers {.}}{%
{\protect \APACyear {2019}}%
}]{%
Durkan2019}
\APACinsertmetastar {%
Durkan2019}%
\begin{APACrefauthors}%
{Durkan}, C.%
, {Bekasov}, A.%
, {Murray}, I.%
\BCBL {}\ \BBA {} {Papamakarios}, G.%
\end{APACrefauthors}%
\unskip\
\newblock
\APACrefYearMonthDay{2019}{{\APACmonth{06}}}{}.
\newblock
{\BBOQ}\APACrefatitle {{Neural Spline Flows}} {{Neural Spline Flows}}.{\BBCQ}
\newblock
\APACjournalVolNumPages{arXiv e-prints}{}{}{arXiv:1906.04032}.
\newblock
\begin{APACrefDOI} \doi{10.48550/arXiv.1906.04032} \end{APACrefDOI}
\PrintBackRefs{\CurrentBib}

\bibitem [\protect \citeauthoryear {%
Dutrieux%
\ \protect \BOthers {.}}{%
Dutrieux%
\ \protect \BOthers {.}}{%
{\protect \APACyear {2014}}%
}]{%
Dutrieux2014}
\APACinsertmetastar {%
Dutrieux2014}%
\begin{APACrefauthors}%
Dutrieux, P.%
, Stewart, C.%
, Jenkins, A.%
, Nicholls, K\BPBI W.%
, Corr, H\BPBI F.%
, Rignot, E.%
\BCBL {}\ \BBA {} Steffen, K.%
\end{APACrefauthors}%
\unskip\
\newblock
\APACrefYearMonthDay{2014}{8}{}.
\newblock
{\BBOQ}\APACrefatitle {Basal terraces on melting ice shelves} {Basal terraces
  on melting ice shelves}.{\BBCQ}
\newblock
\APACjournalVolNumPages{Geophysical Research Letters}{41}{}{5506-5513}.
\newblock
\begin{APACrefDOI} \doi{10.1002/2014GL060618} \end{APACrefDOI}
\PrintBackRefs{\CurrentBib}

\bibitem [\protect \citeauthoryear {%
Eisen%
}{%
Eisen%
}{%
{\protect \APACyear {2008}}%
}]{%
Eisen2008Inference}
\APACinsertmetastar {%
Eisen2008Inference}%
\begin{APACrefauthors}%
Eisen, O.%
\end{APACrefauthors}%
\unskip\
\newblock
\APACrefYearMonthDay{2008}{}{}.
\newblock
{\BBOQ}\APACrefatitle {Inference of velocity pattern from isochronous layers in
  firn, using an inverse method} {Inference of velocity pattern from
  isochronous layers in firn, using an inverse method}.{\BBCQ}
\newblock
\APACjournalVolNumPages{Journal of Glaciology}{54}{187}{613–630}.
\newblock
\begin{APACrefDOI} \doi{10.3189/002214308786570818} \end{APACrefDOI}
\PrintBackRefs{\CurrentBib}

\bibitem [\protect \citeauthoryear {%
Eisen%
\ \protect \BOthers {.}}{%
Eisen%
\ \protect \BOthers {.}}{%
{\protect \APACyear {2008}}%
}]{%
Eisen2008Review}
\APACinsertmetastar {%
Eisen2008Review}%
\begin{APACrefauthors}%
Eisen, O.%
, Frezzotti, M.%
, Genthon, C.%
, Isaksson, E.%
, Magand, O.%
, van~den Broeke, M\BPBI R.%
\BDBL {}Vaughan, D\BPBI G.%
\end{APACrefauthors}%
\unskip\
\newblock
\APACrefYearMonthDay{2008}{}{}.
\newblock
{\BBOQ}\APACrefatitle {Ground-based measurements of spatial and temporal
  variability of snow accumulation in East Antarctica} {Ground-based
  measurements of spatial and temporal variability of snow accumulation in east
  antarctica}.{\BBCQ}
\newblock
\APACjournalVolNumPages{Reviews of Geophysics}{46}{2}{}.
\newblock
\begin{APACrefDOI} \doi{https://doi.org/10.1029/2006RG000218} \end{APACrefDOI}
\PrintBackRefs{\CurrentBib}

\bibitem [\protect \citeauthoryear {%
Eisen%
, Nixdorf%
, Wilhelms%
\BCBL {}\ \BBA {} Miller%
}{%
Eisen%
\ \protect \BOthers {.}}{%
{\protect \APACyear {2004}}%
}]{%
Eisen2004}
\APACinsertmetastar {%
Eisen2004}%
\begin{APACrefauthors}%
Eisen, O.%
, Nixdorf, U.%
, Wilhelms, F.%
\BCBL {}\ \BBA {} Miller, H.%
\end{APACrefauthors}%
\unskip\
\newblock
\APACrefYearMonthDay{2004}{4}{}.
\newblock
{\BBOQ}\APACrefatitle {Age estimates of isochronous reflection horizons by
  combining ice core, survey, and synthetic radar data} {Age estimates of
  isochronous reflection horizons by combining ice core, survey, and synthetic
  radar data}.{\BBCQ}
\newblock
\APACjournalVolNumPages{Journal of Geophysical Research: Solid
  Earth}{109}{}{4106}.
\newblock
\begin{APACrefDOI} \doi{10.1029/2003JB002858} \end{APACrefDOI}
\PrintBackRefs{\CurrentBib}

\bibitem [\protect \citeauthoryear {%
Gallagher%
, Charvin%
, Nielsen%
, Sambridge%
\BCBL {}\ \BBA {} Stephenson%
}{%
Gallagher%
\ \protect \BOthers {.}}{%
{\protect \APACyear {2009}}%
}]{%
Gallagher2009}
\APACinsertmetastar {%
Gallagher2009}%
\begin{APACrefauthors}%
Gallagher, K.%
, Charvin, K.%
, Nielsen, S.%
, Sambridge, M.%
\BCBL {}\ \BBA {} Stephenson, J.%
\end{APACrefauthors}%
\unskip\
\newblock
\APACrefYearMonthDay{2009}{}{}.
\newblock
{\BBOQ}\APACrefatitle {Markov chain Monte Carlo (MCMC) sampling methods to
  determine optimal models, model resolution and model choice for Earth Science
  problems} {Markov chain monte carlo (mcmc) sampling methods to determine
  optimal models, model resolution and model choice for earth science
  problems}.{\BBCQ}
\newblock
\APACjournalVolNumPages{Marine and Petroleum Geology}{26}{4}{525-535}.
\newblock
\begin{APACrefDOI} \doi{https://doi.org/10.1016/j.marpetgeo.2009.01.003}
  \end{APACrefDOI}
\PrintBackRefs{\CurrentBib}

\bibitem [\protect \citeauthoryear {%
Gallée%
\ \BBA {} Schayes%
}{%
Gallée%
\ \BBA {} Schayes%
}{%
{\protect \APACyear {1994}}%
}]{%
Gallee1994}
\APACinsertmetastar {%
Gallee1994}%
\begin{APACrefauthors}%
Gallée, H.%
\BCBT {}\ \BBA {} Schayes, G.%
\end{APACrefauthors}%
\unskip\
\newblock
\APACrefYearMonthDay{1994}{}{}.
\newblock
{\BBOQ}\APACrefatitle {Development of a Three-Dimensional Meso-{$\gamma$}
  Primitive Equation Model: Katabatic Winds Simulation in the Area of Terra
  Nova Bay, Antarctica} {Development of a three-dimensional meso-{$\gamma$}
  primitive equation model: Katabatic winds simulation in the area of terra
  nova bay, antarctica}.{\BBCQ}
\newblock
\APACjournalVolNumPages{Monthly Weather Review}{122}{4}{671 - 685}.
\newblock
\begin{APACrefDOI}
  \doi{https://doi.org/10.1175/1520-0493(1994)122<0671:DOATDM>2.0.CO;2}
  \end{APACrefDOI}
\PrintBackRefs{\CurrentBib}

\bibitem [\protect \citeauthoryear {%
Gardner%
, Fahnestock%
\BCBL {}\ \BBA {} Scambos.%
}{%
Gardner%
\ \protect \BOthers {.}}{%
{\protect \APACyear {2022}}%
}]{%
ITS_LIVE}
\APACinsertmetastar {%
ITS_LIVE}%
\begin{APACrefauthors}%
Gardner, A.%
, Fahnestock, M.%
\BCBL {}\ \BBA {} Scambos., T.%
\end{APACrefauthors}%
\unskip\
\newblock
\APACrefYearMonthDay{2022}{}{}.
\newblock
\APACrefbtitle {MEaSUREs ITS{\_}LIVE Landsat Image-Pair Glacier and Ice Sheet
  Surface Velocities, Version 1.} {Measures its{\_}live landsat image-pair
  glacier and ice sheet surface velocities, version 1.}
\newblock
\APACaddressPublisher{}{NASA National Snow and Ice Data Center Distributed
  Active Archive Center}.
\newblock
\begin{APACrefDOI} \doi{10.5067/IMR9D3PEI28U} \end{APACrefDOI}
\PrintBackRefs{\CurrentBib}

\bibitem [\protect \citeauthoryear {%
{Geffner}%
, {Papamakarios}%
\BCBL {}\ \BBA {} {Mnih}%
}{%
{Geffner}%
\ \protect \BOthers {.}}{%
{\protect \APACyear {2022}}%
}]{%
Geffner2022}
\APACinsertmetastar {%
Geffner2022}%
\begin{APACrefauthors}%
{Geffner}, T.%
, {Papamakarios}, G.%
\BCBL {}\ \BBA {} {Mnih}, A.%
\end{APACrefauthors}%
\unskip\
\newblock
\APACrefYearMonthDay{2022}{{\APACmonth{09}}}{}.
\newblock
{\BBOQ}\APACrefatitle {{Compositional Score Modeling for Simulation-based
  Inference}} {{Compositional Score Modeling for Simulation-based
  Inference}}.{\BBCQ}
\newblock
\APACjournalVolNumPages{arXiv e-prints}{}{}{arXiv:2209.14249}.
\newblock
\begin{APACrefDOI} \doi{10.48550/arXiv.2209.14249} \end{APACrefDOI}
\PrintBackRefs{\CurrentBib}

\bibitem [\protect \citeauthoryear {%
Gladstone%
\ \protect \BOthers {.}}{%
Gladstone%
\ \protect \BOthers {.}}{%
{\protect \APACyear {2021}}%
}]{%
Gladstone2021}
\APACinsertmetastar {%
Gladstone2021}%
\begin{APACrefauthors}%
Gladstone, R.%
, Galton-Fenzi, B.%
, Gwyther, D.%
, Zhou, Q.%
, Hattermann, T.%
, Zhao, C.%
\BDBL {}Moore, J.%
\end{APACrefauthors}%
\unskip\
\newblock
\APACrefYearMonthDay{2021}{2}{}.
\newblock
{\BBOQ}\APACrefatitle {The Framework for Ice Sheet-Ocean Coupling (FISOC) V1.1}
  {The framework for ice sheet-ocean coupling (fisoc) v1.1}.{\BBCQ}
\newblock
\APACjournalVolNumPages{Geoscientific Model Development}{14}{}{889-905}.
\newblock
\begin{APACrefDOI} \doi{10.5194/GMD-14-889-2021} \end{APACrefDOI}
\PrintBackRefs{\CurrentBib}

\bibitem [\protect \citeauthoryear {%
Goelzer%
, Huybrechts%
, Loutre%
\BCBL {}\ \BBA {} Fichefet%
}{%
Goelzer%
\ \protect \BOthers {.}}{%
{\protect \APACyear {2016}}%
}]{%
Goezler2016}
\APACinsertmetastar {%
Goezler2016}%
\begin{APACrefauthors}%
Goelzer, H.%
, Huybrechts, P.%
, Loutre, M\BHBI F.%
\BCBL {}\ \BBA {} Fichefet, T.%
\end{APACrefauthors}%
\unskip\
\newblock
\APACrefYearMonthDay{2016}{}{}.
\newblock
{\BBOQ}\APACrefatitle {Last Interglacial climate and sea-level evolution from a
  coupled ice sheet--climate model} {Last interglacial climate and sea-level
  evolution from a coupled ice sheet--climate model}.{\BBCQ}
\newblock
\APACjournalVolNumPages{Climate of the Past}{12}{12}{2195--2213}.
\newblock
\begin{APACrefDOI} \doi{10.5194/cp-12-2195-2016} \end{APACrefDOI}
\PrintBackRefs{\CurrentBib}

\bibitem [\protect \citeauthoryear {%
Goldberg%
, Gourmelen%
, Kimura%
, Millan%
\BCBL {}\ \BBA {} Snow%
}{%
Goldberg%
\ \protect \BOthers {.}}{%
{\protect \APACyear {2019}}%
}]{%
Goldberg2019}
\APACinsertmetastar {%
Goldberg2019}%
\begin{APACrefauthors}%
Goldberg, D\BPBI N.%
, Gourmelen, N.%
, Kimura, S.%
, Millan, R.%
\BCBL {}\ \BBA {} Snow, K.%
\end{APACrefauthors}%
\unskip\
\newblock
\APACrefYearMonthDay{2019}{}{}.
\newblock
{\BBOQ}\APACrefatitle {How Accurately Should We Model Ice Shelf Melt Rates?}
  {How accurately should we model ice shelf melt rates?}{\BBCQ}
\newblock
\APACjournalVolNumPages{Geophysical Research Letters}{46}{1}{189-199}.
\newblock
\begin{APACrefDOI} \doi{https://doi.org/10.1029/2018GL080383} \end{APACrefDOI}
\PrintBackRefs{\CurrentBib}

\bibitem [\protect \citeauthoryear {%
Goldberg%
\ \BBA {} Holland%
}{%
Goldberg%
\ \BBA {} Holland%
}{%
{\protect \APACyear {2022}}%
}]{%
Goldberg2022}
\APACinsertmetastar {%
Goldberg2022}%
\begin{APACrefauthors}%
Goldberg, D\BPBI N.%
\BCBT {}\ \BBA {} Holland, P\BPBI R.%
\end{APACrefauthors}%
\unskip\
\newblock
\APACrefYearMonthDay{2022}{}{}.
\newblock
{\BBOQ}\APACrefatitle {The Relative Impacts of Initialization and Climate
  Forcing in Coupled Ice Sheet-Ocean Modeling: Application to Pope, Smith, and
  Kohler Glaciers} {The relative impacts of initialization and climate forcing
  in coupled ice sheet-ocean modeling: Application to pope, smith, and kohler
  glaciers}.{\BBCQ}
\newblock
\APACjournalVolNumPages{Journal of Geophysical Research: Earth
  Surface}{127}{5}{e2021JF006570}.
\newblock
\begin{APACrefDOI} \doi{https://doi.org/10.1029/2021JF006570} \end{APACrefDOI}
\PrintBackRefs{\CurrentBib}

\bibitem [\protect \citeauthoryear {%
Gow%
\ \BBA {} Rowland%
}{%
Gow%
\ \BBA {} Rowland%
}{%
{\protect \APACyear {1965}}%
}]{%
Gow1965}
\APACinsertmetastar {%
Gow1965}%
\begin{APACrefauthors}%
Gow, A\BPBI J.%
\BCBT {}\ \BBA {} Rowland, R.%
\end{APACrefauthors}%
\unskip\
\newblock
\APACrefYearMonthDay{1965}{}{}.
\newblock
{\BBOQ}\APACrefatitle {On the Relationship of Snow Accumulation to Surface
  Topography at “Byrd Station”, Antarctica} {On the relationship of snow
  accumulation to surface topography at “byrd station”, antarctica}.{\BBCQ}
\newblock
\APACjournalVolNumPages{Journal of Glaciology}{5}{42}{843–847}.
\newblock
\begin{APACrefDOI} \doi{10.3189/S0022143000018906} \end{APACrefDOI}
\PrintBackRefs{\CurrentBib}

\bibitem [\protect \citeauthoryear {%
{Greenberg}%
, {Nonnenmacher}%
\BCBL {}\ \BBA {} {Macke}%
}{%
{Greenberg}%
\ \protect \BOthers {.}}{%
{\protect \APACyear {2019}}%
}]{%
Greenberg2019}
\APACinsertmetastar {%
Greenberg2019}%
\begin{APACrefauthors}%
{Greenberg}, D\BPBI S.%
, {Nonnenmacher}, M.%
\BCBL {}\ \BBA {} {Macke}, J\BPBI H.%
\end{APACrefauthors}%
\unskip\
\newblock
\APACrefYearMonthDay{2019}{{\APACmonth{05}}}{}.
\newblock
{\BBOQ}\APACrefatitle {{Automatic Posterior Transformation for Likelihood-Free
  Inference}} {{Automatic Posterior Transformation for Likelihood-Free
  Inference}}.{\BBCQ}
\newblock
\APACjournalVolNumPages{arXiv e-prints}{}{}{arXiv:1905.07488}.
\newblock
\begin{APACrefDOI} \doi{10.48550/arXiv.1905.07488} \end{APACrefDOI}
\PrintBackRefs{\CurrentBib}

\bibitem [\protect \citeauthoryear {%
Greene%
, Gwyther%
\BCBL {}\ \BBA {} Blankenship%
}{%
Greene%
\ \protect \BOthers {.}}{%
{\protect \APACyear {2017}}%
}]{%
AntarcticMappingTools}
\APACinsertmetastar {%
AntarcticMappingTools}%
\begin{APACrefauthors}%
Greene, C\BPBI A.%
, Gwyther, D\BPBI E.%
\BCBL {}\ \BBA {} Blankenship, D\BPBI D.%
\end{APACrefauthors}%
\unskip\
\newblock
\APACrefYearMonthDay{2017}{jul}{}.
\newblock
{\BBOQ}\APACrefatitle {Antarctic Mapping Tools for Matlab} {Antarctic mapping
  tools for matlab}.{\BBCQ}
\newblock
\APACjournalVolNumPages{Computers {\&} Geosciences}{104}{}{151--157}.
\newblock
\begin{APACrefDOI} \doi{10.1016/j.cageo.2016.08.003} \end{APACrefDOI}
\PrintBackRefs{\CurrentBib}

\bibitem [\protect \citeauthoryear {%
Greve%
\ \BBA {} Blatter%
}{%
Greve%
\ \BBA {} Blatter%
}{%
{\protect \APACyear {2009}}%
}]{%
Greve2009}
\APACinsertmetastar {%
Greve2009}%
\begin{APACrefauthors}%
Greve, R.%
\BCBT {}\ \BBA {} Blatter, H.%
\end{APACrefauthors}%
\unskip\
\newblock
\APACrefYear{2009}.
\newblock
\APACrefbtitle {Dynamics of Ice Sheets and Glaciers} {Dynamics of ice sheets
  and glaciers}.
\newblock
\APACaddressPublisher{}{Springer Berlin Heidelberg}.
\newblock
\begin{APACrefDOI} \doi{10.1007/978-3-642-03415-2} \end{APACrefDOI}
\PrintBackRefs{\CurrentBib}

\bibitem [\protect \citeauthoryear {%
Gudmundsson%
, Paolo%
, Adusumilli%
\BCBL {}\ \BBA {} Fricker%
}{%
Gudmundsson%
\ \protect \BOthers {.}}{%
{\protect \APACyear {2019}}%
}]{%
Gudmundsson2019}
\APACinsertmetastar {%
Gudmundsson2019}%
\begin{APACrefauthors}%
Gudmundsson, G\BPBI H.%
, Paolo, F\BPBI S.%
, Adusumilli, S.%
\BCBL {}\ \BBA {} Fricker, H\BPBI A.%
\end{APACrefauthors}%
\unskip\
\newblock
\APACrefYearMonthDay{2019}{12}{}.
\newblock
{\BBOQ}\APACrefatitle {Instantaneous Antarctic ice sheet mass loss driven by
  thinning ice shelves} {Instantaneous antarctic ice sheet mass loss driven by
  thinning ice shelves}.{\BBCQ}
\newblock
\APACjournalVolNumPages{Geophysical Research Letters}{46}{}{13903-13909}.
\newblock
\begin{APACrefDOI} \doi{10.1029/2019GL085027} \end{APACrefDOI}
\PrintBackRefs{\CurrentBib}

\bibitem [\protect \citeauthoryear {%
Hennig%
, Osborne%
\BCBL {}\ \BBA {} Kersting%
}{%
Hennig%
\ \protect \BOthers {.}}{%
{\protect \APACyear {2022}}%
}]{%
HennigPN}
\APACinsertmetastar {%
HennigPN}%
\begin{APACrefauthors}%
Hennig, P.%
, Osborne, M\BPBI A.%
\BCBL {}\ \BBA {} Kersting, H\BPBI P.%
\end{APACrefauthors}%
\unskip\
\newblock
\APACrefYear{2022}.
\newblock
\APACrefbtitle {Probabilistic Numerics: Computation as Machine Learning}
  {Probabilistic numerics: Computation as machine learning}.
\newblock
\APACaddressPublisher{}{Cambridge University Press}.
\newblock
\begin{APACrefDOI} \doi{10.1017/9781316681411} \end{APACrefDOI}
\PrintBackRefs{\CurrentBib}

\bibitem [\protect \citeauthoryear {%
Henry%
\ \protect \BOthers {.}}{%
Henry%
\ \protect \BOthers {.}}{%
{\protect \APACyear {2023}}%
}]{%
Henry2023}
\APACinsertmetastar {%
Henry2023}%
\begin{APACrefauthors}%
Henry, A\BPBI C\BPBI J.%
, Schannwell, C.%
, Vi{\v{s}}njevi{\'{c}}, V.%
, Millstein, J.%
, Bons, P\BPBI D.%
, Eisen, O.%
\BCBL {}\ \BBA {} Drews, R.%
\end{APACrefauthors}%
\unskip\
\newblock
\APACrefYearMonthDay{2023}{aug}{}.
\newblock
{\BBOQ}\APACrefatitle {Predicting the three-dimensional age-depth field of an
  ice rise} {Predicting the three-dimensional age-depth field of an ice
  rise}.{\BBCQ}
\newblock
\APACjournalVolNumPages{Authorea}{}{}{}.
\newblock
\begin{APACrefDOI} \doi{10.22541/essoar.169230234.44865946/v1} \end{APACrefDOI}
\PrintBackRefs{\CurrentBib}

\bibitem [\protect \citeauthoryear {%
Holschuh%
, Parizek%
, Alley%
\BCBL {}\ \BBA {} Anandakrishnan%
}{%
Holschuh%
\ \protect \BOthers {.}}{%
{\protect \APACyear {2017}}%
}]{%
Holschuh2017}
\APACinsertmetastar {%
Holschuh2017}%
\begin{APACrefauthors}%
Holschuh, N.%
, Parizek, B\BPBI R.%
, Alley, R\BPBI B.%
\BCBL {}\ \BBA {} Anandakrishnan, S.%
\end{APACrefauthors}%
\unskip\
\newblock
\APACrefYearMonthDay{2017}{}{}.
\newblock
{\BBOQ}\APACrefatitle {Decoding ice sheet behavior using englacial layer
  slopes} {Decoding ice sheet behavior using englacial layer slopes}.{\BBCQ}
\newblock
\APACjournalVolNumPages{Geophysical Research Letters}{44}{11}{5561-5570}.
\newblock
\begin{APACrefDOI} \doi{https://doi.org/10.1002/2017GL073417} \end{APACrefDOI}
\PrintBackRefs{\CurrentBib}

\bibitem [\protect \citeauthoryear {%
Howat%
, Porter%
, Smith%
, Noh%
\BCBL {}\ \BBA {} Morin%
}{%
Howat%
\ \protect \BOthers {.}}{%
{\protect \APACyear {2019}}%
}]{%
Howat2019}
\APACinsertmetastar {%
Howat2019}%
\begin{APACrefauthors}%
Howat, I\BPBI M.%
, Porter, C.%
, Smith, B\BPBI E.%
, Noh, M\BHBI J.%
\BCBL {}\ \BBA {} Morin, P.%
\end{APACrefauthors}%
\unskip\
\newblock
\APACrefYearMonthDay{2019}{}{}.
\newblock
{\BBOQ}\APACrefatitle {The Reference Elevation Model of Antarctica} {The
  reference elevation model of antarctica}.{\BBCQ}
\newblock
\APACjournalVolNumPages{The Cryosphere}{13}{2}{665--674}.
\newblock
\begin{APACrefDOI} \doi{10.5194/tc-13-665-2019} \end{APACrefDOI}
\PrintBackRefs{\CurrentBib}

\bibitem [\protect \citeauthoryear {%
Hubbard%
\ \protect \BOthers {.}}{%
Hubbard%
\ \protect \BOthers {.}}{%
{\protect \APACyear {2013}}%
}]{%
Hubbard2013}
\APACinsertmetastar {%
Hubbard2013}%
\begin{APACrefauthors}%
Hubbard, B.%
, Tison, J\BHBI L.%
, Philippe, M.%
, Heene, B.%
, Pattyn, F.%
, Malone, T.%
\BCBL {}\ \BBA {} Freitag, J.%
\end{APACrefauthors}%
\unskip\
\newblock
\APACrefYearMonthDay{2013}{}{}.
\newblock
{\BBOQ}\APACrefatitle {Ice shelf density reconstructed from optical televiewer
  borehole logging} {Ice shelf density reconstructed from optical televiewer
  borehole logging}.{\BBCQ}
\newblock
\APACjournalVolNumPages{Geophysical Research Letters}{40}{22}{5882-5887}.
\newblock
\begin{APACrefDOI} \doi{https://doi.org/10.1002/2013GL058023} \end{APACrefDOI}
\PrintBackRefs{\CurrentBib}

\bibitem [\protect \citeauthoryear {%
Hull%
\ \protect \BOthers {.}}{%
Hull%
\ \protect \BOthers {.}}{%
{\protect \APACyear {2022}}%
}]{%
Hull2022}
\APACinsertmetastar {%
Hull2022}%
\begin{APACrefauthors}%
Hull, R.%
, Leonarduzzi, E.%
, De~La~Fuente, L.%
, Tran, H\BPBI V.%
, Bennett, A.%
, Melchior, P.%
\BDBL {}Condon, L\BPBI E.%
\end{APACrefauthors}%
\unskip\
\newblock
\APACrefYearMonthDay{2022}{}{}.
\newblock
{\BBOQ}\APACrefatitle {Using simulation-based inference to determine the
  parameters of an integrated hydrologic model: a case study from the upper
  Colorado River basin} {Using simulation-based inference to determine the
  parameters of an integrated hydrologic model: a case study from the upper
  colorado river basin}.{\BBCQ}
\newblock
\APACjournalVolNumPages{Hydrology and Earth System Sciences
  Discussions}{2022}{}{1--38}.
\newblock
\begin{APACrefDOI} \doi{10.5194/hess-2022-345} \end{APACrefDOI}
\PrintBackRefs{\CurrentBib}

\bibitem [\protect \citeauthoryear {%
Kobyzev%
, Prince%
\BCBL {}\ \BBA {} Brubaker%
}{%
Kobyzev%
\ \protect \BOthers {.}}{%
{\protect \APACyear {2019}}%
}]{%
Kobyzev2019}
\APACinsertmetastar {%
Kobyzev2019}%
\begin{APACrefauthors}%
Kobyzev, I.%
, Prince, S\BPBI J.%
\BCBL {}\ \BBA {} Brubaker, M\BPBI A.%
\end{APACrefauthors}%
\unskip\
\newblock
\APACrefYearMonthDay{2019}{8}{}.
\newblock
{\BBOQ}\APACrefatitle {Normalizing Flows: An Introduction and Review of Current
  Methods} {Normalizing flows: An introduction and review of current
  methods}.{\BBCQ}
\newblock
\APACjournalVolNumPages{IEEE Transactions on Pattern Analysis and Machine
  Intelligence}{43}{}{3964-3979}.
\newblock
\begin{APACrefDOI} \doi{10.1109/tpami.2020.2992934} \end{APACrefDOI}
\PrintBackRefs{\CurrentBib}

\bibitem [\protect \citeauthoryear {%
Kr\"amer%
, Schmidt%
\BCBL {}\ \BBA {} Hennig%
}{%
Kr\"amer%
\ \protect \BOthers {.}}{%
{\protect \APACyear {2022}}%
}]{%
Kramer2022}
\APACinsertmetastar {%
Kramer2022}%
\begin{APACrefauthors}%
Kr\"amer, N.%
, Schmidt, J.%
\BCBL {}\ \BBA {} Hennig, P.%
\end{APACrefauthors}%
\unskip\
\newblock
\APACrefYearMonthDay{2022}{28--30 Mar}{}.
\newblock
{\BBOQ}\APACrefatitle {Probabilistic Numerical Method of Lines for
  Time-Dependent Partial Differential Equations} {Probabilistic numerical
  method of lines for time-dependent partial differential equations}.{\BBCQ}
\newblock
\BIn{} G.~Camps-Valls, F\BPBI J\BPBI R.~Ruiz\BCBL {}\ \BBA {} I.~Valera\
  (\BEDS), \APACrefbtitle {Proceedings of The 25th International Conference on
  Artificial Intelligence and Statistics} {Proceedings of the 25th
  international conference on artificial intelligence and statistics}\
  (\BVOL~151, \BPGS\ 625--639).
\newblock
\APACaddressPublisher{}{PMLR}.
\PrintBackRefs{\CurrentBib}

\bibitem [\protect \citeauthoryear {%
Lenaerts%
, Medley%
, van~den Broeke%
\BCBL {}\ \BBA {} Wouters%
}{%
Lenaerts%
\ \protect \BOthers {.}}{%
{\protect \APACyear {2019}}%
}]{%
Lenaerts2019}
\APACinsertmetastar {%
Lenaerts2019}%
\begin{APACrefauthors}%
Lenaerts, J\BPBI T\BPBI M.%
, Medley, B.%
, van~den Broeke, M\BPBI R.%
\BCBL {}\ \BBA {} Wouters, B.%
\end{APACrefauthors}%
\unskip\
\newblock
\APACrefYearMonthDay{2019}{}{}.
\newblock
{\BBOQ}\APACrefatitle {Observing and Modeling Ice Sheet Surface Mass Balance}
  {Observing and modeling ice sheet surface mass balance}.{\BBCQ}
\newblock
\APACjournalVolNumPages{Reviews of Geophysics}{57}{2}{376-420}.
\newblock
\begin{APACrefDOI} \doi{https://doi.org/10.1029/2018RG000622} \end{APACrefDOI}
\PrintBackRefs{\CurrentBib}

\bibitem [\protect \citeauthoryear {%
Lilien%
, Hills%
, Driscol%
, Jacobel%
\BCBL {}\ \BBA {} Christianson%
}{%
Lilien%
\ \protect \BOthers {.}}{%
{\protect \APACyear {2020}}%
}]{%
Lilien2020}
\APACinsertmetastar {%
Lilien2020}%
\begin{APACrefauthors}%
Lilien, D\BPBI A.%
, Hills, B\BPBI H.%
, Driscol, J.%
, Jacobel, R.%
\BCBL {}\ \BBA {} Christianson, K.%
\end{APACrefauthors}%
\unskip\
\newblock
\APACrefYearMonthDay{2020}{}{}.
\newblock
{\BBOQ}\APACrefatitle {ImpDAR: an open-source impulse radar processor} {Impdar:
  an open-source impulse radar processor}.{\BBCQ}
\newblock
\APACjournalVolNumPages{Annals of Glaciology}{61}{81}{114–123}.
\newblock
\begin{APACrefDOI} \doi{10.1017/aog.2020.44} \end{APACrefDOI}
\PrintBackRefs{\CurrentBib}

\bibitem [\protect \citeauthoryear {%
Linde%
, Renard%
, Mukerji%
\BCBL {}\ \BBA {} Caers%
}{%
Linde%
\ \protect \BOthers {.}}{%
{\protect \APACyear {2015}}%
}]{%
Linde2015}
\APACinsertmetastar {%
Linde2015}%
\begin{APACrefauthors}%
Linde, N.%
, Renard, P.%
, Mukerji, T.%
\BCBL {}\ \BBA {} Caers, J.%
\end{APACrefauthors}%
\unskip\
\newblock
\APACrefYearMonthDay{2015}{}{}.
\newblock
{\BBOQ}\APACrefatitle {Geological realism in hydrogeological and geophysical
  inverse modeling: A review} {Geological realism in hydrogeological and
  geophysical inverse modeling: A review}.{\BBCQ}
\newblock
\APACjournalVolNumPages{Advances in Water Resources}{86}{}{86-101}.
\newblock
\begin{APACrefDOI} \doi{https://doi.org/10.1016/j.advwatres.2015.09.019}
  \end{APACrefDOI}
\PrintBackRefs{\CurrentBib}

\bibitem [\protect \citeauthoryear {%
Looyenga%
}{%
Looyenga%
}{%
{\protect \APACyear {1965}}%
}]{%
Looyenga1965}
\APACinsertmetastar {%
Looyenga1965}%
\begin{APACrefauthors}%
Looyenga, H.%
\end{APACrefauthors}%
\unskip\
\newblock
\APACrefYearMonthDay{1965}{}{}.
\newblock
{\BBOQ}\APACrefatitle {Dielectric constants of heterogeneous mixtures}
  {Dielectric constants of heterogeneous mixtures}.{\BBCQ}
\newblock
\APACjournalVolNumPages{Physica}{31}{3}{401-406}.
\newblock
\begin{APACrefDOI} \doi{https://doi.org/10.1016/0031-8914(65)90045-5}
  \end{APACrefDOI}
\PrintBackRefs{\CurrentBib}

\bibitem [\protect \citeauthoryear {%
Lueckmann%
\ \protect \BOthers {.}}{%
Lueckmann%
\ \protect \BOthers {.}}{%
{\protect \APACyear {2021}}%
}]{%
Lueckmann2021}
\APACinsertmetastar {%
Lueckmann2021}%
\begin{APACrefauthors}%
Lueckmann, J\BHBI M.%
, Boelts, J.%
, Greenberg, D\BPBI S.%
, Gonçalves, P\BPBI J.%
, Macke, J\BPBI H.%
, Lueckmann, J\BHBI M.%
\BDBL {}Macke, J\BPBI H.%
\end{APACrefauthors}%
\unskip\
\newblock
\APACrefYearMonthDay{2021}{1}{}.
\newblock
{\BBOQ}\APACrefatitle {Benchmarking Simulation-Based Inference} {Benchmarking
  simulation-based inference}.{\BBCQ}
\newblock
\APACjournalVolNumPages{arXiv}{}{}{arXiv:2101.04653}.
\newblock
\begin{APACrefDOI} \doi{10.48550/ARXIV.2101.04653} \end{APACrefDOI}
\PrintBackRefs{\CurrentBib}

\bibitem [\protect \citeauthoryear {%
Lueckmann%
\ \protect \BOthers {.}}{%
Lueckmann%
\ \protect \BOthers {.}}{%
{\protect \APACyear {2017}}%
}]{%
Lueckmann2017}
\APACinsertmetastar {%
Lueckmann2017}%
\begin{APACrefauthors}%
Lueckmann, J\BHBI M.%
, Goncalves, P\BPBI J.%
, Bassetto, G.%
, \"{O}cal, K.%
, Nonnenmacher, M.%
\BCBL {}\ \BBA {} Macke, J\BPBI H.%
\end{APACrefauthors}%
\unskip\
\newblock
\APACrefYearMonthDay{2017}{}{}.
\newblock
{\BBOQ}\APACrefatitle {Flexible statistical inference for mechanistic models of
  neural dynamics} {Flexible statistical inference for mechanistic models of
  neural dynamics}.{\BBCQ}
\newblock
\BIn{} \APACrefbtitle {Advances in Neural Information Processing Systems}
  {Advances in neural information processing systems}\ (\BVOL~30).
\PrintBackRefs{\CurrentBib}

\bibitem [\protect \citeauthoryear {%
MacGregor%
\ \protect \BOthers {.}}{%
MacGregor%
\ \protect \BOthers {.}}{%
{\protect \APACyear {2009}}%
}]{%
Macgregor2009}
\APACinsertmetastar {%
Macgregor2009}%
\begin{APACrefauthors}%
MacGregor, J\BPBI A.%
, Matsuoka, K.%
, Koutnik, M\BPBI R.%
, Waddington, E\BPBI D.%
, Studinger, M.%
\BCBL {}\ \BBA {} Winebrenner, D\BPBI P.%
\end{APACrefauthors}%
\unskip\
\newblock
\APACrefYearMonthDay{2009}{}{}.
\newblock
{\BBOQ}\APACrefatitle {Millennially averaged accumulation rates for the Vostok
  Subglacial Lake region inferred from deep internal layers} {Millennially
  averaged accumulation rates for the vostok subglacial lake region inferred
  from deep internal layers}.{\BBCQ}
\newblock
\APACjournalVolNumPages{Annals of Glaciology}{50}{51}{25–34}.
\newblock
\begin{APACrefDOI} \doi{10.3189/172756409789097441} \end{APACrefDOI}
\PrintBackRefs{\CurrentBib}

\bibitem [\protect \citeauthoryear {%
Marsh%
\ \protect \BOthers {.}}{%
Marsh%
\ \protect \BOthers {.}}{%
{\protect \APACyear {2016}}%
}]{%
Marsh2016}
\APACinsertmetastar {%
Marsh2016}%
\begin{APACrefauthors}%
Marsh, O\BPBI J.%
, Fricker, H\BPBI A.%
, Siegfried, M\BPBI R.%
, Christianson, K.%
, Nicholls, K\BPBI W.%
, Corr, H\BPBI F\BPBI J.%
\BCBL {}\ \BBA {} Catania, G.%
\end{APACrefauthors}%
\unskip\
\newblock
\APACrefYearMonthDay{2016}{}{}.
\newblock
{\BBOQ}\APACrefatitle {High basal melting forming a channel at the grounding
  line of Ross Ice Shelf, Antarctica} {High basal melting forming a channel at
  the grounding line of ross ice shelf, antarctica}.{\BBCQ}
\newblock
\APACjournalVolNumPages{Geophysical Research Letters}{43}{1}{250-255}.
\newblock
\begin{APACrefDOI} \doi{https://doi.org/10.1002/2015GL066612} \end{APACrefDOI}
\PrintBackRefs{\CurrentBib}

\bibitem [\protect \citeauthoryear {%
Mengert%
}{%
Mengert%
}{%
{\protect \APACyear {2018}}%
}]{%
Mengert2018}
\APACinsertmetastar {%
Mengert2018}%
\begin{APACrefauthors}%
Mengert, M.%
\end{APACrefauthors}%
\unskip\
\newblock
\APACrefYear{2018}.
\unskip\
\newblock
\APACrefbtitle {Spatial and temporal variation of snow accumulation along
  Kottas traverse, Dronning Maud Land, East Antarctica} {Spatial and temporal
  variation of snow accumulation along kottas traverse, dronning maud land,
  east antarctica}\ \APACtypeAddressSchool {\BPhD}{}{Universit{\"a}t Bremen,
  Fachbereich Geowissenschaften}.
\unskip\
\newblock
\begin{APACrefURL} \url{https://epic.awi.de/id/eprint/49129/} \end{APACrefURL}
\PrintBackRefs{\CurrentBib}

\bibitem [\protect \citeauthoryear {%
Morland%
}{%
Morland%
}{%
{\protect \APACyear {1984}}%
}]{%
Morland2007}
\APACinsertmetastar {%
Morland2007}%
\begin{APACrefauthors}%
Morland, L\BPBI W.%
\end{APACrefauthors}%
\unskip\
\newblock
\APACrefYearMonthDay{1984}{}{}.
\newblock
{\BBOQ}\APACrefatitle {Thermomechanical balances of ice sheet flows}
  {Thermomechanical balances of ice sheet flows}.{\BBCQ}
\newblock
\APACjournalVolNumPages{Geophysical \& Astrophysical Fluid
  Dynamics}{29}{1-4}{237-266}.
\newblock
\begin{APACrefDOI} \doi{10.1080/03091928408248191} \end{APACrefDOI}
\PrintBackRefs{\CurrentBib}

\bibitem [\protect \citeauthoryear {%
Morlighem%
\ \protect \BOthers {.}}{%
Morlighem%
\ \protect \BOthers {.}}{%
{\protect \APACyear {2017}}%
}]{%
Morlighem2017}
\APACinsertmetastar {%
Morlighem2017}%
\begin{APACrefauthors}%
Morlighem, M.%
, Williams, C\BPBI N.%
, Rignot, E.%
, An, L.%
, Arndt, J\BPBI E.%
, Bamber, J\BPBI L.%
\BDBL {}Zinglersen, K\BPBI B.%
\end{APACrefauthors}%
\unskip\
\newblock
\APACrefYearMonthDay{2017}{nov}{}.
\newblock
{\BBOQ}\APACrefatitle {{BedMachine} v3: Complete Bed Topography and Ocean
  Bathymetry Mapping of Greenland From Multibeam Echo Sounding Combined With
  Mass Conservation} {{BedMachine} v3: Complete bed topography and ocean
  bathymetry mapping of greenland from multibeam echo sounding combined with
  mass conservation}.{\BBCQ}
\newblock
\APACjournalVolNumPages{Geophysical Research Letters}{44}{21}{11,051--11,061}.
\newblock
\begin{APACrefDOI} \doi{10.1002/2017gl074954} \end{APACrefDOI}
\PrintBackRefs{\CurrentBib}

\bibitem [\protect \citeauthoryear {%
Neckel%
, Drews%
, Rack%
\BCBL {}\ \BBA {} Steinhage%
}{%
Neckel%
\ \protect \BOthers {.}}{%
{\protect \APACyear {2012}}%
}]{%
Neckel2012}
\APACinsertmetastar {%
Neckel2012}%
\begin{APACrefauthors}%
Neckel, N.%
, Drews, R.%
, Rack, W.%
\BCBL {}\ \BBA {} Steinhage, D.%
\end{APACrefauthors}%
\unskip\
\newblock
\APACrefYearMonthDay{2012}{11}{}.
\newblock
{\BBOQ}\APACrefatitle {Basal melting at the Ekström Ice Shelf, Antarctica,
  estimated from mass flux divergence} {Basal melting at the ekström ice
  shelf, antarctica, estimated from mass flux divergence}.{\BBCQ}
\newblock
\APACjournalVolNumPages{Annals of Glaciology}{53}{}{294-302}.
\newblock
\begin{APACrefDOI} \doi{10.3189/2012AOG60A167} \end{APACrefDOI}
\PrintBackRefs{\CurrentBib}

\bibitem [\protect \citeauthoryear {%
Nicholls%
\ \protect \BOthers {.}}{%
Nicholls%
\ \protect \BOthers {.}}{%
{\protect \APACyear {2015}}%
}]{%
Nicholls2015}
\APACinsertmetastar {%
Nicholls2015}%
\begin{APACrefauthors}%
Nicholls, K\BPBI W.%
, Corr, H\BPBI F.%
, Stewart, C\BPBI L.%
, Lok, L\BPBI B.%
, Brennan, P\BPBI V.%
\BCBL {}\ \BBA {} Vaughan, D\BPBI G.%
\end{APACrefauthors}%
\unskip\
\newblock
\APACrefYearMonthDay{2015}{12}{}.
\newblock
{\BBOQ}\APACrefatitle {A ground-based radar for measuring vertical strain rates
  and time-varying basal melt rates in ice sheets and shelves} {A ground-based
  radar for measuring vertical strain rates and time-varying basal melt rates
  in ice sheets and shelves}.{\BBCQ}
\newblock
\APACjournalVolNumPages{Journal of Glaciology}{61}{}{1079-1087}.
\newblock
\begin{APACrefDOI} \doi{10.3189/2015JOG15J073} \end{APACrefDOI}
\PrintBackRefs{\CurrentBib}

\bibitem [\protect \citeauthoryear {%
Omagbon%
\ \protect \BOthers {.}}{%
Omagbon%
\ \protect \BOthers {.}}{%
{\protect \APACyear {2021}}%
}]{%
Omagbon2021}
\APACinsertmetastar {%
Omagbon2021}%
\begin{APACrefauthors}%
Omagbon, J.%
, Doherty, J.%
, Yeh, A.%
, Colina, R.%
, O'Sullivan, J.%
, McDowell, J.%
\BDBL {}O'Sullivan, M.%
\end{APACrefauthors}%
\unskip\
\newblock
\APACrefYearMonthDay{2021}{}{}.
\newblock
{\BBOQ}\APACrefatitle {Case studies of predictive uncertainty quantification
  for geothermal models} {Case studies of predictive uncertainty quantification
  for geothermal models}.{\BBCQ}
\newblock
\APACjournalVolNumPages{Geothermics}{97}{}{102263}.
\newblock
\begin{APACrefDOI} \doi{https://doi.org/10.1016/j.geothermics.2021.102263}
  \end{APACrefDOI}
\PrintBackRefs{\CurrentBib}

\bibitem [\protect \citeauthoryear {%
Overcast%
\ \protect \BOthers {.}}{%
Overcast%
\ \protect \BOthers {.}}{%
{\protect \APACyear {2021}}%
}]{%
Overcast2021}
\APACinsertmetastar {%
Overcast2021}%
\begin{APACrefauthors}%
Overcast, I.%
, Ruffley, M.%
, Rosindell, J.%
, Harmon, L.%
, Borges, P\BPBI A\BPBI V.%
, Emerson, B\BPBI C.%
\BDBL {}Rominger, A.%
\end{APACrefauthors}%
\unskip\
\newblock
\APACrefYearMonthDay{2021}{}{}.
\newblock
{\BBOQ}\APACrefatitle {A unified model of species abundance, genetic diversity,
  and functional diversity reveals the mechanisms structuring ecological
  communities} {A unified model of species abundance, genetic diversity, and
  functional diversity reveals the mechanisms structuring ecological
  communities}.{\BBCQ}
\newblock
\APACjournalVolNumPages{Molecular Ecology Resources}{21}{8}{2782-2800}.
\newblock
\begin{APACrefDOI} \doi{https://doi.org/10.1111/1755-0998.13514}
  \end{APACrefDOI}
\PrintBackRefs{\CurrentBib}

\bibitem [\protect \citeauthoryear {%
{Papamakarios}%
\ \BBA {} {Murray}%
}{%
{Papamakarios}%
\ \BBA {} {Murray}%
}{%
{\protect \APACyear {2016}}%
}]{%
Papamakarios2016}
\APACinsertmetastar {%
Papamakarios2016}%
\begin{APACrefauthors}%
{Papamakarios}, G.%
\BCBT {}\ \BBA {} {Murray}, I.%
\end{APACrefauthors}%
\unskip\
\newblock
\APACrefYearMonthDay{2016}{{\APACmonth{05}}}{}.
\newblock
{\BBOQ}\APACrefatitle {{Fast $\epsilon$-free Inference of Simulation Models
  with Bayesian Conditional Density Estimation}} {{Fast $\epsilon$-free
  Inference of Simulation Models with Bayesian Conditional Density
  Estimation}}.{\BBCQ}
\newblock
\APACjournalVolNumPages{arXiv e-prints}{}{}{arXiv:1605.06376}.
\newblock
\begin{APACrefDOI} \doi{10.48550/arXiv.1605.06376} \end{APACrefDOI}
\PrintBackRefs{\CurrentBib}

\bibitem [\protect \citeauthoryear {%
Papamakarios%
, Nalisnick%
, Rezende%
, Mohamed%
\BCBL {}\ \BBA {} Lakshminarayanan%
}{%
Papamakarios%
\ \protect \BOthers {.}}{%
{\protect \APACyear {2019}}%
}]{%
Papamakarios2019}
\APACinsertmetastar {%
Papamakarios2019}%
\begin{APACrefauthors}%
Papamakarios, G.%
, Nalisnick, E.%
, Rezende, D\BPBI J.%
, Mohamed, S.%
\BCBL {}\ \BBA {} Lakshminarayanan, B.%
\end{APACrefauthors}%
\unskip\
\newblock
\APACrefYearMonthDay{2019}{12}{}.
\newblock
{\BBOQ}\APACrefatitle {Normalizing Flows for Probabilistic Modeling and
  Inference} {Normalizing flows for probabilistic modeling and
  inference}.{\BBCQ}
\newblock
\APACjournalVolNumPages{Journal of Machine Learning Research}{22}{}{1-64}.
\newblock
\begin{APACrefDOI} \doi{10.48550/arxiv.1912.02762} \end{APACrefDOI}
\PrintBackRefs{\CurrentBib}

\bibitem [\protect \citeauthoryear {%
Pattyn%
, Favier%
, Sun%
\BCBL {}\ \BBA {} Durand%
}{%
Pattyn%
\ \protect \BOthers {.}}{%
{\protect \APACyear {2017}}%
}]{%
Pattyn2017}
\APACinsertmetastar {%
Pattyn2017}%
\begin{APACrefauthors}%
Pattyn, F.%
, Favier, L.%
, Sun, S.%
\BCBL {}\ \BBA {} Durand, G.%
\end{APACrefauthors}%
\unskip\
\newblock
\APACrefYearMonthDay{2017}{Sep}{01}.
\newblock
{\BBOQ}\APACrefatitle {Progress in Numerical Modeling of Antarctic Ice-Sheet
  Dynamics} {Progress in numerical modeling of antarctic ice-sheet
  dynamics}.{\BBCQ}
\newblock
\APACjournalVolNumPages{Current Climate Change Reports}{3}{3}{174-184}.
\newblock
\begin{APACrefDOI} \doi{10.1007/s40641-017-0069-7} \end{APACrefDOI}
\PrintBackRefs{\CurrentBib}

\bibitem [\protect \citeauthoryear {%
Pratap%
\ \protect \BOthers {.}}{%
Pratap%
\ \protect \BOthers {.}}{%
{\protect \APACyear {2022}}%
}]{%
Pratap2022}
\APACinsertmetastar {%
Pratap2022}%
\begin{APACrefauthors}%
Pratap, B.%
, Dey, R.%
, Matsuoka, K.%
, Moholdt, G.%
, Lindbäck, K.%
, Goel, V.%
\BDBL {}Thamban, M.%
\end{APACrefauthors}%
\unskip\
\newblock
\APACrefYearMonthDay{2022}{2}{}.
\newblock
{\BBOQ}\APACrefatitle {Three-decade spatial patterns in surface mass balance of
  the Nivlisen Ice Shelf, central Dronning Maud Land, East Antarctica}
  {Three-decade spatial patterns in surface mass balance of the nivlisen ice
  shelf, central dronning maud land, east antarctica}.{\BBCQ}
\newblock
\APACjournalVolNumPages{Journal of Glaciology}{68}{}{174-186}.
\newblock
\begin{APACrefDOI} \doi{10.1017/JOG.2021.93} \end{APACrefDOI}
\PrintBackRefs{\CurrentBib}

\bibitem [\protect \citeauthoryear {%
Ramesh%
\ \protect \BOthers {.}}{%
Ramesh%
\ \protect \BOthers {.}}{%
{\protect \APACyear {2022}}%
}]{%
Ramesh2022}
\APACinsertmetastar {%
Ramesh2022}%
\begin{APACrefauthors}%
Ramesh, P.%
, Lueckmann, J\BHBI M.%
, Boelts, J.%
, Tejero-Cantero, {\'A}.%
, Greenberg, D\BPBI S.%
, Goncalves, P\BPBI J.%
\BCBL {}\ \BBA {} Macke, J\BPBI H.%
\end{APACrefauthors}%
\unskip\
\newblock
\APACrefYearMonthDay{2022}{}{}.
\newblock
{\BBOQ}\APACrefatitle {{GATSBI}: Generative Adversarial Training for
  Simulation-Based Inference} {{GATSBI}: Generative adversarial training for
  simulation-based inference}.{\BBCQ}
\newblock
\BIn{} \APACrefbtitle {International Conference on Learning Representations.}
  {International conference on learning representations.}
\PrintBackRefs{\CurrentBib}

\bibitem [\protect \citeauthoryear {%
Rasmussen%
\ \BBA {} Williams%
}{%
Rasmussen%
\ \BBA {} Williams%
}{%
{\protect \APACyear {2005}}%
}]{%
RasmussenGPs}
\APACinsertmetastar {%
RasmussenGPs}%
\begin{APACrefauthors}%
Rasmussen, C\BPBI E.%
\BCBT {}\ \BBA {} Williams, C\BPBI K\BPBI I.%
\end{APACrefauthors}%
\unskip\
\newblock
\APACrefYear{2005}.
\newblock
\APACrefbtitle {{Gaussian Processes for Machine Learning}} {{Gaussian Processes
  for Machine Learning}}.
\newblock
\APACaddressPublisher{}{The MIT Press}.
\newblock
\begin{APACrefDOI} \doi{10.7551/mitpress/3206.001.0001} \end{APACrefDOI}
\PrintBackRefs{\CurrentBib}

\bibitem [\protect \citeauthoryear {%
Reese%
, Gudmundsson%
, Levermann%
\BCBL {}\ \BBA {} Winkelmann%
}{%
Reese%
\ \protect \BOthers {.}}{%
{\protect \APACyear {2017}}%
}]{%
Reese2017}
\APACinsertmetastar {%
Reese2017}%
\begin{APACrefauthors}%
Reese, R.%
, Gudmundsson, G\BPBI H.%
, Levermann, A.%
\BCBL {}\ \BBA {} Winkelmann, R.%
\end{APACrefauthors}%
\unskip\
\newblock
\APACrefYearMonthDay{2017}{12}{}.
\newblock
{\BBOQ}\APACrefatitle {The far reach of ice-shelf thinning in Antarctica} {The
  far reach of ice-shelf thinning in antarctica}.{\BBCQ}
\newblock
\APACjournalVolNumPages{Nature Climate Change 2017 8:1}{8}{}{53-57}.
\newblock
\begin{APACrefDOI} \doi{10.1038/s41558-017-0020-x} \end{APACrefDOI}
\PrintBackRefs{\CurrentBib}

\bibitem [\protect \citeauthoryear {%
Rubin%
}{%
Rubin%
}{%
{\protect \APACyear {1984}}%
}]{%
Rubin1984}
\APACinsertmetastar {%
Rubin1984}%
\begin{APACrefauthors}%
Rubin, D\BPBI B.%
\end{APACrefauthors}%
\unskip\
\newblock
\APACrefYearMonthDay{1984}{}{}.
\newblock
{\BBOQ}\APACrefatitle {Bayesianly Justifiable and Relevant Frequency
  Calculations for the Applied Statistician} {Bayesianly justifiable and
  relevant frequency calculations for the applied statistician}.{\BBCQ}
\newblock
\APACjournalVolNumPages{The Annals of Statistics}{12}{4}{1151--1172}.
\PrintBackRefs{\CurrentBib}

\bibitem [\protect \citeauthoryear {%
Sandip%
, R\"ass%
\BCBL {}\ \BBA {} Morlighem%
}{%
Sandip%
\ \protect \BOthers {.}}{%
{\protect \APACyear {2023}}%
}]{%
Sandip2023}
\APACinsertmetastar {%
Sandip2023}%
\begin{APACrefauthors}%
Sandip, A.%
, R\"ass, L.%
\BCBL {}\ \BBA {} Morlighem, M.%
\end{APACrefauthors}%
\unskip\
\newblock
\APACrefYearMonthDay{2023}{}{}.
\newblock
{\BBOQ}\APACrefatitle {Graphics processing unit accelerated ice flow solver for
  unstructured meshes using the Shallow Shelf Approximation (FastIceFlo v1.0)}
  {Graphics processing unit accelerated ice flow solver for unstructured meshes
  using the shallow shelf approximation (fasticeflo v1.0)}.{\BBCQ}
\newblock
\APACjournalVolNumPages{Geoscientific Model Development
  Discussions}{2023}{}{1--17}.
\newblock
\begin{APACrefDOI} \doi{10.5194/gmd-2023-32} \end{APACrefDOI}
\PrintBackRefs{\CurrentBib}

\bibitem [\protect \citeauthoryear {%
Schannwell%
\ \protect \BOthers {.}}{%
Schannwell%
\ \protect \BOthers {.}}{%
{\protect \APACyear {2019}}%
}]{%
Schanwell2019}
\APACinsertmetastar {%
Schanwell2019}%
\begin{APACrefauthors}%
Schannwell, C.%
, Drews, R.%
, Ehlers, T\BPBI A.%
, Eisen, O.%
, Mayer, C.%
\BCBL {}\ \BBA {} Gillet-Chaulet, F.%
\end{APACrefauthors}%
\unskip\
\newblock
\APACrefYearMonthDay{2019}{}{}.
\newblock
{\BBOQ}\APACrefatitle {Kinematic response of ice-rise divides to changes in
  ocean and atmosphere forcing} {Kinematic response of ice-rise divides to
  changes in ocean and atmosphere forcing}.{\BBCQ}
\newblock
\APACjournalVolNumPages{The Cryosphere}{13}{10}{2673--2691}.
\newblock
\begin{APACrefDOI} \doi{10.5194/tc-13-2673-2019} \end{APACrefDOI}
\PrintBackRefs{\CurrentBib}

\bibitem [\protect \citeauthoryear {%
Shapero%
, Badgeley%
, Hoffman%
\BCBL {}\ \BBA {} Joughin%
}{%
Shapero%
\ \protect \BOthers {.}}{%
{\protect \APACyear {2021}}%
}]{%
icepack}
\APACinsertmetastar {%
icepack}%
\begin{APACrefauthors}%
Shapero, D\BPBI R.%
, Badgeley, J\BPBI A.%
, Hoffman, A\BPBI O.%
\BCBL {}\ \BBA {} Joughin, I\BPBI R.%
\end{APACrefauthors}%
\unskip\
\newblock
\APACrefYearMonthDay{2021}{}{}.
\newblock
{\BBOQ}\APACrefatitle {icepack: a new glacier flow modeling package in Python,
  version 1.0} {icepack: a new glacier flow modeling package in python, version
  1.0}.{\BBCQ}
\newblock
\APACjournalVolNumPages{Geoscientific Model Development}{14}{7}{4593--4616}.
\newblock
\begin{APACrefDOI} \doi{10.5194/gmd-14-4593-2021} \end{APACrefDOI}
\PrintBackRefs{\CurrentBib}

\bibitem [\protect \citeauthoryear {%
{Sisson}%
, {Fan}%
\BCBL {}\ \BBA {} {Beaumont}%
}{%
{Sisson}%
\ \protect \BOthers {.}}{%
{\protect \APACyear {2018}}%
}]{%
Sisson2018}
\APACinsertmetastar {%
Sisson2018}%
\begin{APACrefauthors}%
{Sisson}, S\BPBI A.%
, {Fan}, Y.%
\BCBL {}\ \BBA {} {Beaumont}, M\BPBI A.%
\end{APACrefauthors}%
\unskip\
\newblock
\APACrefYearMonthDay{2018}{{\APACmonth{02}}}{}.
\newblock
{\BBOQ}\APACrefatitle {{Overview of Approximate Bayesian Computation}}
  {{Overview of Approximate Bayesian Computation}}.{\BBCQ}
\newblock
\APACjournalVolNumPages{arXiv e-prints}{}{}{arXiv:1802.09720}.
\newblock
\begin{APACrefDOI} \doi{10.48550/arXiv.1802.09720} \end{APACrefDOI}
\PrintBackRefs{\CurrentBib}

\bibitem [\protect \citeauthoryear {%
Steen-Larsen%
, Waddington%
\BCBL {}\ \BBA {} Koutnik%
}{%
Steen-Larsen%
\ \protect \BOthers {.}}{%
{\protect \APACyear {2010}}%
}]{%
Steen-Larsen2010}
\APACinsertmetastar {%
Steen-Larsen2010}%
\begin{APACrefauthors}%
Steen-Larsen, H\BPBI C.%
, Waddington, E\BPBI D.%
\BCBL {}\ \BBA {} Koutnik, M\BPBI R.%
\end{APACrefauthors}%
\unskip\
\newblock
\APACrefYearMonthDay{2010}{}{}.
\newblock
{\BBOQ}\APACrefatitle {Formulating an inverse problem to infer the
  accumulation-rate pattern from deep internal layering in an ice sheet using a
  Monte Carlo approach} {Formulating an inverse problem to infer the
  accumulation-rate pattern from deep internal layering in an ice sheet using a
  monte carlo approach}.{\BBCQ}
\newblock
\APACjournalVolNumPages{Journal of Glaciology}{56}{196}{318–332}.
\newblock
\begin{APACrefDOI} \doi{10.3189/002214310791968476} \end{APACrefDOI}
\PrintBackRefs{\CurrentBib}

\bibitem [\protect \citeauthoryear {%
Stordal%
, Karlsen%
, N{\ae}vdal%
, Skaug%
\BCBL {}\ \BBA {} Vall{\`e}s%
}{%
Stordal%
\ \protect \BOthers {.}}{%
{\protect \APACyear {2011}}%
}]{%
Stordal2011}
\APACinsertmetastar {%
Stordal2011}%
\begin{APACrefauthors}%
Stordal, A\BPBI S.%
, Karlsen, H\BPBI A.%
, N{\ae}vdal, G.%
, Skaug, H\BPBI J.%
\BCBL {}\ \BBA {} Vall{\`e}s, B.%
\end{APACrefauthors}%
\unskip\
\newblock
\APACrefYearMonthDay{2011}{Mar}{01}.
\newblock
{\BBOQ}\APACrefatitle {Bridging the ensemble Kalman filter and particle
  filters: the adaptive Gaussian mixture filter} {Bridging the ensemble kalman
  filter and particle filters: the adaptive gaussian mixture filter}.{\BBCQ}
\newblock
\APACjournalVolNumPages{Computational Geosciences}{15}{2}{293-305}.
\newblock
\begin{APACrefDOI} \doi{10.1007/s10596-010-9207-1} \end{APACrefDOI}
\PrintBackRefs{\CurrentBib}

\bibitem [\protect \citeauthoryear {%
Sun%
\ \protect \BOthers {.}}{%
Sun%
\ \protect \BOthers {.}}{%
{\protect \APACyear {2019}}%
}]{%
Sun2019}
\APACinsertmetastar {%
Sun2019}%
\begin{APACrefauthors}%
Sun, S.%
, Hattermann, T.%
, Pattyn, F.%
, Nicholls, K\BPBI W.%
, Drews, R.%
\BCBL {}\ \BBA {} Berger, S.%
\end{APACrefauthors}%
\unskip\
\newblock
\APACrefYearMonthDay{2019}{8}{}.
\newblock
{\BBOQ}\APACrefatitle {Topographic Shelf Waves Control Seasonal Melting Near
  Antarctic Ice Shelf Grounding Lines} {Topographic shelf waves control
  seasonal melting near antarctic ice shelf grounding lines}.{\BBCQ}
\newblock
\APACjournalVolNumPages{Geophysical Research Letters}{46}{}{9824-9832}.
\newblock
\begin{APACrefDOI} \doi{10.1029/2019GL083881} \end{APACrefDOI}
\PrintBackRefs{\CurrentBib}

\bibitem [\protect \citeauthoryear {%
Sutter%
, Fischer%
\BCBL {}\ \BBA {} Eisen%
}{%
Sutter%
\ \protect \BOthers {.}}{%
{\protect \APACyear {2021}}%
}]{%
Sutter2021}
\APACinsertmetastar {%
Sutter2021}%
\begin{APACrefauthors}%
Sutter, J.%
, Fischer, H.%
\BCBL {}\ \BBA {} Eisen, O.%
\end{APACrefauthors}%
\unskip\
\newblock
\APACrefYearMonthDay{2021}{8}{}.
\newblock
{\BBOQ}\APACrefatitle {Investigating the internal structure of the Antarctic
  ice sheet: The utility of isochrones for spatiotemporal ice-sheet model
  calibration} {Investigating the internal structure of the antarctic ice
  sheet: The utility of isochrones for spatiotemporal ice-sheet model
  calibration}.{\BBCQ}
\newblock
\APACjournalVolNumPages{Cryosphere}{15}{}{3839-3860}.
\newblock
\begin{APACrefDOI} \doi{10.5194/tc-15-3839-2021} \end{APACrefDOI}
\PrintBackRefs{\CurrentBib}

\bibitem [\protect \citeauthoryear {%
Symes%
}{%
Symes%
}{%
{\protect \APACyear {2009}}%
}]{%
Symes2009}
\APACinsertmetastar {%
Symes2009}%
\begin{APACrefauthors}%
Symes, W\BPBI W.%
\end{APACrefauthors}%
\unskip\
\newblock
\APACrefYearMonthDay{2009}{dec}{}.
\newblock
{\BBOQ}\APACrefatitle {The seismic reflection inverse problem} {The seismic
  reflection inverse problem}.{\BBCQ}
\newblock
\APACjournalVolNumPages{Inverse Problems}{25}{12}{123008}.
\newblock
\begin{APACrefDOI} \doi{10.1088/0266-5611/25/12/123008} \end{APACrefDOI}
\PrintBackRefs{\CurrentBib}

\bibitem [\protect \citeauthoryear {%
Tebaldi%
\ \BBA {} Sansó%
}{%
Tebaldi%
\ \BBA {} Sansó%
}{%
{\protect \APACyear {2008}}%
}]{%
Tebaldi}
\APACinsertmetastar {%
Tebaldi}%
\begin{APACrefauthors}%
Tebaldi, C.%
\BCBT {}\ \BBA {} Sansó, B.%
\end{APACrefauthors}%
\unskip\
\newblock
\APACrefYearMonthDay{2008}{07}{}.
\newblock
{\BBOQ}\APACrefatitle {{Joint Projections of Temperature and Precipitation
  Change from Multiple Climate Models: A Hierarchical Bayesian Approach}}
  {{Joint Projections of Temperature and Precipitation Change from Multiple
  Climate Models: A Hierarchical Bayesian Approach}}.{\BBCQ}
\newblock
\APACjournalVolNumPages{Journal of the Royal Statistical Society Series A:
  Statistics in Society}{172}{1}{83-106}.
\newblock
\begin{APACrefDOI} \doi{10.1111/j.1467-985X.2008.00545.x} \end{APACrefDOI}
\PrintBackRefs{\CurrentBib}

\bibitem [\protect \citeauthoryear {%
Tejero-Cantero%
\ \protect \BOthers {.}}{%
Tejero-Cantero%
\ \protect \BOthers {.}}{%
{\protect \APACyear {2020}}%
}]{%
Tejero-Cantero2020}
\APACinsertmetastar {%
Tejero-Cantero2020}%
\begin{APACrefauthors}%
Tejero-Cantero, A.%
, Boelts, J.%
, Deistler, M.%
, Lueckmann, J\BHBI M.%
, Durkan, C.%
, Gonçalves, P\BPBI J.%
\BDBL {}Macke, J\BPBI H.%
\end{APACrefauthors}%
\unskip\
\newblock
\APACrefYearMonthDay{2020}{}{}.
\newblock
{\BBOQ}\APACrefatitle {sbi: A toolkit for simulation-based inference} {sbi: A
  toolkit for simulation-based inference}.{\BBCQ}
\newblock
\APACjournalVolNumPages{Journal of Open Source Software}{5}{52}{2505}.
\newblock
\begin{APACrefDOI} \doi{10.21105/joss.02505} \end{APACrefDOI}
\PrintBackRefs{\CurrentBib}

\bibitem [\protect \citeauthoryear {%
Theofilopoulos%
\ \BBA {} Born%
}{%
Theofilopoulos%
\ \BBA {} Born%
}{%
{\protect \APACyear {2023}}%
}]{%
Theofilopoulos2023}
\APACinsertmetastar {%
Theofilopoulos2023}%
\begin{APACrefauthors}%
Theofilopoulos, A.%
\BCBT {}\ \BBA {} Born, A.%
\end{APACrefauthors}%
\unskip\
\newblock
\APACrefYearMonthDay{2023}{}{}.
\newblock
{\BBOQ}\APACrefatitle {Sensitivity of isochrones to surface mass balance and
  dynamics} {Sensitivity of isochrones to surface mass balance and
  dynamics}.{\BBCQ}
\newblock
\APACjournalVolNumPages{Journal of Glaciology}{69}{274}{311–323}.
\newblock
\begin{APACrefDOI} \doi{10.1017/jog.2022.62} \end{APACrefDOI}
\PrintBackRefs{\CurrentBib}

\bibitem [\protect \citeauthoryear {%
Ultee%
, Robel%
\BCBL {}\ \BBA {} Castruccio%
}{%
Ultee%
\ \protect \BOthers {.}}{%
{\protect \APACyear {2023}}%
}]{%
Ultee2023}
\APACinsertmetastar {%
Ultee2023}%
\begin{APACrefauthors}%
Ultee, L.%
, Robel, A\BPBI A.%
\BCBL {}\ \BBA {} Castruccio, S.%
\end{APACrefauthors}%
\unskip\
\newblock
\APACrefYearMonthDay{2023}{}{}.
\newblock
{\BBOQ}\APACrefatitle {A stochastic parameterization of ice sheet surface mass
  balance for the Stochastic Ice-Sheet and Sea-Level System Model (StISSM
  v1.0)} {A stochastic parameterization of ice sheet surface mass balance for
  the stochastic ice-sheet and sea-level system model (stissm v1.0)}.{\BBCQ}
\newblock
\APACjournalVolNumPages{EGUsphere}{2023}{}{1--19}.
\newblock
\begin{APACrefDOI} \doi{10.5194/egusphere-2023-635} \end{APACrefDOI}
\PrintBackRefs{\CurrentBib}

\bibitem [\protect \citeauthoryear {%
van Leeuwen%
, Künsch%
, Nerger%
, Potthast%
\BCBL {}\ \BBA {} Reich%
}{%
van Leeuwen%
\ \protect \BOthers {.}}{%
{\protect \APACyear {2019}}%
}]{%
VanLeeuwen2019}
\APACinsertmetastar {%
VanLeeuwen2019}%
\begin{APACrefauthors}%
van Leeuwen, P\BPBI J.%
, Künsch, H\BPBI R.%
, Nerger, L.%
, Potthast, R.%
\BCBL {}\ \BBA {} Reich, S.%
\end{APACrefauthors}%
\unskip\
\newblock
\APACrefYearMonthDay{2019}{}{}.
\newblock
{\BBOQ}\APACrefatitle {Particle filters for high-dimensional geoscience
  applications: A review} {Particle filters for high-dimensional geoscience
  applications: A review}.{\BBCQ}
\newblock
\APACjournalVolNumPages{Quarterly Journal of the Royal Meteorological
  Society}{145}{723}{2335-2365}.
\newblock
\begin{APACrefDOI} \doi{https://doi.org/10.1002/qj.3551} \end{APACrefDOI}
\PrintBackRefs{\CurrentBib}

\bibitem [\protect \citeauthoryear {%
van Wessem%
\ \protect \BOthers {.}}{%
van Wessem%
\ \protect \BOthers {.}}{%
{\protect \APACyear {2018}}%
}]{%
VanWessem2018}
\APACinsertmetastar {%
VanWessem2018}%
\begin{APACrefauthors}%
van Wessem, J\BPBI M.%
, van~de Berg, W\BPBI J.%
, No\"el, B\BPBI P\BPBI Y.%
, van Meijgaard, E.%
, Amory, C.%
, Birnbaum, G.%
\BDBL {}van~den Broeke, M\BPBI R.%
\end{APACrefauthors}%
\unskip\
\newblock
\APACrefYearMonthDay{2018}{}{}.
\newblock
{\BBOQ}\APACrefatitle {Modelling the climate and surface mass balance of polar
  ice sheets using RACMO2 -- Part 2: Antarctica (1979--2016)} {Modelling the
  climate and surface mass balance of polar ice sheets using racmo2 -- part 2:
  Antarctica (1979--2016)}.{\BBCQ}
\newblock
\APACjournalVolNumPages{The Cryosphere}{12}{4}{1479--1498}.
\newblock
\begin{APACrefDOI} \doi{10.5194/tc-12-1479-2018} \end{APACrefDOI}
\PrintBackRefs{\CurrentBib}

\bibitem [\protect \citeauthoryear {%
Vaňková%
\ \BBA {} Nicholls%
}{%
Vaňková%
\ \BBA {} Nicholls%
}{%
{\protect \APACyear {2022}}%
}]{%
Vankova2022}
\APACinsertmetastar {%
Vankova2022}%
\begin{APACrefauthors}%
Vaňková, I.%
\BCBT {}\ \BBA {} Nicholls, K\BPBI W.%
\end{APACrefauthors}%
\unskip\
\newblock
\APACrefYearMonthDay{2022}{10}{}.
\newblock
{\BBOQ}\APACrefatitle {Ocean Variability Beneath the Filchner-Ronne Ice Shelf
  Inferred From Basal Melt Rate Time Series} {Ocean variability beneath the
  filchner-ronne ice shelf inferred from basal melt rate time series}.{\BBCQ}
\newblock
\APACjournalVolNumPages{Journal of Geophysical Research:
  Oceans}{127}{}{e2022JC018879}.
\newblock
\begin{APACrefDOI} \doi{10.1029/2022JC018879} \end{APACrefDOI}
\PrintBackRefs{\CurrentBib}

\bibitem [\protect \citeauthoryear {%
Verjans%
, Robel%
, Seroussi%
, Ultee%
\BCBL {}\ \BBA {} Thompson%
}{%
Verjans%
\ \protect \BOthers {.}}{%
{\protect \APACyear {2022}}%
}]{%
Verjans2022}
\APACinsertmetastar {%
Verjans2022}%
\begin{APACrefauthors}%
Verjans, V.%
, Robel, A\BPBI A.%
, Seroussi, H.%
, Ultee, L.%
\BCBL {}\ \BBA {} Thompson, A\BPBI F.%
\end{APACrefauthors}%
\unskip\
\newblock
\APACrefYearMonthDay{2022}{}{}.
\newblock
{\BBOQ}\APACrefatitle {The Stochastic Ice-Sheet and Sea-Level System Model v1.0
  (StISSM v1.0)} {The stochastic ice-sheet and sea-level system model v1.0
  (stissm v1.0)}.{\BBCQ}
\newblock
\APACjournalVolNumPages{Geoscientific Model Development}{15}{22}{8269--8293}.
\newblock
\begin{APACrefDOI} \doi{10.5194/gmd-15-8269-2022} \end{APACrefDOI}
\PrintBackRefs{\CurrentBib}

\bibitem [\protect \citeauthoryear {%
Virtanen%
\ \protect \BOthers {.}}{%
Virtanen%
\ \protect \BOthers {.}}{%
{\protect \APACyear {2020}}%
}]{%
scipy}
\APACinsertmetastar {%
scipy}%
\begin{APACrefauthors}%
Virtanen, P.%
, Gommers, R.%
, Oliphant, T\BPBI E.%
, Haberland, M.%
, Reddy, T.%
, Cournapeau, D.%
\BDBL {}{SciPy 1.0 Contributors}%
\end{APACrefauthors}%
\unskip\
\newblock
\APACrefYearMonthDay{2020}{}{}.
\newblock
{\BBOQ}\APACrefatitle {{{SciPy} 1.0: Fundamental Algorithms for Scientific
  Computing in Python}} {{{SciPy} 1.0: Fundamental Algorithms for Scientific
  Computing in Python}}.{\BBCQ}
\newblock
\APACjournalVolNumPages{Nature Methods}{17}{}{261--272}.
\newblock
\begin{APACrefDOI} \doi{10.1038/s41592-019-0686-2} \end{APACrefDOI}
\PrintBackRefs{\CurrentBib}

\bibitem [\protect \citeauthoryear {%
Višnjević%
\ \protect \BOthers {.}}{%
Višnjević%
\ \protect \BOthers {.}}{%
{\protect \APACyear {2022}}%
}]{%
Visnjevic2022}
\APACinsertmetastar {%
Visnjevic2022}%
\begin{APACrefauthors}%
Višnjević, V.%
, Drews, R.%
, Schannwell, C.%
, Koch, I.%
, Franke, S.%
, Jansen, D.%
\BCBL {}\ \BBA {} Eisen, O.%
\end{APACrefauthors}%
\unskip\
\newblock
\APACrefYearMonthDay{2022}{}{}.
\newblock
{\BBOQ}\APACrefatitle {Predicting the steady-state isochronal stratigraphy of
  ice shelves using observations and modeling} {Predicting the steady-state
  isochronal stratigraphy of ice shelves using observations and
  modeling}.{\BBCQ}
\newblock
\APACjournalVolNumPages{The Cryosphere}{16}{}{4763-4777}.
\newblock
\begin{APACrefDOI} \doi{10.5194/tc-16-4763-2022} \end{APACrefDOI}
\PrintBackRefs{\CurrentBib}

\bibitem [\protect \citeauthoryear {%
Waddington%
, Neumann%
, Koutnik%
, Marshall%
\BCBL {}\ \BBA {} Morse%
}{%
Waddington%
\ \protect \BOthers {.}}{%
{\protect \APACyear {2007}}%
}]{%
Waddington2007}
\APACinsertmetastar {%
Waddington2007}%
\begin{APACrefauthors}%
Waddington, E\BPBI D.%
, Neumann, T\BPBI A.%
, Koutnik, M\BPBI R.%
, Marshall, H\BPBI P.%
\BCBL {}\ \BBA {} Morse, D\BPBI L.%
\end{APACrefauthors}%
\unskip\
\newblock
\APACrefYearMonthDay{2007}{12}{}.
\newblock
{\BBOQ}\APACrefatitle {Inference of accumulation-rate patterns from deep layers
  in glaciers and ice sheets} {Inference of accumulation-rate patterns from
  deep layers in glaciers and ice sheets}.{\BBCQ}
\newblock
\APACjournalVolNumPages{Journal of Glaciology}{53}{}{694-712}.
\newblock
\begin{APACrefDOI} \doi{10.3189/002214307784409351} \end{APACrefDOI}
\PrintBackRefs{\CurrentBib}

\bibitem [\protect \citeauthoryear {%
{Ward}%
, {Cannon}%
, {Beaumont}%
, {Fasiolo}%
\BCBL {}\ \BBA {} {Schmon}%
}{%
{Ward}%
\ \protect \BOthers {.}}{%
{\protect \APACyear {2022}}%
}]{%
Ward2022}
\APACinsertmetastar {%
Ward2022}%
\begin{APACrefauthors}%
{Ward}, D.%
, {Cannon}, P.%
, {Beaumont}, M.%
, {Fasiolo}, M.%
\BCBL {}\ \BBA {} {Schmon}, S\BPBI M.%
\end{APACrefauthors}%
\unskip\
\newblock
\APACrefYearMonthDay{2022}{{\APACmonth{10}}}{}.
\newblock
{\BBOQ}\APACrefatitle {{Robust Neural Posterior Estimation and Statistical
  Model Criticism}} {{Robust Neural Posterior Estimation and Statistical Model
  Criticism}}.{\BBCQ}
\newblock
\APACjournalVolNumPages{arXiv e-prints}{}{}{arXiv:2210.06564}.
\newblock
\begin{APACrefDOI} \doi{10.48550/arXiv.2210.06564} \end{APACrefDOI}
\PrintBackRefs{\CurrentBib}

\bibitem [\protect \citeauthoryear {%
Wesche%
\ \BBA {} Regnery%
}{%
Wesche%
\ \BBA {} Regnery%
}{%
{\protect \APACyear {2022}}%
}]{%
Wesche2022}
\APACinsertmetastar {%
Wesche2022}%
\begin{APACrefauthors}%
Wesche, C.%
\BCBT {}\ \BBA {} Regnery, J.%
\end{APACrefauthors}%
\unskip\
\newblock
\APACrefYearMonthDay{2022}{December}{}.
\newblock
\APACrefbtitle {Expeditions to Antarctica: ANT-Land 2021/22 Neumayer Station
  III, Kohnen Station, Flight Operations and Field Campaigns.} {Expeditions to
  antarctica: Ant-land 2021/22 neumayer station iii, kohnen station, flight
  operations and field campaigns.}
\newblock
\APACaddressPublisher{Bremerhaven}{Alfred-Wegener-Institut Helmholtz-Zentrum
  f{\"u}r Polar- und Meeresforschung}.
\newblock
\begin{APACrefDOI} \doi{10.57738/bzpm\_0767\_2022} \end{APACrefDOI}
\PrintBackRefs{\CurrentBib}

\bibitem [\protect \citeauthoryear {%
Wesche%
\ \protect \BOthers {.}}{%
Wesche%
\ \protect \BOthers {.}}{%
{\protect \APACyear {2016}}%
}]{%
Neumayer}
\APACinsertmetastar {%
Neumayer}%
\begin{APACrefauthors}%
Wesche, C.%
, Weller, R.%
, K{\"o}nig-Langlo, G.%
, Fromm, T.%
, Eckstaller, A.%
, Nixdorf, U.%
\BCBL {}\ \BBA {} Kohlberg, E.%
\end{APACrefauthors}%
\unskip\
\newblock
\APACrefYearMonthDay{2016}{}{}.
\newblock
{\BBOQ}\APACrefatitle {Neumayer III and Kohnen station in Antarctica operated
  by the Alfred Wegener Institute} {Neumayer iii and kohnen station in
  antarctica operated by the alfred wegener institute}.{\BBCQ}
\newblock
\APACjournalVolNumPages{Journal of large-scale research facilities
  JLSRF}{2}{}{A85--A85}.
\PrintBackRefs{\CurrentBib}

\bibitem [\protect \citeauthoryear {%
Winkelmann%
, Levermann%
, Martin%
\BCBL {}\ \BBA {} Frieler%
}{%
Winkelmann%
\ \protect \BOthers {.}}{%
{\protect \APACyear {2012}}%
}]{%
Winkelmann2012}
\APACinsertmetastar {%
Winkelmann2012}%
\begin{APACrefauthors}%
Winkelmann, R.%
, Levermann, A.%
, Martin, M\BPBI A.%
\BCBL {}\ \BBA {} Frieler, K.%
\end{APACrefauthors}%
\unskip\
\newblock
\APACrefYearMonthDay{2012}{Dec}{01}.
\newblock
{\BBOQ}\APACrefatitle {Increased future ice discharge from Antarctica owing to
  higher snowfall} {Increased future ice discharge from antarctica owing to
  higher snowfall}.{\BBCQ}
\newblock
\APACjournalVolNumPages{Nature}{492}{7428}{239-242}.
\newblock
\begin{APACrefDOI} \doi{10.1038/nature11616} \end{APACrefDOI}
\PrintBackRefs{\CurrentBib}

\bibitem [\protect \citeauthoryear {%
Wolovick%
, Moore%
\BCBL {}\ \BBA {} Zhao%
}{%
Wolovick%
\ \protect \BOthers {.}}{%
{\protect \APACyear {2021}}%
}]{%
Wolovick2021}
\APACinsertmetastar {%
Wolovick2021}%
\begin{APACrefauthors}%
Wolovick, M\BPBI J.%
, Moore, J\BPBI C.%
\BCBL {}\ \BBA {} Zhao, L.%
\end{APACrefauthors}%
\unskip\
\newblock
\APACrefYearMonthDay{2021}{}{}.
\newblock
{\BBOQ}\APACrefatitle {Joint Inversion for Surface Accumulation Rate and
  Geothermal Heat Flow From Ice-Penetrating Radar Observations at Dome A, East
  Antarctica. Part I: Model Description, Data Constraints, and Inversion
  Results} {Joint inversion for surface accumulation rate and geothermal heat
  flow from ice-penetrating radar observations at dome a, east antarctica. part
  i: Model description, data constraints, and inversion results}.{\BBCQ}
\newblock
\APACjournalVolNumPages{Journal of Geophysical Research: Earth
  Surface}{126}{5}{e2020JF005937}.
\newblock
\begin{APACrefDOI} \doi{https://doi.org/10.1029/2020JF005937} \end{APACrefDOI}
\PrintBackRefs{\CurrentBib}

\bibitem [\protect \citeauthoryear {%
Zeising%
\ \protect \BOthers {.}}{%
Zeising%
\ \protect \BOthers {.}}{%
{\protect \APACyear {2022}}%
}]{%
Zeising2022}
\APACinsertmetastar {%
Zeising2022}%
\begin{APACrefauthors}%
Zeising, O.%
, Steinhage, D.%
, Nicholls, K\BPBI W.%
, Corr, H\BPBI F\BPBI J.%
, Stewart, C\BPBI L.%
\BCBL {}\ \BBA {} Humbert, A.%
\end{APACrefauthors}%
\unskip\
\newblock
\APACrefYearMonthDay{2022}{}{}.
\newblock
{\BBOQ}\APACrefatitle {Basal melt of the southern Filchner Ice Shelf,
  Antarctica} {Basal melt of the southern filchner ice shelf,
  antarctica}.{\BBCQ}
\newblock
\APACjournalVolNumPages{The Cryosphere}{16}{4}{1469--1482}.
\newblock
\begin{APACrefDOI} \doi{10.5194/tc-16-1469-2022} \end{APACrefDOI}
\PrintBackRefs{\CurrentBib}

\bibitem [\protect \citeauthoryear {%
Zhang%
, Nawaz%
, Zhao%
\BCBL {}\ \BBA {} Curtis%
}{%
Zhang%
\ \protect \BOthers {.}}{%
{\protect \APACyear {2021}}%
}]{%
Zhang2021}
\APACinsertmetastar {%
Zhang2021}%
\begin{APACrefauthors}%
Zhang, X.%
, Nawaz, M\BPBI A.%
, Zhao, X.%
\BCBL {}\ \BBA {} Curtis, A.%
\end{APACrefauthors}%
\unskip\
\newblock
\APACrefYearMonthDay{2021}{}{}.
\newblock
{\BBOQ}\APACrefatitle {Chapter Two - An introduction to variational inference
  in geophysical inverse problems} {Chapter two - an introduction to
  variational inference in geophysical inverse problems}.{\BBCQ}
\newblock
\BIn{} C.~Schmelzbach\ (\BED), \APACrefbtitle {Inversion of Geophysical Data}
  {Inversion of geophysical data}\ (\BVOL~62, \BPG~73-140).
\newblock
\APACaddressPublisher{}{Elsevier}.
\newblock
\begin{APACrefDOI} \doi{https://doi.org/10.1016/bs.agph.2021.06.003}
  \end{APACrefDOI}
\PrintBackRefs{\CurrentBib}

\end{thebibliography}

\clearpage

\appendix

\section*{\LARGE{Appendix}}

\section{Forward Model Details}
The layer tracing scheme described in section \ref{sec: ice flow model} is equivalent to solving a set of advection equations for a set of layers. Here, we explicitly write down the advection equations solved and the boundary conditions defined. We formalize the advection equations by adding a term for the normal flow into the flow line. We then account for the inflow boundary condition at the grounding line $x=0$.

\subsection{Advection Equation} \label{sec: advection equation}
 A one-dimensional advection equation for a layer on a flow line reads
\begin{equation*}
    \frac{\partial h_{l}}{\partial t} = v_{x}\frac{\partial h_{l}}{\partial x}.
\end{equation*}
In practice, real flow lines have some incoming or outgoing (normal) flux, $q_{y}$, where $y$ denotes the horizontal direction perpendicular to $x$. We account for this normal ice flux and instead solve
\begin{equation*}
\frac{\partial h_{l}}{\partial t} = v_{x}\frac{\partial h_{l}}{\partial x} + r_{l}(x)\frac{\partial q_{y}(x)}{\partial y},
\label{eq: dQdy correction}
\end{equation*}
where $r_{l}(x) = h_{l}(x)/h(x)$ is the ratio of thickness of layer $l$ to the total thickness of the shelf. This equation holds due to the plug flow assumption, which has the corollary that the flux divergence $\partial q_{y}/\partial y$ is independent of the depth $z$. The quantity $\partial q_{y}/\partial y$ is constant for all layers, and independent of the layer thickness. This normal flux component accounts for lateral compression or extension of the flowtube centered on the flow line. Since we are along a flow line, $v_{y}\ll v_{x}$, where $v_{x}$ and $v_{y}$ are the characteristic velocities along flow and across flow respectively. Hence, it is a reasonable assumption that we are inferring the surface accumulation and basal melt rates upstream on the flowline, rather than at a different location not on the flowline.

\subsection{LMI Body}
\label{sec: LMI Body}

\begin{figure}[ht!]
    \centering
    \includegraphics[width=\textwidth]{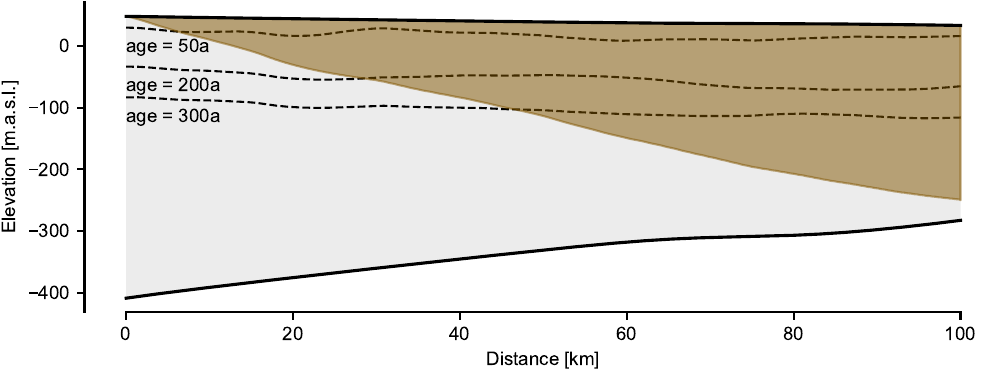}
    \caption{\textbf{Local Meteoric Ice (LMI) body.} The layer elevations in the LMI body (shaded region) are independent of the inflow boundary conditions. Outside the LMI body, the layer elevations are dependent on this boundary condition, and so using IRH observations within this region would require assuming the internal stratigraphy of the incoming ice.}
    \label{fig: LMI Body}
\end{figure}

In order to define the inflow boundary condition, we would need to know the relative thickness of the incoming layers (or alternatively, the vertical age distribution at $x_{0}$). This is typically not available in radar measurements of the stratigraphy. It is therefore important to use only the IRH elevation information within the \textit{Local Meteoric Ice (LMI) body} of the flow line. This is the region of our domain which is independent of the boundary condition we chose at $x_{0}$ (Fig. \ref{fig: LMI Body}). This region can be found by tracking the trajectory traced by a particle initially at $(x=x_{0},z=s(x_{0}))$. The LMI body is the region of the domain above this path. As a consequence of this consideration, we need to discard more of the IRH elevation data for the deeper IRHs in the dataset.

For a complete definition of the simulator, we still need to define some boundary condition at $x_{0}$. Since the true boundary condition is not known, we choose an inflow boundary condition which improves the numerical stability of the layer tracing model,

\begin{equation*}
\frac{\partial h_{i}}{\partial t} = \left.\frac{\partial q_{x}}{\partial x}\right\vert_{x=0} + r_{i}(x)\frac{\partial q_{y}(x)}{\partial y}.
\label{eq: dQdx boundaries}
\end{equation*}
This boundary condition makes the implicit assumption that near the inflow boundary, the layer thickness profile is a scalar multiple of the total thickness,
\begin{equation*}
h(x)\frac{\partial h_{i}(x)}{\partial x} = h_{i}(x)\frac{\partial h(x)}{\partial x}.
\label{eq: boundary dQdx}
\end{equation*} 
We define a similar boundary condition at $x=L_{x}$, however this boundary condition does not have a similar effect on the applicability of the IRH data.

\section{Calibrating the Simulator}
\label{sec: sim calibration}
We calibrate hyperparameters of the noise model using a set of 1000 simulations for the same ice shelf geometry, velocity, and mass balance parameters prior. This set of calibration simulations is not used again to train NPE (Sec. \ref{sec: sbi}), to avoid overfitting. We use the same set of calibration simulations to calibrate the per-layer LMI boundary mask (Sec. \ref{sec: inference definitions}).
\subsection{Noise Distribution}
\label{sec: noise calibration}

Algorithm \ref{alg: noise dist} defines our calibration procedure for the parameters of the noise distribution for Ekström Ice Shelf.
\SetKwComment{Comment}{\# }{}
\begin{algorithm}[ht]

 \textbf{Inputs:} (Noiseless) simulated layer elevations $\mathbf{e}_{l}^{(k)}(\mathbf{x})$ for $l=1,...L$, $k=1...,K_{\text{cal}}$ and simulation grid $\mathbf{x}$, IRH interpolated on the same simulation grid, $\mathbf{e}_{m}(\mathbf{x})$ \\
 \textbf{Outputs:} Mean and standard deviation $\mu_{\beta_{n}},\sigma_{\beta_{n}}$ defining noise model\\

\For{$k\leftarrow 1$ to $K_{\text{cal}}$}{
    $l^{*} \leftarrow \argmin\limits_{l} ||\mathbf{e}_{l}^{(k)}(\mathbf{x})-\mathbf{e}_{m}(\mathbf{x})||_{2}^{2}$   \Comment*[r]{best fitting layer to IRH}
    $\mathbf{\tilde{e}}^{(k)}_{l^{*}}(\mathbf{x}) \leftarrow \mathbf{e}_{l^{*}}^{(k)}(\mathbf{x}) - \texttt{lowpass\textunderscore filter}(\mathbf{e}_{l^{*}}^{(k)}(\mathbf{x}))$  \Comment*[r]{detrend best layer}
    $\mathbf{\tilde{e}}_{m}(\mathbf{x}) \leftarrow \mathbf{e}_{m}((\mathbf{x}) - \texttt{lowpass\textunderscore filter}(\mathbf{e}_{m}(\mathbf{x}))$  \Comment*[r]{detrend IRH}
    $\gamma_{1}^{(k)},...,\gamma_{n}^{(k)} \leftarrow \texttt{PSD}\left( \left[\mathbf{e}_{m}(\mathbf{x})- \mathbf{\tilde{e}}_{m}(\mathbf{x})\right] - \left[ \mathbf{e}_{l^{*}}^{(k)}(\mathbf{x}) - \mathbf{\tilde{e}}^{(k)}_{l^{*}}(\mathbf{x})\right]\right)$  \Comment*[r]{power spectral density of difference}
    $\beta_{1}^{(k)},...,\beta_{n}^{(k)} \leftarrow \log(\gamma_{1}^{(k)}),...,\log(\gamma_{n}^{(k)})$
}
\For{$n \leftarrow 1$ to $N$}{
    $\mu_{\beta_{n}} \leftarrow \frac{1}{K_{\text{cal}}}\sum_{k=1}^{K_{\text{cal}}} \beta_{n}^{(k)}$  \Comment*[r]{empirical mean of coefficients}
    $\sigma_{\beta_{n}} \leftarrow \sqrt{\frac{1}{K_{\text{cal}}-1}\sum_{k=1}^{K_{\text{cal}}}(\beta_{n}^{(k)} - \mu_{\beta_{n}})^{2}}$  \Comment*[r]{empirical standard deviation of coefficients}
}

\textbf{return} $\mu_{\beta_{1}},\sigma_{\beta_{1}},...,\mu_{\beta_{N}},\sigma_{\beta_{N}}$\\
\caption{Noise model calibration}
\label{alg: noise dist}
\end{algorithm}

\subsection{Boundary of LMI Body}
\label{sec: practical LMI}
In Sec. \ref{sec: inference} we used the LMI boundary $i(m)$ for each IRH $m$ in order to define the domain $\mathbf{x}_{i\geq i(m)}$ on which we compare the observed IRH data to the simulated isochronal layers. The physical interpretation of the LMI boundary is found in appendix \ref{sec: LMI Body} , and here we specify the exact definition of $i(m)$.

For each simulation $k$ in the calibration set, we define the trajectory of a particle starting at the surface of the inflow boundary $(x=0,z=s(0))$ by $p^{k}(t) = (x^{k}(t),z^{k}(t))$. In practice, since we are on a flow line, $v_x>0$ everywhere, and thus $x^{k}$ increases monotonically with $t$. Therefore the trajectory traces out a unique curve $z^{k}(x)$ in the domain.

Using this path we define the LMI boundary $i(m,k)$ for simulation $k$ and IRH $m$ to be the first point in the $x$ domain such that the path is below the IRH elevation,
\begin{equation*}
    i(m,k) = \min_{i} \{ i \hspace{1mm} : \hspace{1mm} z^{k}(x_{i}) < e_{m}(x_{i}) \}.
    \label{eq: LMIsim}
\end{equation*}
In the case that the path stays above the IRH for the entire domain, we define $i(m,k) = L_{x}$, meaning that the IRH is entirely outside the LMI body. 

This definition gives a different LMI body for each simulation. In order to perform inference, we require one a fixed LMI body boundary to use across all simulations. We define $i(m)$ "pessimistically" as the $75^{\text{th}}$ percentile of the calculated $i(m,k)$ boundaries in the calibration set. Therefore, some of the simulated layer elevations we use will be dependent on the unknown boundary conditions, but a small amount that should not affect the inferred results.

\section{Implementation Details}
For completeness we give the values of the important hyperparameters involved in the workflow - these are also explicit in the provided code repository.
\label{sec: preprocessing appendix}

\subsection{Ice Flow Model}
We provide the details of the spatial and temporal resolutions of our various simulations, along with the regularization strengths of the smoothing of the Ekström Ice Shelf data (Tables \ref{tab: synthetic setup},\ref{tab: ekstrom setup},\ref{tab: layer tracer setup}).

\begin{table}[ht]
    \centering
    \caption{Hyperparameters for synthetic ice shelf spin up modeling}
    \begin{tabular}{l c }
         \hline
         Parameter& Value\\
         \hline
         mesh resolution $(x,y)$& (310 \unit{m}, 250 \unit{m})\\
         spinup duration& 1000 years\\
         spinup timestep& 1.0 years\\
         Boundary Conditions& Dirichlet inflow and side boundaries\\
         \hline
    \end{tabular}
    \label{tab: synthetic setup}
\end{table}

\begin{table}[ht]
    \caption{Hyperparameters for the pre-processing of the data for Ekström Ice Shelf.}
    \centering
    \begin{tabular}{l c}
         \hline
         Parameter& Value\\
         \hline
         mesh resolution & 300 \unit{m}\\
         Thickness smoothness reg. penalty& 1000 \\
         Log fluidity reg. penalty& 1000\\
         Boundary Conditions& Dirichlet inflow and side boundaries\\
         \hline
    \end{tabular}
    \label{tab: ekstrom setup}
\end{table}

\begin{table}[ht]
    \caption{Layer tracer forward model simulation configuration}
    \centering
    \begin{tabular}{l c}
         \hline
         Parameter& Value\\
         \hline
         simulation time & 1000 years\\
         time step & 0.5 years \\
         \hline
    \end{tabular}

    \label{tab: layer tracer setup}
\end{table}

\subsection{Network Architecture and Training} \label{sec: architecture}
We applied Neural Posterior Estimation (NPE) as implemented in the \texttt{sbi} package \cite{Tejero-Cantero2020} to obtain the results for both the synthetic (Section \ref{sec: synthetic}) and Ekström (Section \ref{sec: Ekström}) ice shelves. We used the Neural Spline Flow (NSF) \cite{Durkan2019} as implemented in \citeA{Lueckmann2021}, and with the same network architecture as the NPE experiment. We also embedded the 500-dimensional observation of layer elevations to a 50-dimensional summary statistic used as the condition for the NSF. The embedding network consisted of two convolutional layers with kernel size 5, each followed by ReLU activations and max pooling with kernel size 2. The number of output channels for the two convolutional layers were 6 and 12 respectively. The output channels of the second convolutional layer were then concatenated and fed through two fully connected linear layers, each followed by ReLU activations. The number of hidden units was set to 50. Training was done as in \citeA{Lueckmann2021}, with the exception that the batch size was set to 1000 (default is 50). For each NPE run, we train 5 networks initialized with different random seeds, and report in our results the run with the best validation loss.

\section{Computational Costs}
\label{sec: computational cost}
We provide a breakdown of the approximate computational costs of the different stages in our workflow for both synthetic and Ekström Ice Shelves in Tables \ref{tab:synthetic costs} and \ref{tab:ekstrom costs} respectively. These are dependent the hardware used, and vary stochastically as a result of random number generators. This section is intended to provide intuition into the relative scales of the different stages of the workflow, rather than exact measurements. We had access to 16-core Intel Xeon Gold 2.9 GHz CPU nodes, and Nvidia RTX 2080ti GPU nodes. While the large number evaluations of the forward layer tracing model was by far the most computationally-intensive section of the workflow in both cases, these simulations were trivially performed in parallel across CPU cores, thus reducing the wall-clock time of the workflow.

This analysis highlights the advantages of our amortized approached to inference. For Ekström Ice Shelf, the total computational time of the noiseless simulations accounted for $\approx 99.8\%$ of the total computational time of the workflow. The simulations need not be repeated when we infer from other IRHs, greatly benefiting the computational efficiency of inference as the number of IRHs in the dataset increases. This advantage is slightly reduced when considering the parallelization we have used (Table \ref{tab:ekstrom costs}), as the training of the probability density estimator is not easily parallelizable across GPU nodes. Accounting for this still results in $\approx 97.6\%$ of the computational cost being amortized.

\begin{table}[!ht]
    \caption{Synthetic ice shelf approximate computational cost breakdown. Some tasks are embarassingly parallelizable - parallel resources and times are shown in square brackets. All times reported in minutes.}
    \centering
    \begin{tabular}{l c c}
        \hline
         Task& Node [parallel]& Time [parallel]\\
         \hline
         Spin up of ice shelf& CPU& 120\\
         1,000 calibration simulations& CPU [100 cores]& 1000 [10]\\
         189,000 noiseless simulations& CPU [100 cores]& $2\times 10^{5}$ [2000]\\
         \hline
         Noise and layer selection& CPU [100 cores]& 400 [20]\\
         Training NPE for 1 IRH& GPU& ~30\\
         \hline
     \multicolumn2l{Only the last two tasks need to be repeated for each IRH measurement.}
    \end{tabular}

    \label{tab:synthetic costs}
\end{table}

\begin{table}[!ht]
     \caption{Ekström Ice Shelf approximate computational cost breakdown. Some tasks are embarassingly parallelizable - parallel resources and times are shown in square brackets. All times reported in minutes.}
    \centering
    \begin{tabular}{l c c}
        \hline
         Task& Node [parallel]& Time [parallel]\\
         \hline
         Generating Mesh& CPU& $<1$\\
         Data preprocessing& CPU& 10\\
          1,000 calibration simulations& CPU [100 cores]& ~1000 [10]\\
         189,000 noiseless simulations& CPU [100 cores]& $2\times 10^{5}$ [2000]\\
         \hline
         Noise and layer selection& CPU [100 cores]& ~400 [20]\\
         Training NPE for 1 IRH& GPU& 30\\
         \hline
     \multicolumn2l{Only the last two tasks need to be repeated for each IRH measurement.}
    \end{tabular}
    \label{tab:ekstrom costs}
\end{table}

\section{Additional Results}
\label{sec: additional results}
\subsection{Synthetic Ice Shelf}
\begin{figure}[ht]
    \centering
    \includegraphics[width=\textwidth]{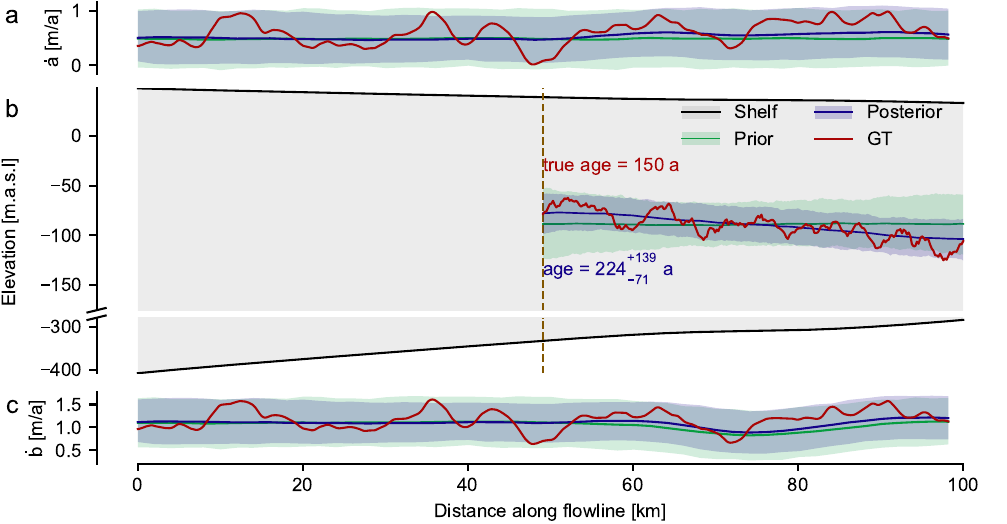}
    \caption{\textbf{Prior and posterior (predictive) for the synthetic dataset. a and c:} Prior and posterior over surface accumulation and basal melt rates respectively for layer 3 of the synthetic ice shelf, of age 150 years. Solid line is the distribution mean, the shaded region represents the 5th and 95th percentiles. The ground truth (GT) parameters used to generate the reference isochronoal layer are also shown. \textbf{b:} Cross section of the ice shelf. Prior and posterior predictive distributions for the layer closest matching the ground truth isochronal layer. The vertical dashed line represents the LMI boundary for this isochronal layer. The posterior predictive reconstructs the observed layer with higher accuracy and lower uncertainty. The posterior predictive distribution of the age of the isochronal layer is $224^{+139}_{-71}$ years.}
    \label{fig: Synthetic layer 3 posterior}
\end{figure}

\begin{figure}
    \centering
    \includegraphics[width=\textwidth]{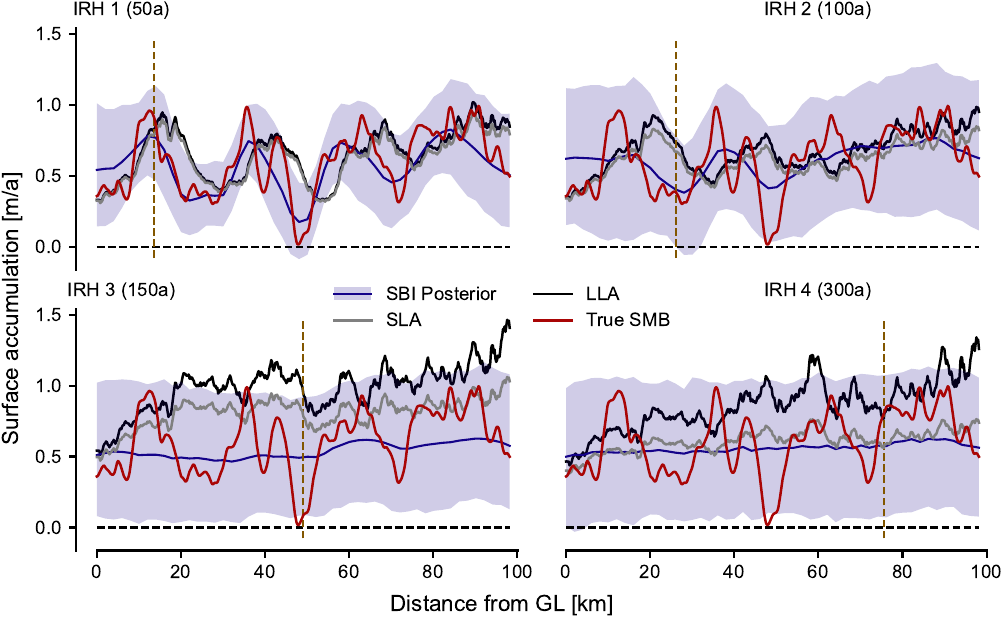}
    \caption{\textbf{Synthetic shelf:} dependence of posterior surface accumulation rate on depth of layer used for inference.}
    \label{fig:Synthetic posterior with layer depth}
\end{figure}

We show the posterior and posterior predictive distributions when inferring from isochronal layer 3 in the synthetic ice shelf dataset, of age 150 years (Fig. \ref{fig: Synthetic layer 3 posterior}. This isochronal layer has average depth of 120~\unit{m}, comparable to the deepest IRH in the Ekström dataset. While the uncertainty is much higher than for the shallow layer, the posterior over the surface accumulation still shows a higher mean than the prior in the downstream section of the ice shelf. Additionally, the posterior predictive reconstructs the ground truth isochronal layer better than the prior predictive. The mean root mean square error (RMSE) of the posterior predictive layers relative to the ground truth is 13.6~\unit{m}, compared to 19.8~\unit{m} for the prior. Therefore we are still able to reconstruct additional information about the mass balance parameters, even from much deeper layers.

We also explore the dependence of the inferred posterior over surface accumulation rate on the depth of the layer used for inference in the synthetic case (Fig. \ref{fig:Synthetic posterior with layer depth}), similar to the analysis done in Section \ref{sec: result comparison} for Ekström Ice Shelf. For the synthetic ice shelf, the surface accumulation and basal melt rates were held constant for the entire simulation time, and hence the increased uncertainty with depth seen in Fig. \ref{fig:Synthetic posterior with layer depth} highlights that information about the mass balance parameters is gradually lost with time as a result of the action of the simulator. Indeed, for the deepest ground truth isochronal layer of average depth 183~\unit{m}, the posterior distribution is almost identical to the prior distribution.

\subsection{Error in Predictive Distributions}
Here we report the Root Mean Square Error (RMSE) in the predicted isochronal layer elevations, relative to the true IRH (for Ekström Ice Shelf) or the ground truth isochronal layer (for the synthetic ice shelf). This is done across simulations from 1000 simulations for each of the prior and posterior distributions, for each IRH. The RMSE is consistently lower for the posterior predictive distribution for all depths, for both the synthetic ice shelf (Table \ref{tab: Synthetic RMSE}) and Ekström Ice Shelf (Table \ref{tab: Ekstrom RMSE}).

\begin{table}[!ht]
     \caption{Synthetic ice shelf - prior and posterior predictive distribution Root Mean Square Error (RMSE) relative to ground truth IRH, estimated from 1000 samples. The mean and standard deviations (SD) in the RMSE are reported. All values are in metres.}
    \centering
    \begin{tabular}{rrrrr}
    
        \hline
                      &     \multicolumn{2}{c}{prior}       &     \multicolumn{2}{c}{posterior}  \\
        \hline
           IRH number &    RMSE mean &    RMSE SD &   RMSE mean &  RMSE SD \\
        \hline
            1 &               11.5 &           3.9 &                    3.9 &               0.5 \\
            2 &               16.0 &           7.6 &                    7.3 &               1.2 \\
            3 &               19.8 &           8.4 &                   13.6 &               3.5 \\
            4 &               22.1 &           7.9 &                   19.8 &               6.8 \\
        \hline
    \end{tabular}
    \label{tab: Synthetic RMSE}

\end{table}

\begin{table}[!ht]
     \caption{Ekström Ice Shelf - prior and posterior predictive distribution Root Mean Square Error (RMSE) relative to ground truth IRH, estimated from 1000 samples. The mean and standard deviations (SD) in the RMSE are reported. All values are in metres.}
    \centering
    \begin{tabular}{rrrrr}
        \hline
                      &     \multicolumn{2}{c}{prior}       &     \multicolumn{2}{c}{posterior}  \\
        \hline
           IRH number &    RMSE mean &    RMSE SD &   RMSE mean &  RMSE SD \\
        \hline
            1 &                6.8 &           2.1 &                    3.0 &               0.9 \\
            2 &               11.8 &           3.4 &                    4.6 &               1.3 \\
            3 &               17.0 &           6.6 &                    6.8 &               1.4 \\
            4 &               16.4 &           7.6 &                   10.0 &               2.1 \\
        \hline
    \end{tabular}
    \label{tab: Ekstrom RMSE}

\end{table}

\section{Kottas Traverse Data}
\label{sec: Kottas data}

Here we describe the mapping of the surface accumulation measurements on Kottas traverse to the flow line transect. For each measurement year, and each location $\tilde{x}_{i}$ on the Kottas traverse, we find the nearest point $x_{i}$ on the flow line transect. We assume the accumulation rate at this point to be normally distributed, with mean $\tilde{\dot{a}}_{i}$ (the Kottas traverse measurement at $\tilde{x}_{i}$), and variance $\sigma_{\dot{a}}^{2}||\tilde{x}_{i}-x_{i}||^{2}_{2}/l_{\dot{a}} $.  We set the length scale $l_{\dot{a}} = 2.5$~\unit{km} and the accumulation rate variance $\sigma_{\dot{a}}^{2} = 0.25^{2}$~\unit{m^{2}}\unit{a^{-2}}. These values are chosen in accordance to the definition of the surface accumulation rate prior distribution (Sec. \ref{sec: prior choice}). 

\begin{figure}[ht]
    \centering
    \includegraphics[width=\textwidth]{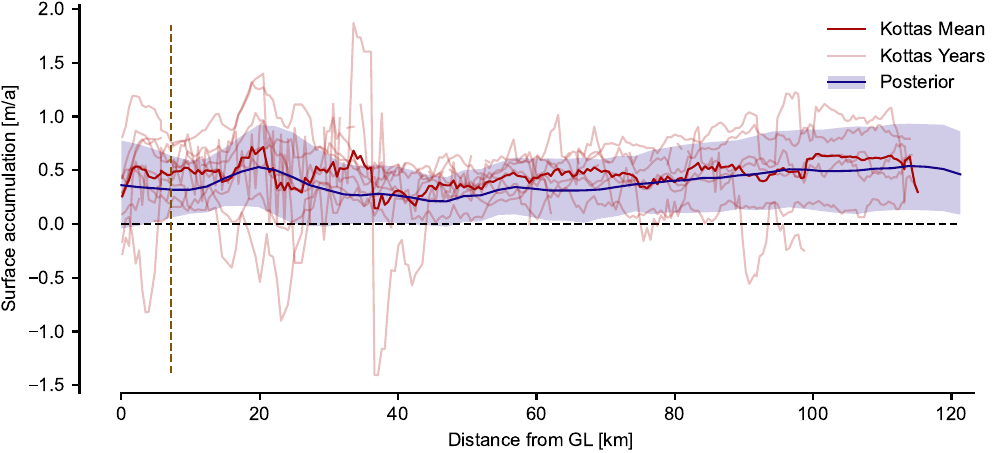}
    \caption{Yearly variations of the Kottas surface accumulation stakes measurement dataset.  Years shown are 1995 to 2005, and 2017 to 2019. These are compared with the posterior distribution inferred using IRH 1 of the Ekström IRH dataset.}
    \label{fig:Kottas Uncertainty}
\end{figure}

The yearly variations of the estimated surface accumulation as measured using the stake line along the Kottas traverse (Fig. \ref{fig:Kottas Uncertainty} reasonably agree with the posterior inferred using IRH 2 (Section \ref{sec: Ekström results}). These show that there is a high year-to-year variability in the surface accumulation (\citeA{Mengert2018}, Fig. 13) even in this steady-state region in East Antarctica. However, the mean of these yearly measurements matches the inferred posterior mean closely. 

\section{Synthetic Results with Miscalibrated Prior}
\label{sec: miscalibrated prior}
\begin{figure}[ht]
    \centering
    \includegraphics[width=\textwidth]{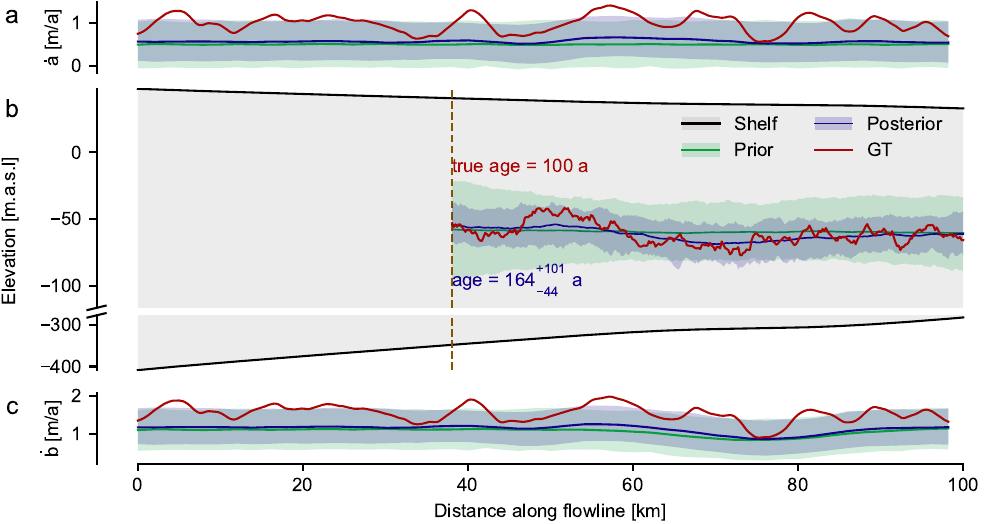}
    \caption{\textbf{Prior and posterior (predictive) for the synthetic ice shelf with the low-probability ground truth. a and c:} Prior and posterior over surface accumulation and basal melt rates respectively for an isochronal layer of age 100 years. Solid line is the distribution mean, the shaded region represents the 5th and 95th percentiles. The ground truth (GT) parameters used to generate the reference isochronal layer are also shown. \textbf{b:} Cross section of the ice shelf. Prior and posterior predictive distributions for the layer closest matching the ground truth isochronal layer. The vertical dashed line represents the LMI boundary for this isochronal layer. The posterior predictive reconstructs the observed layer with higher accuracy and lower uncertainty. The posterior predictive distribution of the age of the isochronal layer is $164^{+101}_{-44}$ years.}
    \label{fig: Synthetic miscalibrated}
\end{figure}

An additional outcome of our approach is the estimation of the age of isochronal layers. However, the validity of this estimate depends on the prior distribution containing the true mass balance parameters. When the true mass balance parameters have low probability under the prior distribution, the resulting estimate for the age of the isochronal layer can be wrong. We test this on the synthetic ice shelf by choosing a ground truth surface accumulation $\mathbf{\dot{a}_{\text{mis}}}$ that has low probability under the prior distribution. We calculate the isochronal layer of age 100 years under this ground truth and obtain the posterior distribution as before, using the same set of simulations as in the main text. The posterior does not capture that the mean surface accumulation rate should be higher than what is defined in the prior (Fig. \ref{fig: Synthetic miscalibrated}a). However, this is not a failure of the inference method, as we can see that the posterior predictive still reconstructs the ground truth isochronal layer at higher fidelity than the prior (Fig. \ref{fig: Synthetic miscalibrated}b). The predicted age of the isochronal layer $164^{+101}_{-44}$ years, which greatly overestimates the true age of 100 years.

\end{document}